\documentclass[fleqn,usenatbib]{mnras}

\usepackage[T1]{fontenc}
\usepackage{ae,aecompl}

\usepackage{graphicx}	
\usepackage{adjustbox}  
\usepackage{amsmath}	
\usepackage{amssymb}	
\usepackage{wasysym}    
\usepackage{bm}         
\usepackage{txfonts}
\usepackage{etoolbox}
\usepackage{float}
\usepackage{enumerate}
\makeatletter 
  \patchcmd{\NAT@citex}
    {\@citea\NAT@hyper@{%
      \NAT@nmfmt{\NAT@nm}%
      \hyper@natlinkbreak{\NAT@aysep\NAT@spacechar}{\@citeb\@extra@b@citeb}%
      \NAT@date}}
    {\@citea\NAT@nmfmt{\NAT@nm}%
    \NAT@aysep\NAT@spacechar\NAT@hyper@{\NAT@date}}{}{}

  \patchcmd{\NAT@citex}
    {\@citea\NAT@hyper@{%
      \NAT@nmfmt{\NAT@nm}%
      \hyper@natlinkbreak{\NAT@spacechar\NAT@@open\if*#1*\else#1\NAT@spacechar\fi}%
        {\@citeb\@extra@b@citeb}%
      \NAT@date}}
    {\@citea\NAT@nmfmt{\NAT@nm}%
    \NAT@spacechar\NAT@@open\if*#1*\else#1\NAT@spacechar\fi\NAT@hyper@{\NAT@date}}
    {}{}
\makeatother

\newcommand\orcid[1]{\href{http://orcid.org/#1}{\adjustbox{trim={-.15\width} {0\height} {-.15\width} {0\height},clip}{\includegraphics[height=12pt]{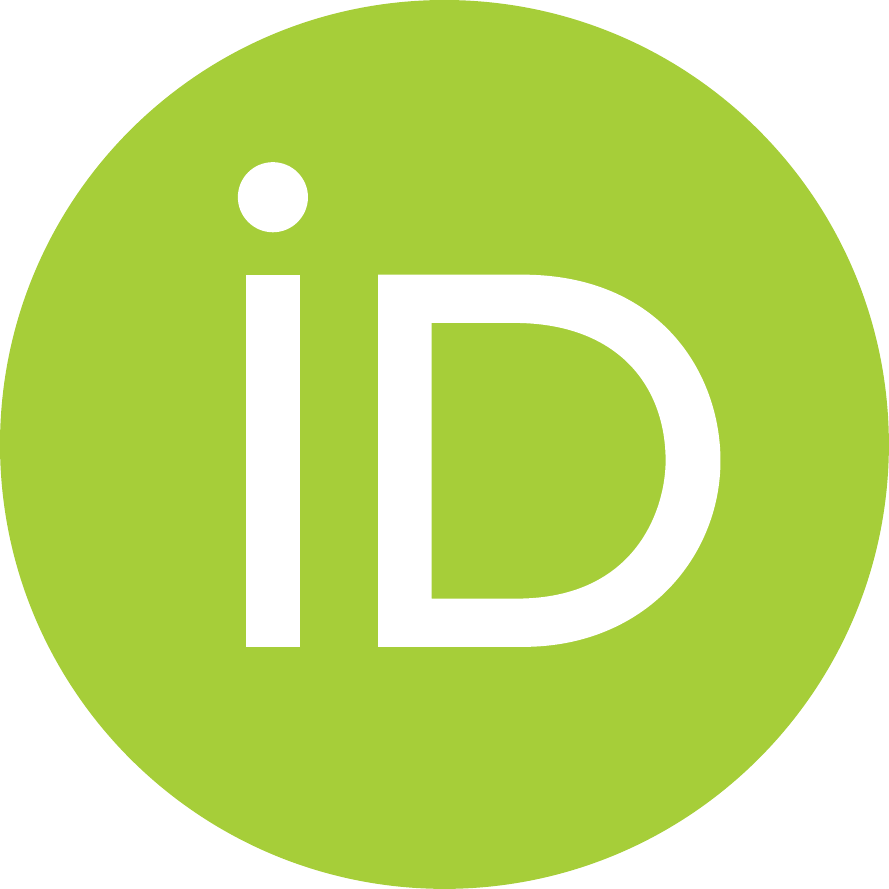}}}}

\title[Resonant-line radiative transfer within power-law density profiles]{Resonant-line radiative transfer within power-law density profiles}

\author[B.\ Lao \& A.\ Smith]{Bing-Xin~Lao$^{1}$\orcid{0000-0002-6320-185X} 
  and
  Aaron~Smith$^{2}$\orcid{0000-0002-2838-9033}\thanks{E-mail: \href{mailto:arsmith@mit.edu}{arsmith@mit.edu}}\thanks{NHFP Einstein Fellow}
  \\
  $^{1}$Department of Physics,
  University of Science and Technology of China, Hefei, 230026, China \\
  $^{2}$Department of Physics,
  Massachusetts Institute of Technology, Cambridge, MA 02139, USA
}

\date{Accepted 2020 July 23. Received 2020 July 21; in original form 2020 May 19}

\pubyear{2020}

\begin{document}
\label{firstpage}
\pagerange{\pageref{firstpage}--\pageref{lastpage}}
\maketitle

\begin{abstract}
  Star-forming regions in galaxies are surrounded by vast reservoirs of gas capable of both emitting and absorbing Lyman-alpha (Ly$\alpha$) radiation. Observations of Ly$\alpha$ emitters and spatially extended Ly$\alpha$ haloes indeed provide insights into the formation and evolution of galaxies. However, due to the complexity of resonant scattering, only a few analytic solutions are known in the literature. We discuss several idealized but physically motivated scenarios to extend the existing formalism to new analytic solutions, enabling quantitative predictions about the transport and diffusion of Ly$\alpha$ photons. This includes a closed form solution for the radiation field and derived quantities including the emergent flux, peak locations, energy density, average internal spectrum, number of scatters, outward force multiplier, trapping time, and characteristic radius. To verify our predictions, we employ a robust gridless Monte Carlo radiative transfer (GMCRT) method, which is straightforward to incorporate into existing ray-tracing codes but requires modifications to opacity-based calculations, including dynamical core-skipping acceleration schemes. We primarily focus on power-law density and emissivity profiles, however both the analytic and numerical methods can be generalized to other cases. Such studies provide additional intuition and understanding regarding the connection between the physical environments and observational signatures of galaxies throughout the Universe.
\end{abstract}

\begin{keywords}
line: profiles -- radiative transfer -- methods: analytical -- methods: numerical
\end{keywords}



\section{Introduction}
\label{sec:intro}
The Lyman-alpha (Ly$\alpha$) line of neutral hydrogen is an important probe of galaxy formation and evolution throughout cosmic history \citep{Partridge1967}. However, due to the complex nature of resonant scattering of Ly$\alpha$ photons in optically thick environments, the necessary radiative transfer modeling and interpretation of observations are often challenging \citep{Dijkstra2014}. Valuable insights into the physical mechanisms regulating Ly$\alpha$ escape can be obtained from back of the envelope calculations \citep{Osterbrock1962,Adams1972,Adams1975,HansenOh2006}. Furthermore, the fundamental physical processes are well studied and a few analytic solutions exist in the literature for idealized cases \citep{Harrington1973,Neufeld1990,LoebRybicki1999,Dijkstra2006,Tasitsiomi2006b,Higgins2012,GeWise2017,Smith2017,Smith2018,Seon2020}. The application of the Fokker-Planck approximation within the radiative transfer equation has played a central role in this analytic progress due to the simplification of local frequency diffusion in the wings of the line profile compared to a full treatment of partial redistribution \citep{Unno1952,Hummer1962,Rybicki1994}. So far, full solutions have been limited to homogeneous media in Cartesian or spherical geometries. In this work, we generalize the solutions by considering the physically motivated case of power-law density and emissivity profiles. We intend to be as thorough and general as possible while highlighting the most salient features of our results in the final summary and discussion section of this work.

The development of Monte Carlo radiative transfer (MCRT) codes with acceleration schemes has allowed for an accurate, universal approach to Ly$\alpha$ calculations \citep[e.g.][]{Auer1968,Ahn2002,Zheng2002}. The most common implementation is to perform ray-tracing assuming constant density within finite volume cells, especially when applied to analyze hydrodynamical simulations in post-processing \citep[e.g.][]{Tasitsiomi2006,Laursen2009,Verhamme2012}. However, in this paper we employ a gridless Monte Carlo radiative transfer (GMCRT) method for exact path integration rather than conforming to a discretized representation of space. In the case of a power-law profile, the gridless method makes our calculations more natural and accurate, which is desirable to ensure robustness despite the additional expense of computing the associated special functions. Furthermore, most Ly$\alpha$ MCRT codes employ a core-skipping technique to accelerate frequency diffusion into the wings of the Ly$\alpha$ profile and a peeling-off technique to construct surface brightness images, so we also provide the necessary modifications to retain this in GMCRT.

This paper is organized as follows. In Section~\ref{sec:Lyman-alpha radiation transport}, we briefly discuss the Ly$\alpha$ radiative transfer process and present the static partial differential equation in arbitrary coordinates. This provides the framework for the analytic calculations presented in the remaining sections, which focus on specific geometries and situations. In Section~\ref{sec:General solution for slab geometry}, we derive a complete solution of the transfer equation for the case of an optically-thick, static, isothermal slab with an arbitrary density and emissivity profiles. In Section~\ref{sec:Sol_for_spherical_geometry}, following the same process as the slab case, we derive analogous solutions for optically-thick, static, isothermal spheres with homogeneous density but with power-law emissivity profiles. We cannot provide a general expression for arbitrary density profiles in spherical geometry because the eigenfunction expansion is coupled to the spatial dependence of the absorption coefficient. The absence of a universal variable transformation to eliminate geometrical effects leads to significant differences between slab and spherical geometry solutions. However, the results are qualitatively similar so we present them together, illustrating the continuous transition between concentrated and extended emissivity-to-opacity configurations. In Section~\ref{sec:spherical-power-law}, we generalize the spherical geometry solutions to allow power-law profiles for both the emissivity and density. In Section~\ref{sec:Gridless Monte-Carlo Method}, we introduce our GMCRT method for gridless transport and its application to power-law density profiles. This enables us to numerically validate each of the new analytic results for resonant-line radiative transfer within power-law profiles. Finally, in Section~\ref{sec:summary}, we provide a summary and perspective on the utility and insights gained by pursuing the idealized models considered in this work.

\section{Resonant-line radiation transport}
\label{sec:Lyman-alpha radiation transport}
We now briefly introduce the problem setup and equations suitable for any coordinate system. The specific intensity $I_\nu(\bmath{r},\bmath{n},t)$ encodes all information about the radiation field taking into account the frequency $\nu$, spatial position $\bmath{r}$, propagation direction unit vector $\bmath{n}$, and time $t$. The general radiative transfer equation is given by:
\begin{equation} \label{eq:general}
  \frac{1}{c} \frac{\partial I_\nu}{\partial t} + \bmath{n} \bmath{\cdot} \bmath{\nabla} I_\nu = j_\nu - k_\nu I_\nu + \iint k_{\nu'} I_{\nu'} R_{\nu', \bmath{n}' \rightarrow \nu, \bmath{n}} \text{d}\Omega' \text{d}\nu' \, ,
\end{equation}
where $k_\nu$ and $j_\nu$ denote the absorption and emission coefficients, and the last term accounts for frequency redistribution due to partially coherent scattering \citep{Dijkstra2014}. The redistribution function $R$ is the differential probability per unit initial photon frequency $\nu'$ and per unit initial directional solid angle $\Omega'$ that the scattering of such a photon traveling in direction $\bmath{n}'$ would place the scattered photon at frequency $\nu$ and directional unit vector $\bmath{n}$. It is convenient to convert to the dimensionless frequency
\begin{equation}
    x \equiv \frac{\nu - \nu_0}{\Delta \nu_\text{D}} \, ,
\end{equation}
where $\nu_0$ denotes the frequency at line centre, $\Delta \nu_\text{D} \equiv (v_\text{th}/c)\nu_0$ the Doppler width of the profile, and $v_\text{th} \equiv (2 k_\text{B} T / m_\text{H})^{1/2}$ the thermal velocity. The frequency dependence of the absorption coefficient is given by the Voigt profile $\phi_\text{Voigt}$. For convenience we define the Hjerting-Voigt function $H(a,x) = \sqrt{\pi} \Delta \nu_\text{D} \phi_\text{Voigt}(\nu)$ as the dimensionless convolution of Lorentzian and Maxwellian distributions,
\begin{equation} \label{eq:H}
  H(a,x) = \frac{a}{\pi} \int_{-\infty}^\infty \frac{e^{-y^2}\text{d}y}{a^2+(y-x)^2} \approx
    \begin{cases}
      e^{-x^2} & \quad \text{`core'} \\
      {\displaystyle \frac{a}{\sqrt{\pi} x^2} } & \quad \text{`wing'}
    \end{cases} \, .
\end{equation}
The `damping parameter', $a \equiv \Delta \nu_L /2 \Delta \nu_D$, describes the relative broadening compared to the natural line width $\Delta \nu_\text{L}$. In isothermal gas, $a$ is simply a parameter representing the temperature.

\subsection{Scaling relations}
Before proceeding further, we review some important scaling relations to provide context for the results that follow. For a spherical cloud of radius $R$ and line centre optical depth $\tau_0$, the escape of resonance photons in extremely optically thick media ($a \tau_0 \gtrsim 10^3$) can be thought of as a diffusion process in both space and frequency \citep{Adams1972}. The resonant scattering is characterized by random walks in the wings of the line profile, with an average drift back to the core of $\langle \Delta x | x \rangle \approx - 1 / x$ and a RMS frequency displacement of $\sqrt{\langle \Delta x^2 | x \rangle} \approx  1$ \citep{Osterbrock1962}. Therefore, wing photons tend to return to the core after approximately $N_\text{scat} \approx x^2$ scattering events. Standard random-walk arguments also give the number of scatterings as $N_\text{scat} \approx (R / \lambda_\text{mfp})^2$, where the mean-free-path between scatterings is $\lambda_\text{mfp} \approx R / \tau_0 H(x) \approx \sqrt{\pi} x^2 / a\tau_0$. Equating these two expressions gives the characteristic escape frequency $x_\text{esc}$, which has the same form as the characteristic optical depth and trapping time in units of the light crossing time, $t_\text{light} = R/c$, \citep{Adams1975}
\begin{equation}
  x_\text{esc} \approx \frac{t_{\text{trap}}}{t_{\text{light}}} \approx \tau_{\text{esc}} \approx \frac{a \tau_0}{\sqrt{\pi} x^2_{\text{esc}}} \approx \left(\frac{a\tau_0}{\sqrt{\pi}}\right)^{1/3} \, .
\end{equation}
In summary, photons can successfully escape in a single excursion when the RMS spatial displacement is comparable to the radius. However, the actual number of scatterings is dominated by the repetitive attempts to escape the core. If the probability for a photon to scatter from the core into a frequency interval $(x, x + \text{d} x)$ for the first time is $\phi(x) \text{d} x / x^2$, then the approximate number of core scatterings before a successful excursion is related to the cumulative escape probability via
\begin{equation}
  N_{\text{scat}}^{\text{core}} \sim P^{-1}_{\text{esc}} \approx \left[ 2 \int_{x_{\text{esc}}}^\infty \frac{\phi(x) }{x^2} \text{d} x \right]^{-1} \approx \frac{3 \pi x_\text{esc}^3}{2 a} \sim \tau_0 \, .
\end{equation}

We now consider the transport of photons in non-uniform media. It is most convenient and illuminating to think in terms of random walks in optical depth space. For simplicity we specialize to plane-parallel slab geometry with power-law profiles for both the opacity and emissivity. Specifically, the (normalized) emissivity is $\eta(z) = (\alpha + 1) |z/Z|^\alpha / (2 Z) = \eta_0 |z|^\alpha$ and the absorption coefficient is $k(z) = k_0 |z|^\beta$ for $z \in [-Z,Z]$ with both zero elsewhere. We require $\{\alpha,\beta\} > -1$ so that path integrals are finite, with the cumulative optical depth defined as $\tau_0 = \int_0^Z k(z')\text{d}z' = k_0 Z^{\beta+1} / (\beta+1)$. The probability that newly emitted photons have an initial optical depth coordinate within $\tau$ is
\begin{equation} \label{eq:P(<tau)}
  P(< \tau) = \int_{-Z \left( \frac{\tau}{\tau_0}\right)^{\frac{1}{\beta+1}}}^{Z \left( \frac{\tau}{\tau_0}\right)^{\frac{1}{\beta+1}}} \eta(z')\,\text{d}z' = \left(\frac{\tau}{\tau_0}\right)^\delta \, ,
\end{equation}
where $\delta \equiv (\alpha + 1) / (\beta + 1)$. In this calculation we utilized both the opacity and emissivity profiles to translate the cumulative optical depth to a corresponding physical coordinate while accounting for all possible photons. The important parameter $\delta$ appears throughout the remainder of this work and captures the relative distribution of emissivity-to-opacity, i.e. as being either concentrated ($\delta = 0$) or extended ($\delta \gtrsim 1$). From equation~(\ref{eq:P(<tau)}) the normalized probability distribution function for the initial optical depth coordinate is then $p(\tau) = \delta (\tau / \tau_0)^{\delta-1} / \tau_0$. So far we have ignored frequency dependence, which can be included by considering that the mean optical depth per step is corrected by a factor of $H(x)$ independent of the opacity and emissivity profiles. General random walk theory then gives the number of scatterings as\footnote{Photons randomly walk in optical depth space with a mean step size of $\langle \tau \rangle \approx 1$. We denote the number of scattering events before escape by $N_\tau$, where the subscript represents the optical depth coordinate. The probabilistic outcomes depend on the sequence of states in the Markov chain defined by the recursion relation $N_\tau = 1 + \frac{1}{2}N_{\tau+1} + \frac{1}{2}N_{\tau-1}$, in combination with the boundary conditions $N_{\tau_0} = N_{-\tau_0} = 0$. If we apply the ansatz that $N_\tau = a + b \tau + c \tau^2$, we find that the number of steps starting from an arbitrary position in the slab is simply $N_\tau = \tau_0^2 - \tau^2$.}
\begin{equation}
  N_\text{scat} \approx H(x)^2 \int_0^{\tau_0} p(\tau) \left( \tau_0^2 - \tau^2 \right) \text{d}\tau \approx \frac{2}{\delta+2} \left(\frac{a \tau_0}{\sqrt{\pi} x^2}\right)^2 \, .
\end{equation}
When combined with the condition that $N_\text{scat} \approx x^2$ we now have
\begin{equation} \label{eq:x_esc_estimate}
  x_\text{esc} \approx \left(\frac{a\tau_0}{\sqrt{\pi (1+\delta/2)}}\right)^{1/3} \, .
\end{equation}
We emphasize that this estimate can be used for a qualitative understanding of resonant line transfer in non-uniform media. However, in the remainder of this work we provide more rigorous analytic solutions verified by numerical simulations. This estimate is in general agreement with those results.

\subsection{Diffusion approximation}
In general, it is only possible to solve equation~(\ref{eq:general}) numerically, so we apply further simplifications before attempting to find analytic solutions. We define angular moments of the radiation intensity as $J_x \equiv \frac{1}{4\pi} \int \text{d} \Omega I_x$ and $\bmath{H}_x \equiv \frac{1}{4\pi} \int \text{d} \Omega I_x \bmath{n}$. The angular-averaged form of equation~(\ref{eq:general}) is the zeroth order moment equation:
\begin{equation} \label{eq:RTE-moment}
  \frac{1}{c} \frac{\partial J_x}{\partial t} + \bmath{\nabla} \bmath{\cdot} \bmath{H}_x = \int \frac{j_x}{4\pi} \text{d} \Omega - k_x J_x + \int k_{x'} J_{x'} R_{x' \rightarrow x} \text{d}x' \, ,
\end{equation}
where $R_{x' \rightarrow x} \equiv (4\pi)^{-2} \iint \text{d}\Omega' \text{d}\Omega R_{x', \bmath{n}' \rightarrow x, \bmath{n}}$.
In optically-thick environments we may apply Fick's law as a closure relation to the moment equations:
\begin{equation} \label{eq:sec2_Ficks-Law}
  \bmath{H}_x \approx -\frac{\bmath{\nabla} J_x}{3 k_x} \, .
\end{equation}
Likewise, we take advantage of the Fokker-Planck approximation to rewrite the redistribution integral \citep{Rybicki1994}:
\begin{equation} \label{eq:Fokker-Planck}
  -k_x J_x + \int k_{x'} J_{x'} R_{x' \rightarrow x} \text{d}x' \approx \frac{\partial}{\partial x} \left( \frac{k_x}{2} \frac{\partial J_x}{\partial x} \right) \, .
\end{equation}
Thus, after incorporating equations~(\ref{eq:sec2_Ficks-Law}) and (\ref{eq:Fokker-Planck}) into (\ref{eq:RTE-moment}) we have
\begin{equation} \label{eq:intial_in_any_coordinate}
    \frac{1}{c}\frac{\partial J_x}{\partial t} = \int \frac{j_x}{4\pi} \text{d} \Omega + \bmath{\nabla} \bmath{\cdot} \left(\frac{\bmath{\nabla} J_x}{3 k_x} \right) + \frac{\partial}{\partial x} \left( \frac{k_x}{2} \frac{\partial J_x}{\partial x} \right) \, .
\end{equation}
In this paper we focus on steady-state solutions with $\partial J_x / \partial t \approx 0$.
Furthermore, we assume a static, isothermal environment, which implies spatial-frequency independence for the absorption coefficient $k_x = k(\bmath{r}) H(x)$. We also break the constant luminosity source into separable components, i.e. $\iiint j_x \text{d}\text{V} \text{d}x \text{d}\Omega = \mathcal{L}$ with the spatial, frequency, and angular dependence isolated as $\eta(\bmath{r})$, $H(x)/\sqrt{\pi}$, and $1/(4\pi)$, respectively (each normalized to unity):
\begin{equation} \label{eq:sec2_static_initial}
  \frac{1}{k(\bmath{r})} \bmath{\nabla}  \bmath{\cdot} \left(\frac{\bmath{\nabla} J}{k(\bmath{r})} \right) + \frac{3}{2} H(x) \frac{\partial}{\partial x} \left( H(x) \frac{\partial J}{\partial x} \right) = -\frac{3\mathcal{L}}{4\pi} \frac{\eta(\bmath{r})}{k(\bmath{r})} \frac{H^2(x)}{\sqrt{\pi}} \, .
\end{equation}
We then apply a change of variables with
\begin{equation}
  \text{d}\tilde{x} = \sqrt{\frac{2}{3}}\frac{\text{d}x}{\tau_0 H(x)} \quad \text{such that} \quad \tilde{x} \approx \sqrt{\frac{2\pi}{27}} \frac{x^3}{a \tau_0} \, ,
\end{equation}
where $\tau_0$ denotes the optical depth at line centre. The transformation is based on the wing approximation from equation~(\ref{eq:H}), and maps onto the same domain $\tilde{x} \in (-\infty, +\infty)$.
We similarly transform from real space to normalized optical depth coordinates according to
\begin{equation} \label{eq:general_r_tilde}
  \tilde{\bmath{\nabla}} \equiv \frac{\tau_0 \bmath{\nabla}}{k(\bmath{r})}
  \quad \text{such that} \quad
  \tilde{\bmath{r}}
  = \min \int_{\bmath{0}}^{\bmath{r}} \frac{k(\bmath{r}')}{\tau_0}\,\text{d}\bmath{r}' \, ,
\end{equation}
which is well-defined if $k(\bmath{r})$ is continuous and the path integration yields the minimum optical depth over all paths. In this paper we focus on slab and spherical geometries.
In terms of the overall width of the line, $H^2(x)$ is sharply peaked at $x = 0$, so we can replace it with a delta function. To preserve normalization, we note that $\int 3 \tau_0 H^2 \text{d}\tilde{x} = \int \sqrt{6} H\,\text{d}x = \sqrt{6 \pi}$, which allows a replacement of $3 \tau_0 H^2(x) \approx \sqrt{6\pi} \delta(\tilde{x})$.
Finally, if we set $J = \tilde{J} \mathcal{L} \tau_0 \sqrt{6} / (4\pi)$ then the final equation describing our general setup is
\begin{equation} \label{eq:final_transfer_eq}
  \tilde{\nabla}^2 \tilde{J} + \frac{\partial^2 \tilde{J}}{\partial \tilde{x}^2} = - \frac{\eta(\bmath{r})}{k(\bmath{r})}\delta(\tilde{x}) \, .
\end{equation}
We will now derive a general analytic solution for resonant-line radiative transfer based on equation~(\ref{eq:final_transfer_eq}). Without loss of generality, the physical domain can be infinite as long as the total optical depth is still finite. Our solution extends the one presented by \citet{Harrington1973} and serves as a prototype for the analytic and numerical methods in the remaining sections.

This second order partial differential equation needs suitable boundary conditions to get a unique and stable solution. Following previous studies we require the solution to be finite throughout the slab, zero as $x \rightarrow \pm\infty$, and the surface intensity to be proportional to the outward flux. For concreteness, if $\bmath{s}$ represents the finite optical depth surface and $\tilde{\partial}_{\bmath{s}}$ is the gradient in the normal direction of the surface then the boundary conditions can be written as
\begin{equation} \label{eq:sec2_boundary condition}
  \left[\tilde{\partial}_{\bmath{s}} \tilde{J} + f \tau_0 H(\tilde{x}) \tilde{J} \right]_{\bmath{s}} = 0 \qquad \text{and} \qquad \lim_{\tilde{x} \rightarrow \pm \infty} \tilde{J} = 0 \, ,
\end{equation}
where $f$ is a positive constant of order unity (see Appendix~\ref{app:derivation_of_f}). The domain of $\tilde{\bmath{r}}$ is compact so we employ an eigenfunction expansion with separable space and frequency components:
\begin{equation} \label{eq:general_decomposition}
  \tilde{J}(\tilde{\bmath{r}},\tilde{x}) = \sum_{n=1}^\infty \vartheta_n(\tilde{\bmath{r}}) \varphi_n(\tilde{x}) \, .
\end{equation}
The solutions of the homogeneous equation
\begin{equation}
  \tilde{\nabla}^2 \vartheta_n + \lambda_n^2 \vartheta_n = 0 \, ,
\end{equation}
form an orthonormal basis with eigenvalues $\lambda_n$, requiring that the volume integrals satisfy the relation $\int \vartheta_n \vartheta_m^\ast\,\text{d}\tilde{V} = \delta_{nm}$. Upon substitution of equation~(\ref{eq:general_decomposition}) into equation~(\ref{eq:final_transfer_eq}), multiplying by $\vartheta_m$, and integrating over the volume $\tilde{V}$ we obtain
\begin{equation} \label{eq:general_x_ODE}
  \frac{\text{d}^2 \varphi_n}{\text{d}\tilde{x}^2} - \lambda_n^2 \varphi_n = -\frac{Q_n}{\tau_0} \delta(\tilde{x}) \, ,
\end{equation}
where the source term coefficients in our convention are
\begin{equation} \label{eq:general_Qn}
  Q_n = \int \eta(\bmath{r}) \vartheta_n(\bmath{r})\,\text{d}V
  = \tau_0 \int \frac{\eta(\tilde{\bmath{r}})}{k(\tilde{\bmath{r}})} \vartheta_n(\tilde{\bmath{r}})\,\text{d}\tilde{V} \, .
\end{equation}
Away from $\tilde{x} = 0$, the solution satisfying the boundary conditions $\lim_{\tilde{x}\rightarrow\pm\infty} \tilde{J} = 0$, and the jump condition $\Delta (\text{d}\varphi_n/\text{d}\tilde{x})_{\tilde{x} = 0} = -Q_n / \tau_0$ derived from integrating equation~(\ref{eq:general_x_ODE}) is
\begin{equation}
  \varphi_n = \frac{Q_n}{2 \tau_0 \lambda_n} e^{-\lambda_n |\tilde{x}|} \, .
\end{equation}
Putting this all together we have a final solution of
\begin{equation} \label{eq:general_full_solution}
  J(\tilde{\bmath{r}},\tilde{x}) = \frac{\mathcal{L} \sqrt{6}}{8 \pi} \sum_{n=1}^\infty \frac{Q_n}{\lambda_n} e^{-\lambda_n |\tilde{x}|} \vartheta_n(\tilde{\bmath{r}}) \, .
\end{equation}
The radiation energy density can be derived as
\begin{equation} \label{eq:sec2_u(r)}
  u(\tilde{\bmath{r}}) = \frac{4\pi}{c} \int J\,\text{d}x = \frac{\mathcal{L}}{c} \Gamma\left(\frac{1}{3}\right) \left(\frac{2 a \tau_0}{\sqrt{\pi}}\right)^{1/3} \sum_{n=1}^\infty \frac{Q_n}{\lambda_n^{4/3}} \vartheta_n(\tilde{\bmath{r}}) \, .
\end{equation}
Furthermore, we define volume-weighted averages as
\begin{equation} \label{eq:sec2_k-weight}
  \langle f \rangle \equiv \frac{\int f(\bmath{r})\,\text{d}V}{\int \text{d}V}
  = \frac{\int f(\tilde{\bmath{r}})\,\text{d} \tilde{V}/k(\tilde{\bmath{r}})}{\int \text{d}\tilde{V}/k(\tilde{\bmath{r}})} \, ,
\end{equation}
and for convenience we define eigenfunction averages by $T_n \equiv \langle \vartheta_n \rangle$. Thus, a general expression for the average internal spectrum is
\begin{equation} \label{eq:sec2_Jxk}
  \langle J(\tilde{x}) \rangle \equiv \frac{\mathcal{L} \sqrt{6}}{8\pi} \sum_{n=1}^\infty \frac{Q_n T_n}{\lambda_n} e^{-\lambda_n |\tilde{x}|} \, ,
\end{equation}
and the average radiation energy density is
\begin{equation} \label{eq:sec2_uk}
  \langle u \rangle \equiv \frac{\mathcal{L}}{c} \Gamma\left(\frac{1}{3}\right) \left(\frac{2 a \tau_0}{\sqrt{\pi}}\right)^{1/3} \sum_{n=1}^\infty \frac{Q_n T_n}{\lambda_n^{4/3}} \, ,
\end{equation}
which is related to the trapping time, defined by the expression $t_\text{trap} = \mathcal{L}^{-1} \int u(\tilde{\bmath{r}})\,\text{d}V$. We can also derive the outward force multiplier by approximating the flux as $F \propto \nabla u$:
\begin{align} \label{eq:sec2_M_F}
  M_\text{F}
  &\equiv \mathcal{L}^{-1} \iint k(\bmath{r}) F\,\text{d}x\,\text{d}V
  = -\frac{c}{3\mathcal{L}} \int \nabla u(\bmath{r})\,\text{d}V \notag \\
  &= -\Gamma\left(\frac{4}{3}\right) \left(\frac{2 a \tau_0}{\sqrt{\pi}}\right)^{1/3} \sum_{n=1}^\infty \frac{Q_n}{\lambda_n^{4/3}} \int \nabla\vartheta_n(\bmath{r})\,\text{d}V \, ,
\end{align}
related to the enhancement of momentum coupling compared to the single scattering limit of $\mathcal{L}/c$.
In addition, we define a characteristic radius as the volume-weighted expectation value of position
\begin{equation} \label{eq:sec2_k-weighted-radius}
  r_c \equiv \frac{\langle r u \rangle}{\langle u \rangle}
  = \frac{\sum_{n=1}^\infty Q_n R_n \lambda_n^{-4/3}}{\sum_{n=1}^\infty Q_n T_n \lambda_n^{-4/3}} \, ,
\end{equation}
where for convenience we let $R_n \equiv \langle r \vartheta_n \rangle$.
Finally, the average number of scatterings photons undergo from emission to escape is
\begin{align} \label{eq:sec2_N_scat}
  N_\text{scat} &\approx 4 \pi^{3/2} \mathcal{L}^{-1} \int J(\bmath{r},0) k(\bmath{r})\,\text{d}V \notag \\
  &= \tau_0 \sqrt{\frac{3 \pi}{2}} \sum_{n=1}^\infty \frac{Q_n}{\lambda_n} \int \vartheta_n(\tilde{\bmath{r}})\,\text{d}\tilde{V} \, .
\end{align}
We must consider specific geometries to make further progress. We first introduce a few special functions that will be employed throughout this paper. The Lerch transcendent is defined as
\begin{equation} \label{eq:Lerch}
  \Phi(z,s,a) = \sum_{n=0}^{\infty} \frac{z^n}{(n + a)^s} \, ,
\end{equation}
which is sufficiently general for most of the solutions in this paper. Further simplifications result in the polylogarithm function defined as $\text{Li}_s(z) = \Phi(z,s,1) = \sum_{n=1}^\infty z^n / n^s$, the Hurwitz zeta function defined as $\zeta(s,a) = \Phi(1,s,a) = \sum_{n=0}^\infty (n + a)^{-s}$, and the Riemann zeta function $\zeta(s) = \Phi(1,s,1) = \text{Li}_s(1) = \zeta(s,1) = \sum_{n=1}^\infty n^{-s}$.

\section{General solution for slab geometry}
\label{sec:General solution for slab geometry}
We now present a general solution for slab geometries. This is possible because any density profile in real space is isomorphic to a homogeneous representation in normalized optical depth space. Specifically, the transformation in equation~(\ref{eq:general_r_tilde}) becomes:
\begin{equation}
  \tilde{z} = \int_0^z \frac{k(z')}{\tau_0}\,\text{d}z' \, .
\end{equation}
At this point, we rewrite equation~(\ref{eq:final_transfer_eq}) for the specific case of a static, isothermal, optically-thick slab, substituting to $\tilde{z}$:
\begin{equation} \label{eq:sec4_intial_eq}
  \frac{\partial^2 \tilde{J}}{\partial \tilde{z}^2} + \frac{\partial^2 \tilde{J}}{\partial \tilde{x}^2} = -\frac{\eta(\tilde{z})}{k(\tilde{z})} \delta(\tilde{x}) \, .
\end{equation}
For simplicity we also assume the functions $\eta$ and $k$ are symmetric (even) about the central plane $\tilde{z} = 0$ so $\tilde{z} \in (-1, 1)$. The solutions of the homogeneous equation, $\vartheta_n''  +\lambda_n^2 \vartheta_n = 0$, are of the form
\begin{equation}
  \vartheta_n = \cos(\lambda_n \tilde{z}) \qquad \text{where} \quad n = 1, 2, \ldots \, ,
\end{equation}
where the length of the eigenfunction is approximately unity,
\begin{equation} \label{eq:sec4_nor_of_eigenfunction}
  \int_{-1}^1 \cos^2(\lambda_n \tilde{z})\,\text{d}\tilde{z} = 1 + \frac{\sin(2 \lambda_n)}{2 \lambda_n} \approx 1 \, ,
\end{equation}
and the boundary conditions require the eigenvalues to satisfy
\begin{equation} \label{eq:BCs_of_eigenfunction}
  \lambda_n \tan(\lambda_n) = f \tau_0 H(\tilde{x}) \, .
\end{equation}
If the optical depth is large out to any frequency with appreciable radiation, photons escape before they diffuse to frequencies where the slab is optically thin, and we can use the approximation:
\begin{equation} \label{eq:sec4_keyapprox1}
  \lambda_n \ll f \tau_0 H(\tilde{x}) \, .
\end{equation}
Thus, the eigenvalue to zeroth order is approximately
\begin{equation}
  \lambda_n = \pi (n - 1) + \tan^{-1}\left(\frac{f \tau_0 H(\tilde{x})}{\lambda_n}\right) \approx \pi \left( n - \frac{1}{2} \right) \, .
\end{equation}
Furthermore, the emission constants from equation~(\ref{eq:general_Qn}) are
\begin{align} \label{eq:sec4_Qn}
  Q_n &= 2 \tau_0 \int_0^1 \frac{\eta(\tilde{z})}{k(\tilde{z})} \cos(\lambda_n \tilde{z})\,\text{d}\tilde{z} \notag \\
  &= 2 \int_0^\infty \eta(z) \cos\left(\lambda_n \int_0^z \frac{k(z')}{\tau_0} \text{d}z'\right)\,\text{d}z \, ,
\end{align}
averages from equation~(\ref{eq:sec2_k-weight}) are
\begin{equation} \label{eq:sec4_Tn}
  T_n = Z^{-1} \int_0^Z \cos\left(\lambda_n \int_0^z \frac{k(z')}{\tau_0} \text{d}z'\right)\,\text{d}z \, ,
\end{equation}
and the factor in characteristic radius from equation~(\ref{eq:sec2_k-weighted-radius}) is
\begin{equation} \label{eq:sec3_Rn}
  R_n = Z^{-1}\int_0^Z z \cos\left(\lambda_n \int_0^z \frac{k(z')}{\tau_0} \text{d}z'\right)\,\text{d}z   \, .
\end{equation}
The final solution from equation~(\ref{eq:general_full_solution}) is given by
\begin{equation} \label{eq:sec4_J(z,x)}
  J(\tilde{z},\tilde{x}) = \frac{\mathcal{L} \sqrt{6}}{8\pi} \sum_{n=1}^\infty \frac{Q_n}{\lambda_n} \cos(\lambda_n \tilde{z})\,e^{-\lambda_n |\tilde{x}|} \, .
\end{equation}
The spectral line profile at the boundary is particularly relevant for observations. We use equation~(\ref{eq:BCs_of_eigenfunction}) to substitute
$\cos(\lambda_n) = \lambda_n \sin(\lambda_n) / [f \tau_0 H(\tilde{x})] \approx \lambda_n (-1)^{n-1} / [f \tau_0 H(\tilde{x})]$, yielding
\begin{equation} \label{eq:sec4_general_solution_J}
  J(\tilde{x}) = \frac{\mathcal{L} \sqrt{6}}{8\pi} \frac{e^{\pi |\tilde{x}| / 2}}{f \tau_0 H(\tilde{x})} \sum_{n=1}^\infty (-1)^{n-1} Q_n e^{-n \pi |\tilde{x}|} \, ,
\end{equation}
with normalization $\int_{-\infty}^{+\infty}J(x) \text{d}x = \frac{3 \mathcal{L}}{4\pi^2 f} \sum_{n=1}^{\infty}(-1)^{n-1} Q_n / (n - \frac{1}{2})$.
The radiation energy density from equation~(\ref{eq:sec2_u(r)}) is
\begin{equation} \label{eq:sec4_u(z)}
  u(\tilde{z}) = \frac{\mathcal{L}}{c} \Gamma\left(\frac{1}{3}\right) \left(\frac{2 a \tau_0}{\sqrt{\pi}}\right)^{1/3} \sum_{n=1}^\infty \frac{Q_n}{\lambda_n^{4/3}} \cos(\lambda_n \tilde{z}) \, ,
\end{equation}
with $2 \int_0^Z \frac{\text{d}}{\text{d}z} \cos\left(\lambda_n (z/Z)^{\beta+1}\right) \text{d}z = 2 [ \cos(\lambda_n) -1 ] \approx -2$ the outward force multiplier from equation~(\ref{eq:sec2_M_F}) becomes
\begin{equation}
  M_\text{F} = 2 \Gamma\left(\frac{4}{3}\right) \left(\frac{2 a \tau_0}{\sqrt{\pi}}\right)^{1/3} \sum_{n=1}^\infty \frac{Q_n}{\lambda_n^{4/3}}  \, ,
\end{equation}
and with $\int_{-1}^1 \cos(\lambda_n \tilde{z})\,\text{d}\tilde{z} = 2 \sin(\lambda_n) \approx 2 (-1)^{n-1}/\lambda_n$ the average number of scatters before escape from equation~(\ref{eq:sec2_N_scat}) becomes
\begin{equation}
  N_\text{scat} = \sqrt{6 \pi}\,\tau_0 \sum_{n=1}^\infty \frac{(-1)^{n-1} Q_n}{\lambda_n^2} \, .
\end{equation}
The volume-averaged internal spectrum, radiation energy density, and characteristic depth each depend on the specific density profile. In the remaining subsections we consider specific cases for $Q_n$, $T_n$, and $R_n$, although we note that for a homogeneous slab $\tilde{z} = z / Z$ so the coefficients simplify to $T_n = (-1)^{n-1}/\lambda_n$ and $R_n = Z (T_n - \lambda_n^{-2})$. The analytic solutions then also allow us to find precise peak positions for a given model. The procedure is to set $\partial J / \partial x = 0$ and numerically solve the resulting transcendental equations. Finally, if we assume constant opacity, $k(z) = k_0$, then the average trapping time normalized to known factors ($t_\text{light} = Z/c$) is simply $t_\text{trap}/t_\text{light} = 2 c \langle u \rangle / \mathcal{L}$.
To allow a more compact notation in this section we introduce the function
\begin{equation} \label{eq:PsiFunction}
  \Psi_s^{\pm}(z) = \pi^{-s} e^{\pi z/2}\,\Phi\left(\pm e^{\pi z}, s, \frac{1}{2}\right) \, .
\end{equation}

\subsection{Central point source} \label{sec4.1}
For a point source $\eta(z) = \delta(z)$ and $Q_n = 1$. In this case the full solution reduces after some mathematical manipulation to
\begin{equation} \label{eq:sec4-J(z,x)-PS}
  J(\tilde{z},\tilde{x}) = \frac{\mathcal{L}\sqrt{6}}{8 \pi^2} \text{tanh}^{-1}\left[ \cos\left(\frac{\pi \tilde{z}}{2}\right) \text{sech}\left(\frac{\pi \tilde{x}}{2}\right)\right] \, .
\end{equation}
The spectral line profile at the boundary is found with the relation $\sum_{n=1}^{\infty} (-1)^{n-1} x^n  = x/(1+x)$, valid for $|x| < 1$, which gives
\begin{equation} \label{eq:sec4_sol_uniform_CPS}
  J(\tilde{x}) = \frac{\mathcal{L}\sqrt{6}}{16\pi f \tau_0 H(\tilde{x})} \text{sech}\left( \frac{\pi \tilde{x}} {2} \right) \, .
\end{equation}
After transforming back to the original frequency notation, we arrive at the final normalized version \citep{Harrington1973}:
\begin{equation}
  \frac{J(x)}{\int_{-\infty}^{+\infty} J(x)\,\text{d}x} = \sqrt{\frac{\pi}{6}} \frac{x^2}{a\tau_0} \text{sech}\left( \sqrt{\frac{\pi^3}{54}}\frac{x^3}{a\tau_0}\right) \, .
\end{equation}
We have shown that the density profile $k(z)$ does not affect the solution at all when the emission follows a central point source. Also, the peaks are located at $x_\text{p} = \pm1.07\,(a \tau_0)^{1/3}$, calculated from the equation $\bar{x}\,\text{tanh}\bar{x} = 2/3$, where $\bar{x} = \sqrt{\pi^3/54} x^3 / a\tau_0$.

From equation~(\ref{eq:sec4_u(z)}) the radiation energy density is
\begin{equation}
  u(\tilde{z}) = \frac{\mathcal{L}}{c} \Gamma\left(\frac{1}{3}\right) \left(\frac{2 a \tau_0}{\sqrt{\pi}}\right)^{1/3} \text{Re}\left[\Psi_{4/3}^{+}(i\tilde{z})\right] \, ,
\end{equation}
where $\Psi$ is the special function defined in equation~(\ref{eq:PsiFunction}). We note that the outward force multiplier in this case is $M_\text{F} \approx 2.2\,(a\tau_0)^{1/3}$ and the number of scatters is $N_\text{scat} \approx 1.6\,\tau_0$. Furthermore, in a uniform density environment the trapping time is $t_\text{trap} / t_\text{light} \approx 1.8\,(a\tau_0)^{1/3}$ and the characteristic depth is $z_c \approx 0.3\,Z$.

\subsection{Uniform source}
If the emissivity traces the absorption coefficient then $k(z) = 2 \tau_0 \eta(z)$. Therefore, the constants from equation~(\ref{eq:sec4_Qn}) are
\begin{equation}
  Q_n = \int_0^1 \cos(\lambda_n \tilde{z})\,\text{d}\tilde{z}
  = \frac{\sin(\lambda_n)}{\lambda_n} \approx \frac{(-1)^{n-1}}{\lambda_n} \, .
\end{equation}
The general expression from equation~(\ref{eq:sec4_J(z,x)}) reduces to
\begin{equation}
  J(\tilde{z},\tilde{x}) = \frac{\mathcal{L}\sqrt{6}}{8 \pi} \text{Re}\left[\Psi_2^{-}\left(-|\tilde{x}| + i \tilde{z}\right)\right] \, .
\end{equation}
The line profile at the boundary from equation~(\ref{eq:sec4_general_solution_J}) becomes
\begin{equation} \label{eq:sec4_J(x)_uniform}
  J(\tilde{x}) = \frac{\mathcal{L}\sqrt{6}}{4\pi^2 f \tau_0 H(\tilde{x})} \text{tanh}^{-1}\left(e^{-\pi |\tilde{x}| / 2}\right) \, .
\end{equation}
After transforming back to the original frequency notation, we arrive at the final normalized version \citep{Harrington1973}:
\begin{equation} \label{eq:sec4_sol_uniform_source}
  \frac{J(x)}{\int_{-\infty}^{+\infty}J(x)\,\text{d}x} = \sqrt{\frac{8}{3\pi}} \frac{x^2}{a\tau_0} \tanh^{-1}\left[ \exp\left(-\sqrt{\frac{\pi^3}{54}}\frac{|x^3|}{a\tau_0}\right) \right] \, .
\end{equation}
Therefore, when the emission follows the density, i.e. $\eta(z) \propto k(z)$, the spectra is the same as a uniform source in a homogeneous slab. Also, the peaks are located at $x_\text{p} = \pm0.86\,(a \tau_0)^{1/3}$, calculated from $4\,\text{coth}^{-1}(\exp\bar{x}) = 3 \bar{x}\,\text{csch}\bar{x}$, with the same $\bar{x}$ as before. The energy density is
\begin{equation} \label{eq:sec4_u(z)_uniform}
  u(\tilde{z}) = \frac{\mathcal{L}}{c} \Gamma\left(\frac{1}{3}\right) \left(\frac{2 a \tau_0}{\sqrt{\pi}}\right)^{1/3} \text{Re}\left[\Psi_{7/3}^{-}\left(i \tilde{z}\right) \right] \, .
\end{equation}
In this case the outward force multiplier is $M_\text{F} \approx 0.61\,(a\tau_0)^{1/3}$ and the number of scatters is $N_\text{scat} \approx 1.2\,\tau_0$. Furthermore, in a uniform density environment the trapping time is $t_\text{trap} / t_\text{light} \approx 1.3\,(a\tau_0)^{1/3}$ and the characteristic depth is $z_c \approx 0.4\,Z$.

\begin{figure}
  \centering
  \includegraphics[width=\columnwidth]{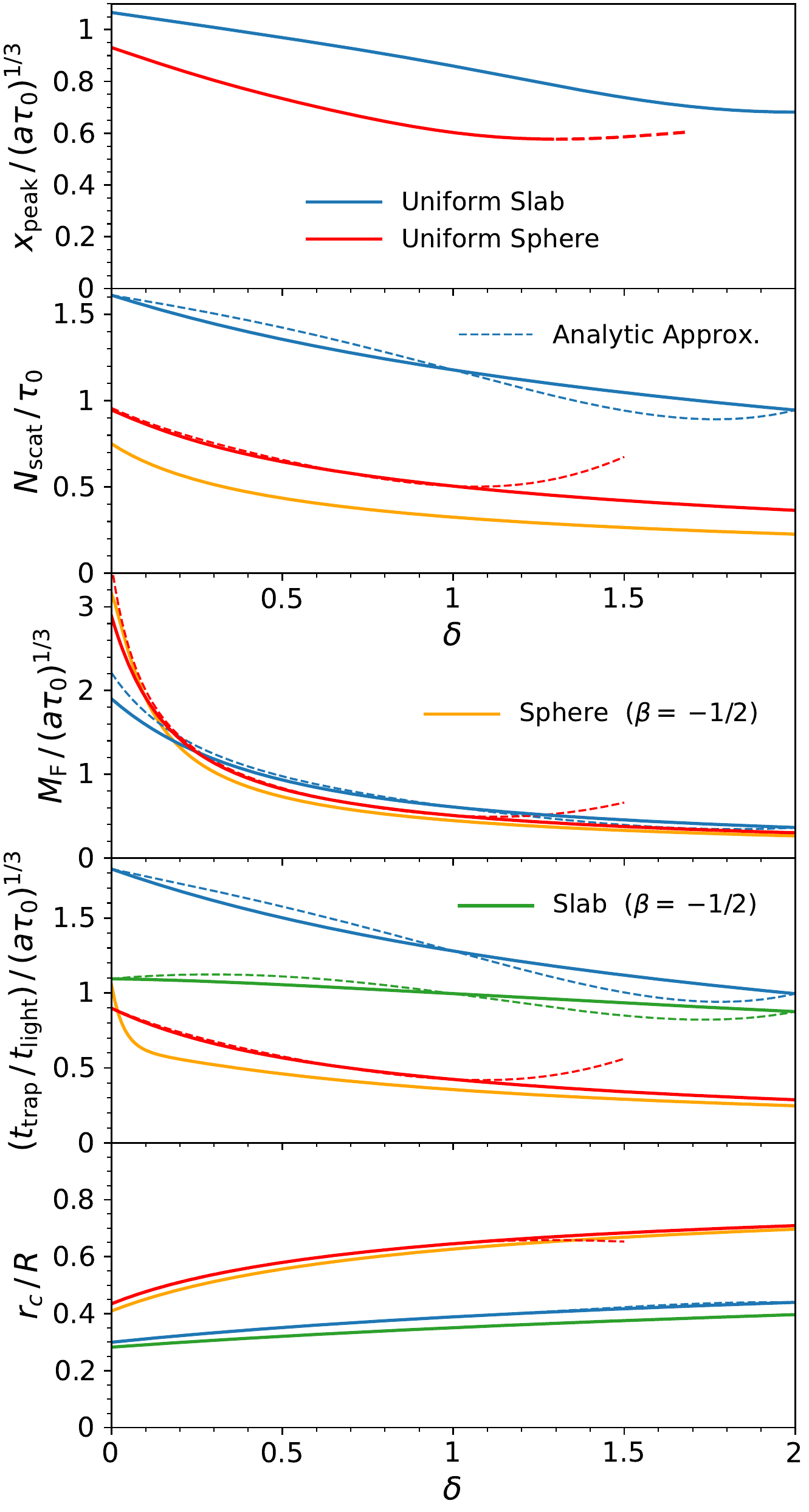}
  \caption{Peak frequency $x_\text{peak}$, number of scatterings $N_\text{scat}$, outward force multiplier $M_\text{F}$, light trapping time $t_\text{trap}$, and characteristic radius $r_c$ as a function of the emissivity-to-opacity parameter $\delta$ for both slab and spherical geometries, illustrating the transition between central point ($\delta = 0$) and uniform ($\delta = 1$) sources. The solid curves are calculated with the full series expansion, while the dashed curves reveal regions where the analytic solutions are less reliable due to the approximations in the derivations. As $\delta$ increases, the peak position shifts towards line centre and in the slab case is well described by a slope of $\text{d}x_\text{peak}/\text{d}\delta \approx -0.2$ (see Figure~\ref{fig:combination_Jx}). Similarly,
  $t_\text{trap}$, $N_\text{scat}$, and $M_\text{F}$ decrease while $r_c$ becomes more extended (see Figure~\ref{fig:sec5_energy_density}). The blue (red) curve is for a uniform slab (sphere) while the green (yellow) curve corresponds to a steeper profile of $k \propto r^{-1/2}$.}
  \label{fig:sec5_peak_pos_trapping_time}
\end{figure}

\begin{figure}
  \centering
  \includegraphics[width=\columnwidth]{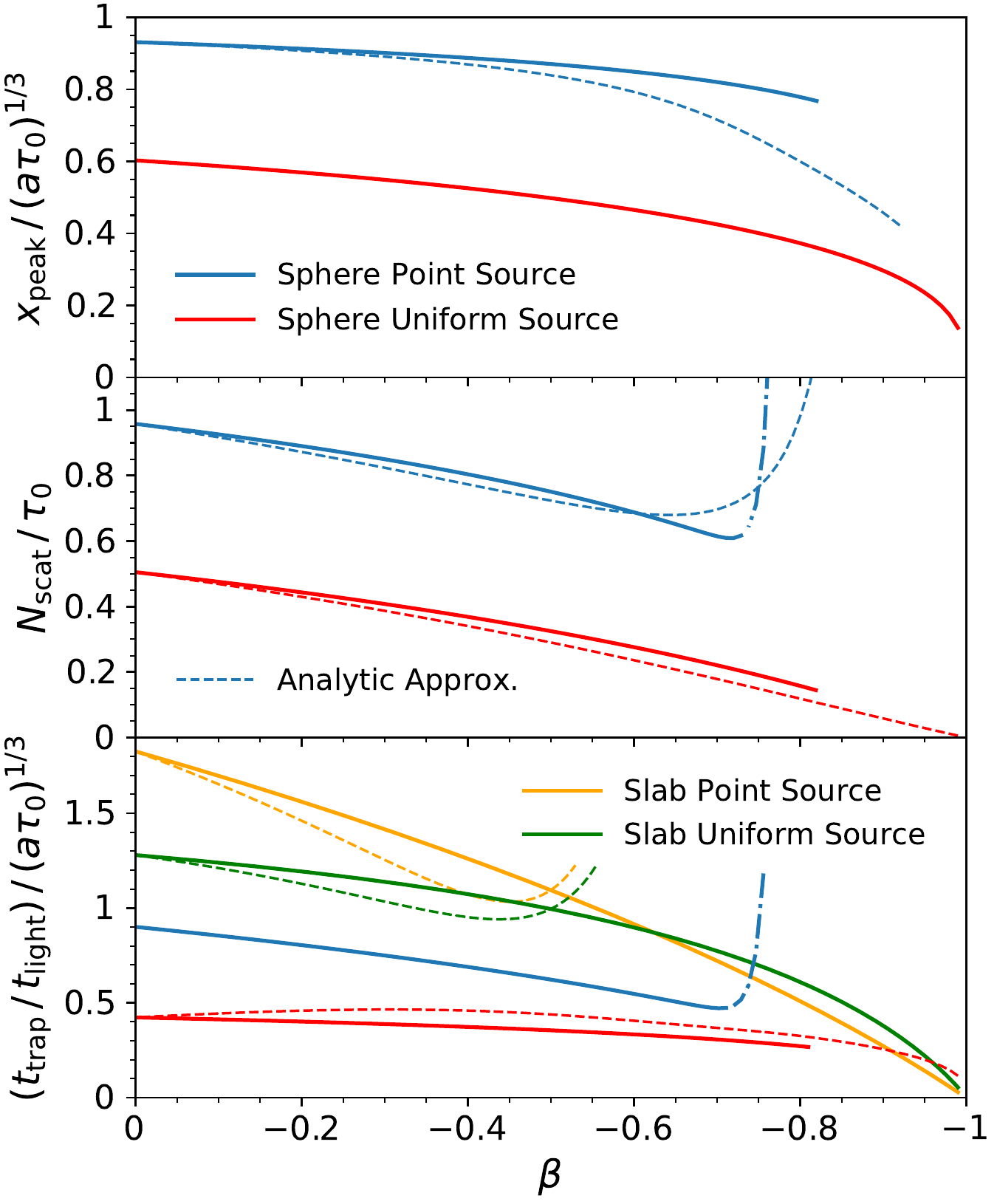}
  \caption{The evolution of the peak frequency $x_\text{peak}$, number of scatters $N_\text{scat}$, and trapping time $t_\text{trap}$ as a function of the power-law opacity exponent $\beta$. Results are shown for both slab and spherical geometries with central point ($\delta = 0$) and uniform ($\delta = 1$) emissivities. The solid curves are calculated with the full series expansion, while the dashed curves are from the approximate analytic solutions. The solutions are accurate when $\beta$ is close to zero, but has clear deviations as $\beta \rightarrow -1$. Intuitively, as the profiles steepen $x_\text{peak}$ shifts towards line centre and both $N_\text{scat}$ and $t_\text{trap}$ are reduced. Due to the positive exponent of $\lambda_n$ in $Q_n$ for the point source (see equation~\ref{eq:sec6_Qn_exact}), both $N_\text{scat}$ and $t_\text{trap}$ are expected to diverge at some point below $\beta \lesssim -1/2$. We also note the the force multiplier $M_\text{F}$ and characteristic radius $r_c$ do not depend strongly on $\beta$ so for simplicity are not shown.}
  \label{fig:sec5_peak_trapping_time}
\end{figure}

\subsection{Power-law profiles}
We now consider power-law profiles for both the emissivity and absorption coefficient. This results in a single parameter representing the continuous transition from inward to outward opacity relative to the sources. We give the (normalized) emissivity as $\eta(z) = (\alpha + 1) |z/Z|^\alpha / (2 Z) = \eta_0 |z|^\alpha$ and absorption coefficient as $k(z) = k_0 |z|^\beta$ for $z \in [-Z,Z]$ with both zero elsewhere. We require $\{\alpha,\beta\} > -1$ so that path integrals are finite, i.e. the cumulative optical depths are $\tau_0 = \int_0^Z k(z')\text{d}z' = k_0 Z^{\beta+1} / (\beta+1)$ and $\tilde{z} = \int_0^z k(z')\,\text{d}z' / \tau_0 = (z/Z) |z/Z|^\beta$. Therefore, the unknown constant from equation~(\ref{eq:sec4_Qn}) is
\begin{equation} \label{eq:sec4_Qn_generalization}
  Q_n = \delta \int_0^1 \tilde{z}^{\delta-1} \cos\left(\lambda_n \tilde{z} \right)\,\text{d}\tilde{z}
  = {}_1 F_2 \left(\frac{\delta}{2};\frac{1}{2},\frac{\delta }{2}+1;-\frac{\lambda_n^2}{4}\right) \, ,
\end{equation}
where $\delta \equiv (\alpha + 1) / (\beta + 1) > 0$ and ${}_1F_2$ is a generalized hypergeometric function. In order to move forward, we use a Taylor expansion to approximate the hypergeometric function, considering the special value of $\lambda_n \approx \pi (n - 1/2)$. The expression is as follows:
\begin{equation} \label{eq:sec4_Qn_generalization_approximation}
  Q_n \approx \frac{C(\delta)}{\lambda_n^\delta} + \frac{\delta (-1)^{n-1}}{\lambda_n} + \mathcal{O}\left(\frac{1}{\lambda_n^3}\right) \, ,
\end{equation}
where $C(\delta) = \cos (\pi \delta/2) \Gamma(1+\delta)$. This approximation is good when $\delta < 2$, otherwise we require higher-order correction terms. The error of the approximation is shown in Figure~\ref{fig:appendix_slab_error}. After substitution into equation~(\ref{eq:sec4_J(z,x)}) the final expression is
\begin{equation} \label{eq:sec3_J(z,x)_power_law}
  J(\tilde{z},\tilde{x}) = \frac{\mathcal{L}\sqrt{6}}{8} C(\delta) \text{Re}\left[\Psi^{+}_{\delta+1}(-|\tilde{x}| + i \tilde{z})\right] + \delta J_{\text{uni}}(\tilde{z},\tilde{x}) \, .
\end{equation}
The spectral line profile at the boundary is
\begin{equation} \label{eq:sec4_generalized_sol}
  J(\tilde{x}) = \frac{\mathcal{L}\sqrt{6} C(\delta)}{8\pi^2 f \tau_0 H(\tilde{x})} \Psi_\delta^{-}\left( -|\tilde{x}| \right)
  + \delta J_\text{uni}(\tilde{x}) \, ,
\end{equation}
where $J_\text{uni}(\tilde{x})$ denotes the result from equation~(\ref{eq:sec4_J(x)_uniform}). The normalization factor is given by the following expression:
\begin{equation}
  \int_{-\infty}^{+\infty} J(x)\,\text{d}x = \frac{3\mathcal{L}}{8 \pi f} \left( 2 C(\delta) \Psi_{\delta+1}^{-}(0) + \delta \right) \, .
\end{equation}
In this case the peak locations satisfy
\begin{equation}
  \frac{2 C(\delta)}{\delta} \left( \Psi^{-}_{\delta}(\bar{y}) - \frac{3 \bar{x}}{\pi} \Psi^{-}_{\delta-1}(\bar{y}) \right) = 3 \bar{x} \text{csch}\left(\bar{x} \right) - 4 \text{coth}^{-1}\left(e^{\bar{x}}\right) \, ,
\end{equation}
with $\bar{y} = -2 \bar{x}/\pi$, the solutions of which are shown in Figure~\ref{fig:sec5_peak_pos_trapping_time}, demonstrating that the peak locations are within an order unity factor from $(a \tau_0)^{1/3}$. The radiation energy density is
\begin{equation} \label{eq:sec4_u(z)_PL}
  u(\tilde{z}) = \frac{\mathcal{L}}{c} \Gamma\left(\frac{1}{3}\right) \left(\frac{2 a \tau_0}{\sqrt{\pi}}\right)^{1/3} C(\delta) \text{Re}\left[\Psi_{\delta+4/3}^{+}\left(i\tilde{z}\right) \right] + \delta u_\text{uni}(\tilde{z}) \, ,
\end{equation}
the outward force multiplier is
\begin{equation}
  M_\text{F} \approx -2\Gamma\left(\frac{4}{3}\right) \left(\frac{2 a \tau_0}{\sqrt{\pi}}\right)^{1/3} \left( C(\delta) \Psi^+_{\delta + 4/3}(0) + \pi \delta \Psi^-_{7/3}(0)\right) \, ,
\end{equation}
and the average number of scattering events is
\begin{equation}
  N_\text{scat} \approx \tau_0 \sqrt{6\pi} \left( C(\delta) \Psi^-_{\delta + 2}(0) + \delta \Psi^+_{3}(0)\right) \, .
\end{equation}
In this case $T_n$ and $R_n$ are similar to $Q_n$ but with $\delta$ replaced by factors of $\kappa = 2/(\beta+1)$. Although the approximations are only valid for a limited range in $\beta$, for completeness we provide approximate solutions. From equation~(\ref{eq:sec2_Jxk}) the average internal spectrum is
\begin{align}
  \langle J(\tilde{x}) \rangle = \frac{\sqrt{6}\mathcal{L}}{8 \pi} \bigg( &C(\delta) C(\kappa/2) \Psi^+_{\delta+\frac{\kappa}{2}+1}\left(-|\tilde{x}|\right) + \frac{\kappa C(\delta)}{2} \Psi^-_{\delta+2}\left(-|\tilde{x}|\right) \notag \\
  &+ \delta C(\kappa/2) \Psi^-_{2+\kappa/2} \left(-|\tilde{x}|\right) + \frac{\kappa \delta}{2} \Psi^+_3 \left(-|\tilde{x}|\right) \bigg) \, ,
\end{align}
and from equation~(\ref{eq:sec2_uk}) the average energy density is
\begin{align}
  \langle u \rangle =& \frac{\mathcal{L}}{c} \Gamma\left(\frac{1}{3}\right) \left(\frac{2 a \tau_0}{\sqrt{\pi}}\right)^{1/3} \bigg(C(\kappa/2)\delta \Psi^-_{\frac{7}{3}+\kappa/2}(0) + \frac{\kappa \delta}{2} \Psi^+_{\frac{10}{3}}(0) \notag \\
  &+ C(\delta) C(\kappa/2) \Psi^+_{\delta+\frac{\kappa}{2}+\frac{4}{3}}(0) + \frac{\kappa C(\delta)}{2} \Psi^-_{\delta+\frac{7}{3}}(0) \bigg) \, .
\end{align}
The approximation for the characteristic depth is only accurate when $\beta \approx 0$. The homogeneous solution from equation~(\ref{eq:sec2_k-weighted-radius}) is
\begin{equation}
  \frac{r_c}{Z} = \left( C(\delta) \bar{\Psi}^{-}_{\delta+\frac{7}{3}} + \delta \bar{\Psi}^{+}_{\frac{10}{3}} \right) \bigg/ \left( C(\delta) \Psi^-_{\delta+\frac{7}{3}}(0) + \delta \Psi^+_{\frac{10}{3}}(0) \right) \, ,
\end{equation}
where for compactness of notation we let $\bar{\Psi}^{\pm}_s = \Psi^{\pm}_s(0) - \Psi^{\mp}_{s+1}(0)$. Figures~\ref{fig:sec5_peak_pos_trapping_time} and \ref{fig:sec5_peak_trapping_time} illustrate the dependence of these quantities on the power-law parametrizations $\delta$ and $\beta$. To evaluate the accuracy of these approximate analytic solutions we also show the exact results based on numerical calculations of the full series solution.

We note that $\delta = 0$ corresponds to a point source, $\delta = 1$ to a uniform source ($\alpha = \beta$), and $\delta = 2$ to a linear source ($\alpha = 2 \beta + 1$). In these cases, $C(0) = 1$ ($\Psi$ reduces to the hyperbolic secant), $C(1) = 0$, and $C(2) = -2$, so the solutions are as expected. Therefore, we have found a general solution representing a continuous transition between central ($\delta = 0$) and uniform ($\delta = 1$) sources. It is also worth mentioning that $\beta = 0$ corresponds to a power-law emission profile coupled to a homogeneous environment, and the solution in this case is given by setting $\delta = \alpha + 1$. Finally, motivated by two-body emission processes, such as recombination and collisional excitation emission, it is interesting to consider the relation $\eta \propto k^2$, which corresponds to $\alpha = 2 \beta$ or more explicitly $\delta = 2 - 1 / (\beta+1)$. We simply note that the range of solutions is limited to being between the central and uniform cases, as $\beta \in (-1/2,0)$ implies $\delta \in (0,1)$.

\section{Spherical Geometry: Homogeneous Case}
\label{sec:Sol_for_spherical_geometry}
Previously, \citet{Dijkstra2006} generalized the uniform slab solution to an equivalent uniform sphere solution. Similarly, now that we have discussed slab geometry we proceed to investigate spherically symmetric solutions. In this case the radiation field is represented as a function of radius and frequency, $J = J(r,x)$, and as in previous sections we transform to normalized optical depth coordinates
\begin{equation}
  \tilde{r} = \int_0^r \frac{k(r')}{\tau_0}\,\text{d}r' \, ,
\end{equation}
with $\tilde{r} \in (0,1)$. At this point we rewrite equation~(\ref{eq:final_transfer_eq}) for the specific case of a static, isothermal, optically-thick sphere
\begin{equation} \label{eq:sec5_initial_eq}
  r^{-2} \frac{\partial}{\partial \tilde{r}} \left( r^2 \frac{\partial \tilde{J}}{\partial \tilde{r}} \right) + \frac{\partial^2 \tilde{J}}{\partial \tilde{x}^2} = -\frac{\eta(\tilde{r})}{k(\tilde{r})}\delta(\tilde{x}) \, .
\end{equation}
This time the radial dependence of the absorption coefficient cannot be transformed out. In other words, we cannot derive a solution without first specifying $k(r)$ to determine the inverse relation $r(\tilde{r})$. We also require an additional boundary condition that the eigenfunctions remain finite in the limit as $r \rightarrow 0$.

For simplicity, we first consider the solution for a sphere of constant density, such that $k(r) = k_0$, $\tau_0 = k_0 R$, and $\tilde{r} = r/R$. This condition is relaxed in Section~\ref{sec:spherical-power-law} where we generalize to power-law dependence for the absorption coefficient. The geometric weight factor $r^2$ is related to the determinant of the Jacobian matrix for the coordinate system. We are free to choose the constant of proportionality, so for notational simplicity we let $\text{d}\tilde{V} \rightarrow 2 \tilde{r}^2\,\text{d}\tilde{r}$ and reserve an extra factor of $2\pi R^2$ for the $N_\text{scat}$ and $M_\text{F}$ volume integrals.
Equation~(\ref{eq:sec5_initial_eq}) reduces to the homogeneous equation, $\vartheta_n'' + 2 \vartheta_n'/\tilde{r} + \lambda_n^2 \vartheta_n = 0$, which has solutions of the form
\begin{equation} \label{eq:sec5_homogeneous_sphere_eigenfunction}
  \vartheta_n = \frac{\sin(\lambda_n \tilde{r})}{\tilde{r}} \qquad \text{where} \quad n = 1, 2, \ldots \, ,
\end{equation}
where the length of the eigenfunction is approximately unity,
\begin{equation} \label{eq:sec5_nor_of_eigenfunction}
  \int_0^1 2 \tilde{r}^2 \left(\frac{\sin(\lambda_n \tilde{r})}{\tilde{r}}\right)^2\,\text{d}\tilde{r} = 1 - \frac{\sin(2 \lambda_n)}{2 \lambda_n} \approx 1 \, ,
\end{equation}
and the boundary conditions require the eigenvalues to satisfy
\begin{equation} \label{eq:sec5_BCs_of_eigenfunction_sphere}
  \lambda_n \cot(\lambda_n) = 1 - f \tau_0 H(\tilde{x}) \approx -f \tau_0 H(\tilde{x}) \, .
\end{equation}
Thus, the eigenvalue to zeroth order is approximately
\begin{equation} \label{eq:sec5_homogeneous_sphere_eigenvalue}
  \lambda_n = \pi n + \tan^{-1}\left(\frac{\lambda_n}{1 - f \tau_0 H(\tilde{x})}\right) \approx \pi n \, .
\end{equation}
Furthermore, the emission constants from equation~(\ref{eq:general_Qn}) are
\begin{equation} \label{eq:sec5_Qn}
  Q_n = 2 R \int_0^1 \tilde{r} \eta(\tilde{r}) \sin(\lambda_n \tilde{r})\,\text{d}\tilde{r} \, ,
\end{equation}
volume-weighted averages from equation~(\ref{eq:sec2_k-weight}) are
\begin{equation} \label{eq:sec5_Kn}
  T_n = 3 \int_0^1 \tilde{r} \sin(\lambda_n \tilde{r})\,\text{d}\tilde{r}
  \approx -\frac{3 \cos(\lambda_n)}{\lambda_n} \approx \frac{3 (-1)^{n-1}}{\lambda_n} \, ,
\end{equation}
and the factor for the characteristic radius is
\begin{equation} \label{eq:sec5_Rn}
  R_n = 3R\int_0^1 \tilde{r}^2 \sin(\lambda_n \tilde{r})\,\text{d}\tilde{r}
  \approx R\left[ K_n - \frac{6(1-(-1)^n)}{\lambda_n^3} \right] \, .
\end{equation}
The final solution is given by
\begin{equation} \label{eq:sec5_J(r,x)}
  J(\tilde{r},\tilde{x}) = \frac{\mathcal{L} \sqrt{6}}{8\pi} \sum_{n=1}^\infty \frac{Q_n}{\lambda_n} \frac{\sin(\lambda_n \tilde{r})}{\tilde{r}} e^{-\lambda_n |\tilde{x}|} \, .
\end{equation}
Using equation~(\ref{eq:sec5_BCs_of_eigenfunction_sphere}), i.e. $\sin(\lambda_n) \approx (-1)^{n-1} \lambda_n/ f \tau_0 H(\tilde{x})$, the spectral line profile at the boundary is
\begin{equation} \label{eq:sec5_general_solution_J}
  J(\tilde{x}) = \frac{\mathcal{L} \sqrt{6}}{8\pi} \frac{1}{f \tau_0 H(\tilde{x})} \sum_{n=1}^\infty (-1)^{n-1} Q_n e^{-n \pi |\tilde{x}|} \, \, ,
\end{equation}
with normalization $\int_{-\infty}^{+\infty} J(x) \text{d}x  = \frac{3 \mathcal{L}}{4 \pi^2 f} \sum_{n=1}^{\infty}(-1)^{n-1} Q_n / n$. From equation~(\ref{eq:sec2_u(r)}), the radiation energy density is
\begin{equation} \label{eq:sec5_u(r)}
  u(\tilde{r}) = \frac{\mathcal{L}}{c} \Gamma\left(\frac{1}{3}\right) \left(\frac{2 a \tau_0}{\sqrt{\pi}}\right)^{1/3} \sum_{n=1}^\infty \frac{Q_n}{\lambda_n^{4/3}} \frac{\sin(\lambda_n \tilde{r})}{\tilde{r}} \, ,
\end{equation}
from equation~(\ref{eq:sec2_Jxk}) the average internal spectrum is
\begin{equation} \label{eq:sec5_Jxk}
  \langle J(\tilde{x}) \rangle = \frac{3\mathcal{L}\sqrt{6}}{8\pi} \sum_{n=1}^\infty (-1)^{n-1} \frac{Q_n}{\lambda_n^2} e^{-\lambda_n |\tilde{x}|} \, ,
\end{equation}
from equation~(\ref{eq:sec2_uk}) the average radiation energy density is
\begin{equation} \label{eq:sec5_uk}
  \langle u \rangle = \frac{3\mathcal{L}}{c} \Gamma\left(\frac{1}{3}\right) \left(\frac{2 a \tau_0}{\sqrt{\pi}}\right)^{1/3} \sum_{n=1}^\infty (-1)^{n-1} \frac{Q_n}{\lambda_n^{7/3}} \, ,
\end{equation}
and from equations~(\ref{eq:sec2_k-weighted-radius}) and (\ref{eq:sec5_Rn}) the characteristic radius is
\begin{equation}
  \frac{r_c}{R} = 1 - 2 \frac{\sum_{n=1}^{\infty}Q_n \left(1-(-1)^n\right) \lambda_n^{-13/3}}{\sum_{n=1}^{\infty}Q_n(-1)^{n-1}\lambda_n^{-7/3}} \, .
\end{equation}
With $4\pi R^2 \int \tilde{r}^2 \frac{\text{d}}{\text{d}\tilde{r}}[\sin(\lambda_n \tilde{r}) / \tilde{r}] \text{d}\tilde{r} \approx -16 \pi R^2 [1 - (-1)^n] / \lambda_n$, the force multiplier from equation~(\ref{eq:sec2_M_F}) becomes
\begin{equation}
  M_\text{F} = 8\pi R^2 \Gamma\left(\frac{4}{3}\right) \left(\frac{2 a \tau_0}{\sqrt{\pi}}\right)^{1/3} \sum_{n=1}^\infty  \frac{\left(1 - (-1)^n\right)Q_n}{\lambda_n^{7/3}} \, ,
\end{equation}
and with $4\pi R^2 \int \tilde{r} \sin(\lambda_n \tilde{r}) \text{d}\tilde{r} \approx 4\pi R^2 (-1)^{n-1} / \lambda_n$ the average number of scatterings before escapes from equation~(\ref{eq:sec2_N_scat}) is
\begin{equation}
  N_\text{scat} = \sqrt{24} \pi^{3/2}\,\tau_0 R^2 \sum_{n=1}^\infty \frac{(-1)^{n-1} Q_n}{\lambda_n^2} \, .
\end{equation}
In the remaining subsections we consider specific cases for $Q_n$. In this case the average trapping time normalized to known factors ($t_\text{light} = R/c$) is simply $t_\text{trap}/t_\text{light} = 4 \pi c R^2 \langle u \rangle / 3 \mathcal{L}$.

\begin{figure}
  \centering
  \includegraphics[width=\columnwidth]{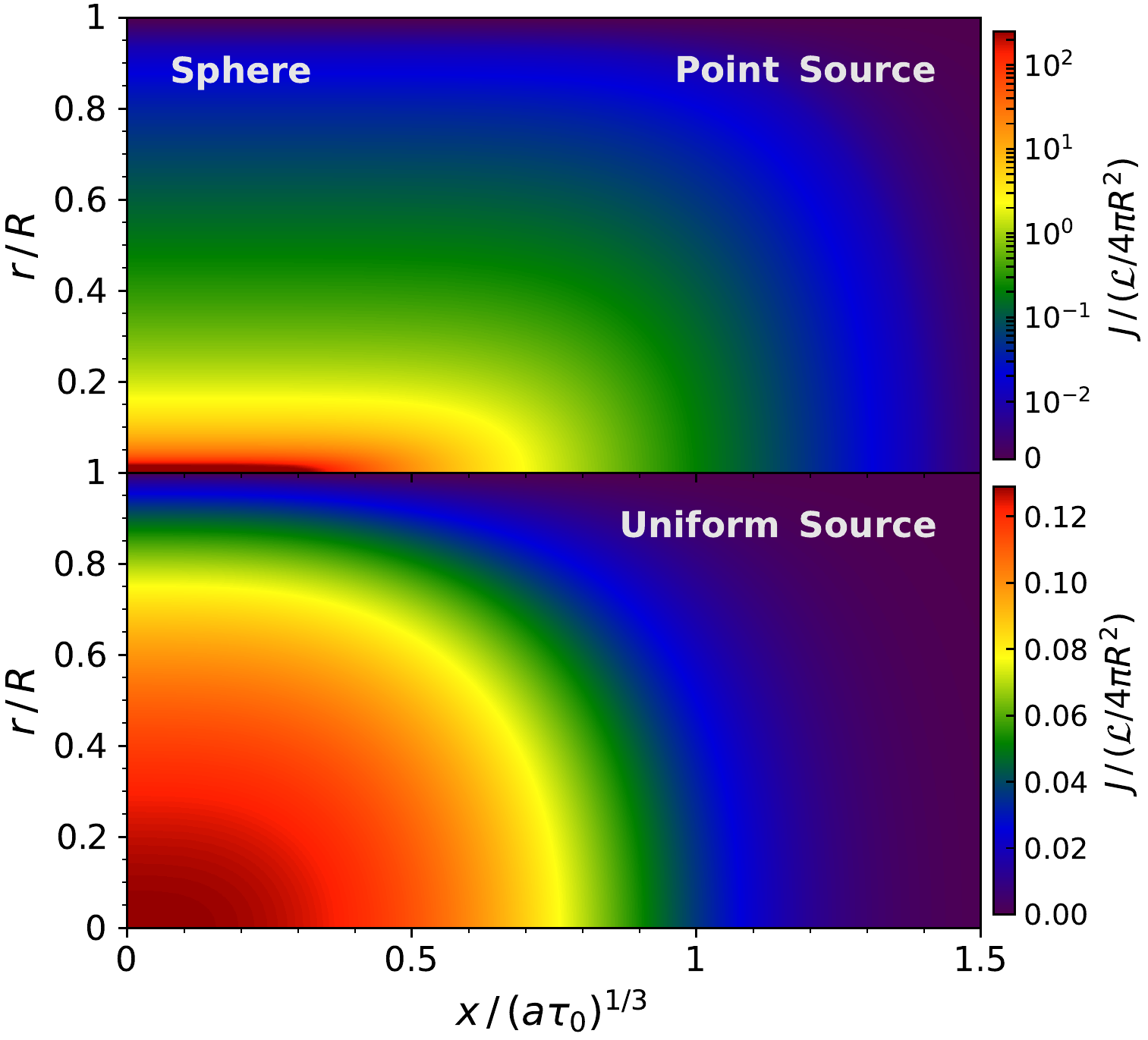}
  \caption{The internal radiation field $J$ as a function of frequency $x$ and radius $r$ for a central point source and a uniform source in a homogeneous sphere (equations~\ref{eq:sec5-J(r,x)-PS} and \ref{eq:sec5-J(r,x)-US}). The point source leads to a singularity at the origin, even though the total energy is finite. The uniform source distributes the emission throughout the volume so the intensity flattens off with much more gradual changes. All results in this study are symmetric with respect to frequency so we only plot the positive half ($x > 0$).}
  \label{fig:sec5_inertial_spectrum}
\end{figure}

\subsection{Central point source}
For a point source we have $\eta(r) = \delta(r) / (4 \pi r^2) = \delta(\tilde{r}) / (4 \pi \tilde{r}^2 R^3)$ and $Q_n = \lambda_n / (2 \pi R^2)$. Therefore, the full solution reduces to
\begin{equation} \label{eq:sec5-J(r,x)-PS}
  J(\tilde{r},\tilde{x}) = \frac{\mathcal{L}\sqrt{6}}{32 \pi^2 R^2 \tilde{r}} \frac{\sin(\pi \tilde{r})}{\cosh(\pi \tilde{x}) - \cos(\pi \tilde{r})} \, ,
\end{equation}
as shown in Figure~\ref{fig:sec5_inertial_spectrum}. The spectral line profile at the boundary is
\begin{equation} \label{eq:sec5_point_source}
  J(\tilde{x}) = \frac{\mathcal{L}\sqrt{6}}{64\pi R^2 f \tau_0 H(\tilde{x})} \text{sech}^2 \left(\frac{\pi \tilde{x}}{2}\right) \, .
\end{equation}
After transforming back to the original frequency notation, we arrive at the final normalized version \citep{Dijkstra2006}:
\begin{equation} \label{eq:sec4_CPS}
  \frac{J(x)}{\int_{-\infty}^{+\infty}J(x)\,\text{d}x} = \sqrt{\frac{\pi^3}{24}} \frac{x^2}{a\tau_0} \text{sech}^2\left( \sqrt{\frac{\pi^3}{54}}\frac{x^3}{a\tau_0}\right) \, .
\end{equation}
Thus, for a point source in a homogeneous sphere the frequency peak positions are located at $x_\text{p} = \pm 0.93099\,(a\tau_0)^{1/3}$, calculated from the equation $\bar{x}\tanh \bar{x} = 1/3$, where $\bar{x} = \sqrt{\pi^3/54}x^3/a\tau_0$.

From equation~(\ref{eq:sec5_u(r)}) the radiation energy density is
\begin{equation}
  u(\tilde{r}) = \frac{\mathcal{L}}{c R^2} \Gamma\left(\frac{1}{3}\right) \frac{(2 a \tau_0)^{1/3}}{2 \pi^{3/2}} \tilde{r}^{-1} \text{Im}\left[\text{Li}_{1/3}\left(e^{i \pi \tilde{r}}\right)\right] \, ,
\end{equation}
where $\text{Li}_s(z)$ is the polylogarithm function defined as $\text{Li}_s(z)=\sum_{n=1}^{\infty}z^n/n^s$.
From equation~(\ref{eq:sec5_Jxk}) the average internal spectrum is
\begin{equation}
  \langle J(\tilde{x}) \rangle = \frac{3\sqrt{6}\mathcal{L}}{16 \pi^3 R^2}\ln\left(1+e^{-\pi |\tilde{x}|}\right) \, ,
\end{equation}
from equation~(\ref{eq:sec5_uk}) the average radiation energy density is
\begin{equation}
  \langle u \rangle = \frac{3 \mathcal{L} (a\tau_0)^{1/3}}{2 \pi^{5/2} c R^2} \Gamma\left(\frac{1}{3}\right) \left(\sqrt[3]{2} - 1\right) \zeta\left(\frac{4}{3}\right) \, ,
\end{equation}
or an equivalent trapping time of $t_\text{trap} / t_\text{light} \approx 0.901\,(a\tau_0)^{1/3}$. Likewise, the outward force multiplier is
\begin{equation}
  \frac{M_\text{F}}{(a\tau_0)^{1/3}} = \frac{4}{\pi^{3/2}} \left(2^{4/3} - 1\right) \Gamma\left(\frac{4}{3}\right) \zeta\left(\frac{4}{3}\right) \approx 3.51 \, ,
\end{equation}
and the number of scatters is $N_\text{scat} / \tau_0 = \ln(2) \sqrt{6/\pi} \approx 0.958$.
From the expressions for $R_n$ and $T_n$, we derive a characteristic radius of
\begin{equation}
  \frac{r_c}{R} = 1 - \left(\frac{15}{7} + 2^{1/3} + 2^{2/3}\right) \frac{7 \zeta(10/3)}{2 \pi^2 \zeta(4/3)} \approx 0.44 \, .
\end{equation}

\subsection{Uniform source}
If the emissivity traces the absorption coefficient then $\eta(r) = 3 / (4\pi R^3)$. Therefore, the constants from equation~(\ref{eq:sec5_Qn}) are
\begin{equation}
  Q_n = \frac{3}{2\pi R^2} \int_0^1 \tilde{r} \sin(\lambda_n \tilde{r})\,\text{d}\tilde{r} \approx -\frac{3\cos(\lambda_n)}{2\pi R^2 \lambda_n} \approx \frac{3 (-1)^{n-1}}{2\pi R^2 \lambda_n} \, .
\end{equation}
The general expression from equation~(\ref{eq:sec5_J(r,x)}) reduces to
\begin{equation} \label{eq:sec5-J(r,x)-US}
  J(\tilde{r},\tilde{x}) = \frac{3\sqrt{6}\mathcal{L}}{16 \pi^4 R^2 \tilde{r}} \text{Im}\left[\text{Li}_2\left(-e^{-\pi (|\tilde{x}| + i \tilde{r})}\right)\right] \, ,
\end{equation}
which is shown in Figure~\ref{fig:sec5_inertial_spectrum}. The line profile from equation~(\ref{eq:sec5_general_solution_J}) is
\begin{equation} \label{eq:sec5_J(x)_uniform}
  J(\tilde{x}) = \frac{3\sqrt{6}\mathcal{L}}{16 \pi^3 R^2 f \tau_0 H(\tilde{x})} \text{ln}\left(\frac{1}{1 - e^{-\pi |\tilde{x}|}}\right) \, .
\end{equation}
After transforming back to the original frequency notation, we arrive at the final normalized version:
\begin{equation} \label{eq:sec4_US}
  \frac{J(x)}{\int_{-\infty}^{+\infty}J(x)\,\text{d}x} = -\sqrt{\frac{6}{\pi}} \frac{x^2}{a\tau_0} \ln\left( 1 - e^{-\sqrt{\frac{2\pi^3}{27}}\frac{|x^3|}{a\tau_0}}\right) \, .
\end{equation}
In this case the peaks are located at $x_\text{p} = \pm0.6026\,(a \tau_0)^{1/3}$, calculated from $3\bar{x}\,[1 - \text{coth}(\bar{x})] = 2 \text{ln}[1 - \exp(-2\bar{x})]$, with $\bar{x}$ as before. We derive the following spatial and spectral integrated quantities:
\begin{equation} \label{eq:sec5_u(z)_uniform}
  u(\tilde{r}) = \frac{3\mathcal{L}}{c R^2} \Gamma\left(\frac{1}{3}\right) \frac{(2 a \tau_0)^{1/3}}{2 \pi^{7/2}} \tilde{r}^{-1} \text{Im}\left[\text{Li}_{7/3}\left(-e^{-i \pi \tilde{r}}\right)\right] \, ,
\end{equation}
\begin{equation} \label{eq:sec5_J_avg_uniform}
  \langle J(\tilde{x}) \rangle= \frac{9\sqrt{6}\mathcal{L}}{16\pi^5 R^2} \text{Li}_3\left(e^{-\pi |\tilde{x}|}\right) \, ,
\end{equation}
and
\begin{equation} \label{eq:sec5_u_avg_uniform}
  \langle u \rangle = \frac{9\mathcal{L}}{c R^2} \Gamma\left(\frac{1}{3}\right) \frac{(2 a \tau_0)^{1/3}}{2 \pi^{9/2}} \zeta\left(\frac{10}{3}\right) \, ,
\end{equation}
such that the trapping time is $t_\text{trap} / t_\text{light} \approx 0.423\,(a\tau_0)^{1/3}$,
\begin{equation}
  \frac{M_\text{F}}{(a\tau_0)^{1/3}} = \frac{(8 \sqrt[3]{2} - 1)}{\pi^{7/2}} \Gamma\left(\frac{1}{3}\right) \zeta\left(\frac{10}{3}\right) \approx 0.51 \, ,
\end{equation}
the number of scatters is $N_\text{scat} / \tau_0 = 3\sqrt{6} \zeta(3)/\pi^{5/2} \approx 0.505$, and
\begin{equation}
  \frac{r_c}{R} = 1 + \left(2^{2/3} - 64\right) \frac{\zeta(16/3)}{16 \pi^2 \zeta(10/3)}
  \approx 0.65 \, .
\end{equation}

\begin{figure}
  \centering
  \includegraphics[width=\columnwidth]{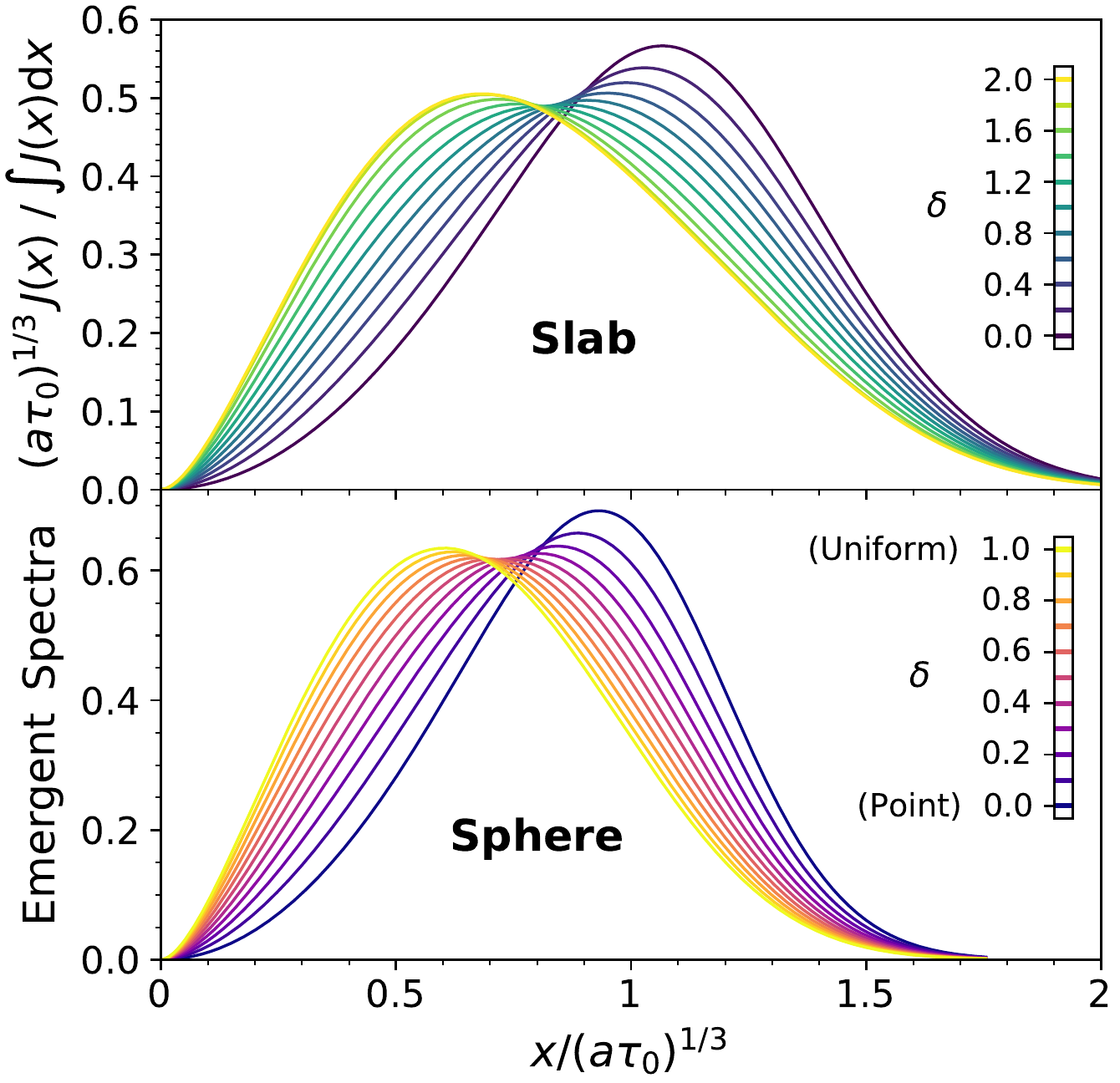}
  \caption{The emergent spectral line profile $J(x)$ for different values of $\delta$ for the homogeneous slab and sphere cases (equations~\ref{eq:sec4_generalized_sol} and \ref{eq:sec5_generalized_sol}). The curves illustrate the continuous transition between a central point source ($\delta = 0$) and a uniform source ($\delta = 1$). This is not a trivial interpolation between solutions but is a nonlinear family of functions. As delta increases the peak shifts towards line centre and the profile shape becomes more skewed.}
  \label{fig:combination_Jx}
\end{figure}

\subsection{Power-law profiles}
We now consider a power-law profile for the emissivity term. This again results in a single parameter representing the continuous transition from inward to outward opacity. We give the (normalized) emissivity as $\eta(r) = \eta_0 r^\alpha$, where $\eta_0 = (\alpha+3)/4\pi R^{\alpha+3}$ with $\alpha \geq -3$. Therefore, the constant from equation~(\ref{eq:sec5_Qn}) is
\begin{align} \label{eq:sec5_1F2}
  Q_n &= \frac{(\alpha + 3)}{2 \pi R^2} \int_0^1 \tilde{r}^{\alpha+1} \sin(\lambda_n \tilde{r})\,\text{d}\tilde{r} \notag \\
  &= \frac{\lambda_n}{2 \pi R^2}\,_1F_2\left(\frac{3\delta
    }{2};\frac{3}{2},\frac{3\delta}{2}+1;-\frac{\lambda_n^2}{4}\right) \, .
\end{align}
which mirrors the result from equation~(\ref{eq:sec4_Qn_generalization}) in Section~\ref{sec:General solution for slab geometry}. In this case $\delta = (\alpha+3)/3$ with $\delta > 0$, chosen to represent the transition from a point source ($\delta = 0$) to a uniform source ($\delta = 1$). To make further progress, we again use a Taylor expansion of $\lambda_n$ at $\infty$ to approximate the hypergeometric function noting that $\lambda_n \approx n\pi$:
\begin{equation} \label{eq:sec5_approx_1F2}
  Q_n \approx \frac{\lambda_n}{2 \pi R^2} \left[ \frac{C(\delta)}{\lambda_n^{3\delta}} + \frac{3 \delta (-1)^{n-1}}{\lambda_n^2} + \mathcal{O} \left(\frac{1}{\lambda_n^4}\right) \right] \, ,
\end{equation}
where $C(\delta) = \cos(3\pi \delta/2) \Gamma(1+3\delta)/(1 - 3 \delta)$.
Although the approximation will introduce some error, it still provides accurate expressions describing the transition between point and uniform sources. The behaviour of $Q_n$ and the relative error introduced by equation~(\ref{eq:sec5_approx_1F2}) is illustrated in Figure~\ref{fig:appendix_sphere_error}.
After substitution into equation~(\ref{eq:sec5_J(r,x)}) the final expression becomes
\begin{equation} \label{eq;sec5_general_J}
  J(\tilde{r},\tilde{x}) = \frac{\mathcal{L}\sqrt{6} C(\delta)}{16 \pi^{3 \delta + 2} R^2} \tilde{r}^{-1} \text{Im}\left[\text{Li}_{3\delta}\left(e^{-\pi \psi}\right)\right] + \delta J_\text{uni}(\tilde{r},\tilde{x}) \, ,
\end{equation}
where $\psi = |\tilde{x}| - i\tilde{r}$ and we have neglected the higher order term in the equation~(\ref{eq:sec5_approx_1F2}).
The spectral line profile at the boundary is
\begin{equation} \label{eq:sec5_generalized_sol}
  J(\tilde{x}) = \frac{-\mathcal{L}\sqrt{6} C(\delta)}{16\pi^{3\delta+1} R^2 f \tau_0 H(\tilde{x})} \text{Li}_{3\delta-1}\left(-e^{-\pi |\tilde{x}|}\right) + \delta J_\text{uni}(\tilde{x}) \, ,
\end{equation}
where $J_\text{uni}(\tilde{x})$ is the result from equation~(\ref{eq:sec5_J(x)_uniform}) with normalization
\begin{equation} \label{eq:sec5_Normalization_general_sol}
  \int_{-\infty}^{+\infty} J(\tilde{x})\,\text{d}x = \frac{3\mathcal{L}}{8 f R^2} \left[\frac{(1-2^{1-3\delta}) C(\delta)\zeta(3\delta)}{\pi^{3\delta+2}} + \frac{\delta}{2\pi^2}\right] \, .
\end{equation}
This solution is a generalization of the point source solution in equation~(\ref{eq:sec5_point_source}) mixed with the uniform source solution from equation~(\ref{eq:sec5_J(x)_uniform}). We now derive the peak positions from equation~(\ref{eq:sec5_generalized_sol}). For simplicity we set $\bar{x} = \sqrt{2\pi^3/27} x^3 / a\tau_0$ with $x > 0$, such that the peaks satisfy the following transcendental equation
\begin{align}
  C(\delta) &\left[3 \bar{x} \text{Li}_{3 \delta -2}\left(-e^{-\bar{x}}\right)-2 \text{Li}_{3 \delta  -1}\left(-e^{-\bar{x}}\right)\right] = \notag \\
    &3 \pi^{3\delta-2} \delta\left[  2\ln \left(1-e^{-\bar{x}}\right) +\frac{3 \bar{x}}{e^{\bar{x}}-1}\right] \, ,
\end{align}
the solutions of which are shown in Figure~\ref{fig:sec5_peak_pos_trapping_time}.

\begin{figure}
  \centering
  \includegraphics[width=\columnwidth]{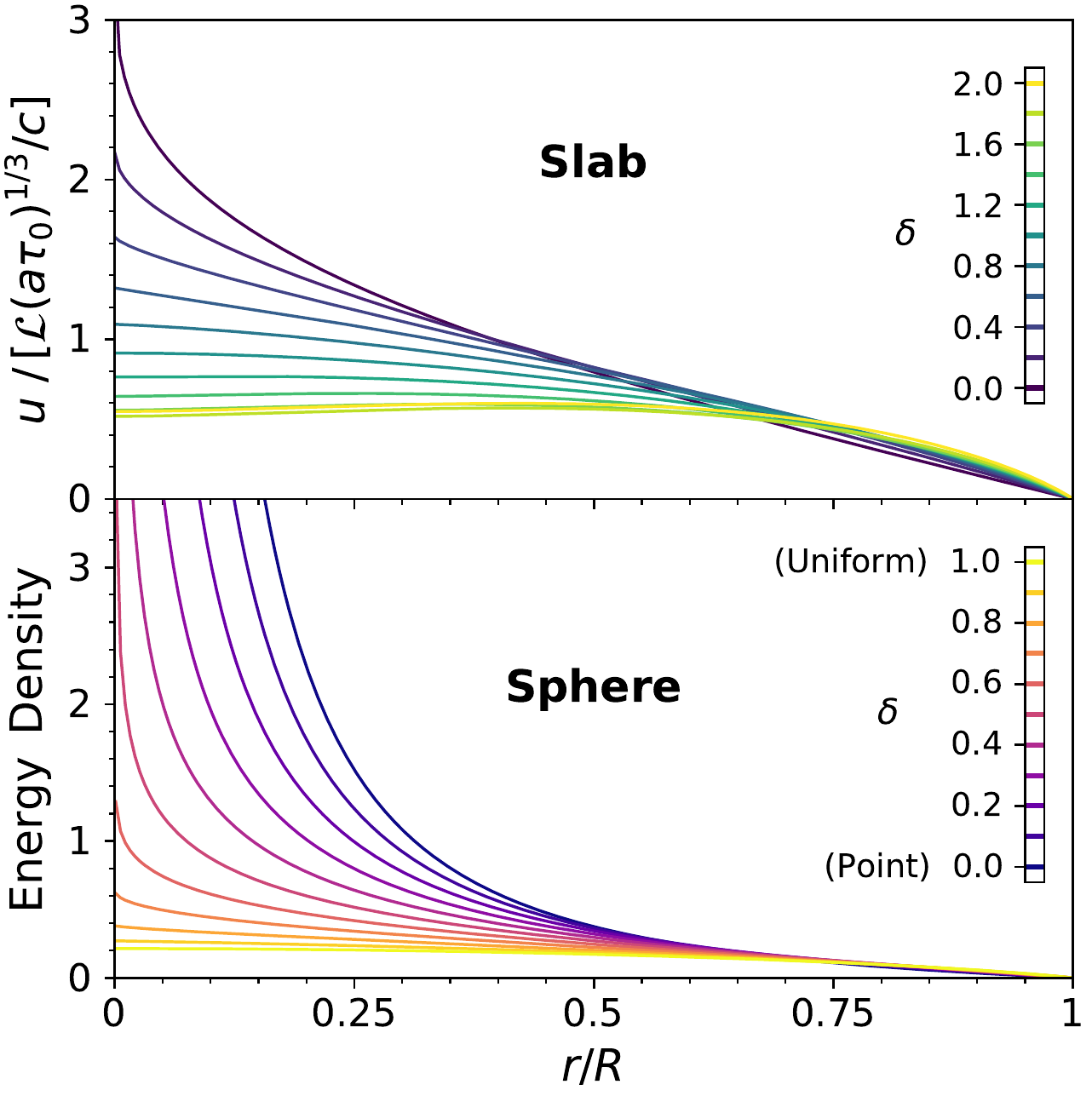}
  \caption{The internal energy density $u(r)$ for different values of $\delta$ for the homogeneous slab and sphere cases (equations~\ref{eq:sec4_u(z)_PL} and \ref{eq:sec5_u(r)_PL}). A higher value of $\delta$ results in flatter radiation profiles around the geometric centre. The spherical case contains a singularity at the origin for all $\delta > 0$ due to the presence of $r$ in the denominator.}
  \label{fig:sec5_energy_density}
\end{figure}

The spatial and spectral integrated quantities are: the radiation energy density,
\begin{equation} \label{eq:sec5_u(r)_PL}
  u(\tilde{r}) = \frac{\mathcal{L} C(\delta) \Gamma(\frac{1}{3}) \left(2 a \tau_0\right)^{\frac{1}{3}}}{2 \pi^{3\delta+3/2} c R^2 \tilde{r}} \text{Im}\left[\text{Li}_{3\delta+\frac{1}{3}}\left(e^{i \pi \tilde{r}}\right) \right] + \delta u_\text{uni}(\tilde{r}) \, ,
\end{equation}
with $u_\text{uni}(\tilde{z})$ from equation~(\ref{eq:sec5_J(x)_uniform}), the internal spectrum,
\begin{equation} \label{eq:sec5_J_avg_PL}
  \langle J(\tilde{x})\rangle = \frac{-3\sqrt{6}C(\delta)\mathcal{L}}{16 \pi^{3\delta+3} R^2} \text{Li}_{3\delta+1}\left(-e^{-\pi |\tilde{x}|}\right) + \delta \langle J_\text{uni}(\tilde{x})\rangle \, ,
\end{equation}
with $\langle J_\text{uni}(\tilde{x})\rangle$ from equation~(\ref{eq:sec5_J_avg_uniform}), and finally
\begin{equation}
  \langle u \rangle = \frac{3\mathcal{L} C(\delta) \Gamma(\frac{1}{3}) \left(a \tau_0\right)^{\frac{1}{3}}}{2 \pi^{3\delta+5/2} c R^2} \frac{\zeta\left(3\delta + 4/3\right)}{\left(2^{1/3} - 2^{-3 \delta}\right)^{-1}} + \delta\langle u_\text{uni} \rangle \, ,
\end{equation}
with $\langle u_\text{uni} \rangle$ from equation~(\ref{eq:sec5_u_avg_uniform}). The characteristic radius is
\begin{equation}
  \frac{r_c}{R} =  1 + \frac{2}{\pi^2} \frac{C(\delta) \Theta(3\delta+10/3) + 3\pi^{3\delta-2} \delta \Theta(16/3)}{C(\delta) \text{Li}_{3\delta + 4/3}(-1) - 3\pi^{3\delta-2} \delta \zeta(10/3)} \, ,
\end{equation}
where $\Theta(z) \equiv \zeta(z) - \text{Li}_z(-1)$. The outward force multiplier is
\begin{equation}
  \frac{M_\text{F}}{(a\tau_0)^{1/3}} = \frac{4 \sqrt[3]{2}}{\pi^{3/2}} \Gamma\left(\frac{4}{3}\right) \left[\frac{C(\delta)}{\pi^{3\delta}}\Theta\left(3\delta+\frac{4}{3}\right) + \frac{3\delta}{\pi^2}\Theta\left(\frac{10}{3}\right)\right] \, ,
\end{equation}
and the number of scatters is
\begin{equation}
  \frac{N_\text{scat}}{\tau_0} = \sqrt{6\pi} \left(\left(1 - 8^{-\delta}\right) \frac{C(\delta)}{\pi^{3\delta+1}} \zeta(3\delta+1) + \frac{3\delta}{\pi^3} \zeta(3)\right) \, .
\end{equation}
These quantities are also shown in Figures~\ref{fig:sec5_peak_pos_trapping_time} and \ref{fig:sec5_peak_trapping_time} to illustrate the dependence on the power-law parameterization $\delta$. The behaviour in spherical geometry is qualitatively similar to the slab case, but with slightly different structure.

We note that $\delta = 0$ and $\delta = 1$ again correspond to point ($\alpha = -3$) and uniform ($\alpha = 0$) sources, respectively. In these cases, $C(0) = 1$ (Li reduces to the hyperbolic secant) and $C(1) = 0$, so the solutions are as expected. It is particularly interesting that the non-integrable power-law profiles with $\delta = 0$ exhibit the same physical meaning as the Dirac-delta function and lead to the same solutions.
There are other special values for $\delta$ that also deserve our attention. First is the case when $\delta = 1/3$, or $\alpha = -2$, for which the zeta function in the normalization factor of equation~(\ref{eq:sec5_Normalization_general_sol}) is not convergent. However, it is relatively straightforward to find that the emergent line profile from equation~(\ref{eq:sec5_generalized_sol}) in this case can be reduced to the following simple form, noting that $\lim_{\delta \rightarrow 1/3} C(\delta) = \pi/2$:
\begin{equation}
  J(\tilde{x}) = \frac{\mathcal{L}\sqrt{6}}{32 \pi^3 R^2 f \tau_0 H(\tilde{x})} \left[ \frac{\pi^2}{1+e^{\pi \left|x\right|}} - 2 \ln\left(1-e^{-\pi \left|\tilde{x}\right|}\right)\right] \, ,
\end{equation}
with a normalization of $\int_{-\infty}^{+\infty} J(\tilde{x})\,\text{d}x = \mathcal{L} (1 + \ln 8) / (16 f \pi^2 R^2)$.
Finally, noticing that when $\delta = 2/3$, or $\alpha = -1$, the approximation in equation~(\ref{eq:sec5_approx_1F2}) has no error, we find the emergent spectra in this case simplifies to the following, noting that $C(2/3) = 2$:
\begin{equation}
  J(\tilde{x}) = \frac{\mathcal{L}\sqrt{6}}{4 \pi^3 R^2 f \tau_0 H(\tilde{x})} \text{coth}^{-1}\left(e^{\pi \left|\tilde{x}\right|}\right) \, ,
\end{equation}
with a normalization of $\int_{-\infty}^{+\infty} J(\tilde{x})\,\text{d}x = 3 \mathcal{L} / (16 f \pi^2 R^2)$.

\begin{figure}
  \centering
  \includegraphics[width=\columnwidth]{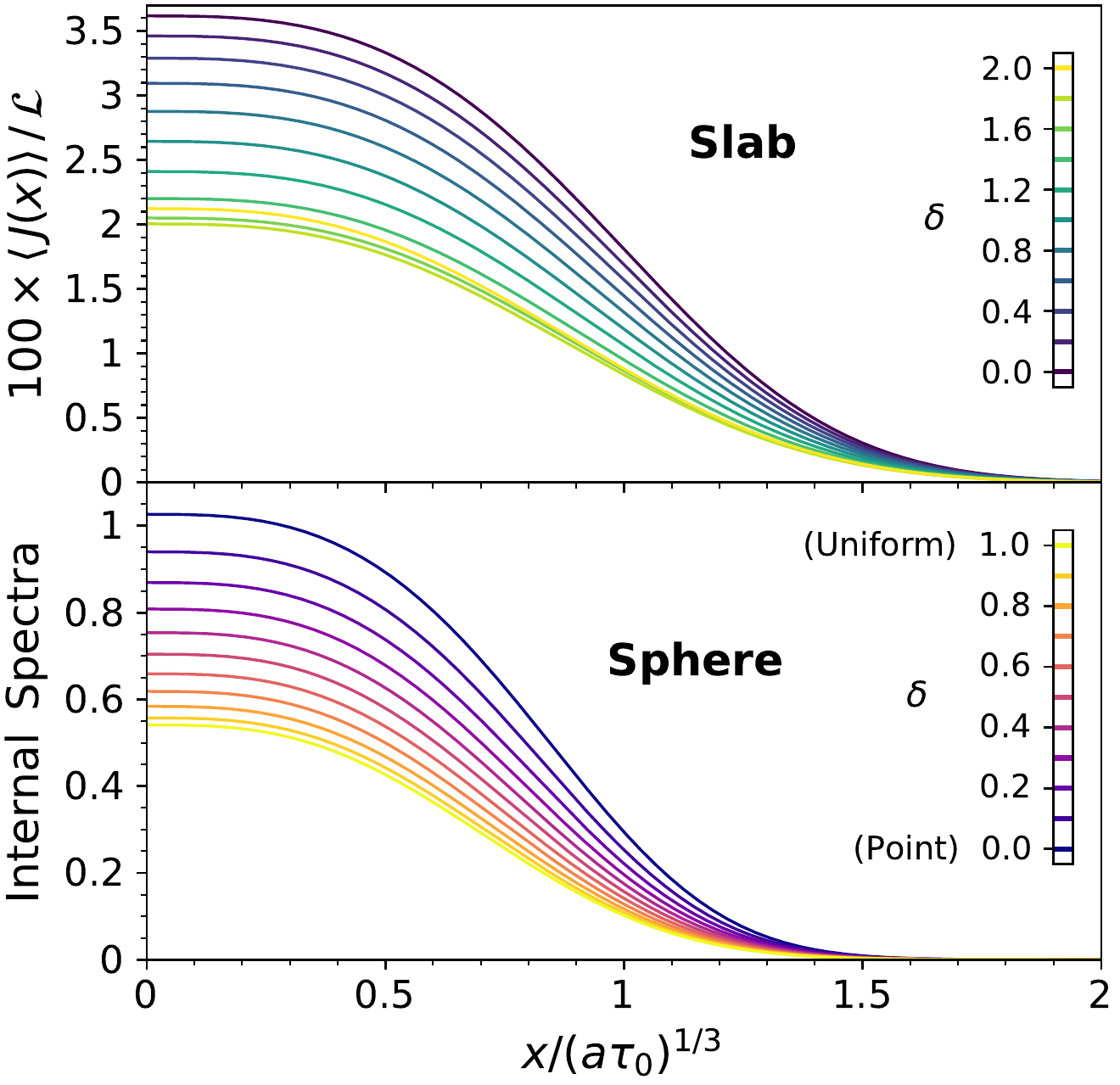}
  \caption{The volume-averaged internal spectral profile $\langle J \rangle$ for different values of $\delta$ for the homogeneous slab and sphere cases. The shapes are fairly similar due to the frequency diffusion process but the heights reflect the longer trapping times for more concentrated sources.}
  \label{fig:sec5_internal_spectrum}
\end{figure}

\section{Spherical geometry: power-law case}
\label{sec:spherical-power-law}
Unlike the plane parallel slab case, in spherical geometry the solutions are sensitive to changes in the density environment. This means that it is hopeful to infer some information about the structure of galaxies from spatial and spectral observational data. In the previous section we focused on homogeneous spheres, but we now generalize the derivations to include power-law profiles with $k(r) = k_0 r^\beta$. In this case we require $\beta \in (-1,0]$ so the total optical depth at line centre is finite, i.e. $\tau_0 = k_0 R^{\beta+1} / (\beta+1)$. As in previous sections we transform to normalized optical depth coordinates such that $\tilde{r} = (r/R)^{\beta+1}$. In this coordinate system the general equation~(\ref{eq:final_transfer_eq}) within a power-law density profile reduces to
\begin{equation} \label{eq:sec6_initial_eq}
  \tilde{r}^{-\kappa} \frac{\partial}{\partial \tilde{r}}\left( \tilde{r}^\kappa \frac{\partial \tilde{J}}{\partial \tilde{r}} \right) + \frac{\partial^2 \tilde{J}}{\partial \tilde{x}^2} = -\frac{\eta(\tilde{r})}{k(\tilde{r})} \delta(\tilde{x}) \, ,
\end{equation}
where $\kappa \equiv 2 / (\beta+1) > 2$. In this case we let $\text{d}\tilde{V} \rightarrow 2 \tilde{r}^\kappa \text{d}\tilde{r}$ and again reserve an extra factor of $2\pi R^2$ for the $N_\text{scat}$ and $M_\text{F}$ volume integrals. Therefore, equation~(\ref{eq:sec6_initial_eq}) reduces to the homogeneous equation, $\vartheta_n'' + \kappa \vartheta_n'/\tilde{r} + \lambda_n^2 \vartheta_n = 0$, which has solutions of the form
\begin{equation} \label{eq:sec6_homogeneous_sphere_eigenfunction}
  \vartheta_n = \tilde{r}^{1-\gamma} \frac{J_{\gamma-1}(\lambda_n \tilde{r})}{J_\gamma(\lambda_n)} \qquad \text{where} \quad n = 1, 2, \ldots \, ,
\end{equation}
where $\gamma \equiv (\kappa+1)/2 > 3/2$ and $J_\gamma$ denotes the Bessel function of the first kind. We note that we have discarded Bessel functions of the second kind to avoid singularities at the origin. From the boundary condition in equation~(\ref{eq:sec2_boundary condition}) the eigenvalues are required to satisfy
\begin{equation} \label{eq:sec6_eq_of_eigenvalue}
  \lambda_n J_\gamma(\lambda_n) = f \tau_0 H(\tilde{x}) J_{\gamma-1}(\lambda_n) \, .
\end{equation}
Recalling the discussion related to equation~(\ref{eq:sec4_keyapprox1}), we are considering optically thick environments for which $\lambda_n \ll f \tau_0 H(\tilde{x})$. Therefore, the length of the eigenfunction is approximately unity,
\begin{equation} \label{eq:sec6_nor_of_eigenfunction}
  \int_0^1 2 \tilde{r} \left(\frac{J_{\gamma-1}(\lambda_n \tilde{r})}{J_\gamma(\lambda_n)}\right)^2 \text{d}\tilde{r} = 1 - \frac{2(\gamma-1)}{f \tau_0 H(\tilde{x})} + \left(\frac{\lambda_n}{f \tau_0 H(\tilde{x})}\right)^2 \, .
\end{equation}
There are no accurate elementary approximations for the Bessel function eigenvalues in equation~(\ref{eq:sec6_eq_of_eigenvalue}) valid across the entire domain. However, we can capture the essential physics by appealing to an asymptotic expansion for large values, specifically,
\begin{equation} \label{eq:sec6_eq_of_approx_Bessel}
  J_\gamma(\lambda_n) = \sqrt{\frac{2}{\pi \lambda_n}} \cos\left(\lambda_n - \frac{\gamma\pi}{2} - \frac{\pi}{4}\right) + \mathcal{O}\left(\lambda_n^{-3/2}\right) \, ,
\end{equation}
which is valid when $\lambda_n \gg |(\gamma-1)^2 - 1/4| = \kappa (\kappa - 2) / 4$.
Therefore, the equation for the eigenvalues reduces to
\begin{equation} \label{eq:sec6_BCs_of_eigenfunction_sphere}
  \lambda_n \tan\left(\lambda_n - \frac{\kappa\pi}{4}\right) \approx f \tau_0 H(\tilde{x}) \, ,
\end{equation}
such that the eigenvalues to zeroth order are approximately
\begin{equation} \label{eq:sec6_power_law_eigenvalue}
  \lambda_n \approx \pi \left(n - 1 + \frac{\kappa}{4}\right) + \tan^{-1}\left(\frac{f \tau_0 H(\tilde{x})}{\lambda_n}\right) \approx \pi \left(n - \frac{1}{2} + \frac{\kappa}{4}\right) \, .
\end{equation}
As the first terms dominate the error in the asymptotic expansion, this implies a constraint on the accuracy of our analytic solutions to at best until $\beta \gtrsim -1/2$, which is obtained by comparison with numerical solutions.
Continuing the calculations, we find the emission constants from equation~(\ref{eq:general_Qn}) are
\begin{equation} \label{eq:sec6_Qn}
  Q_n = \kappa R \int_0^1 \tilde{r}^{\kappa - 1/2} \eta(\tilde{r}) \frac{J_{\gamma-1}(\lambda_n \tilde{r})}{J_\gamma(\lambda_n)}\,\text{d}\tilde{r} \, ,
\end{equation}
the averages related to equation~(\ref{eq:sec2_k-weight}) are
\begin{equation} \label{eq:sec6_Kn}
  T_n = \frac{2 \gamma}{\lambda_n} \frac{{}_1F_2\left(\frac{3}{4}\kappa; \gamma, \frac{3}{4}\kappa + 1; -\lambda_n^2/4\right)}{{}_0F_1\left(\gamma+1; -\lambda_n^2/4\right)} \, ,
\end{equation}
and the averages related to the characteristic radius are
\begin{equation} \label{eq:sec6_Rn}
  R_n = \frac{3 R \gamma}{2 \lambda_n} \frac{{}_1F_2\left(\kappa; \gamma, 2\gamma; -\lambda_n^2/4\right)}{{}_0F_1\left(\gamma+1, -\lambda_n^2/4\right)} \, .
\end{equation}
The final solution from equation~(\ref{eq:general_full_solution}) is given by
\begin{equation}
  J(\tilde{r},\tilde{x}) = \frac{\mathcal{L} \sqrt{6}}{8 \pi} \sum_{n=1}^\infty \frac{Q_n}{\lambda_n \tilde{r}^{\gamma-1}} \frac{J_{\gamma-1}(\lambda_n \tilde{r})}{J_\gamma(\lambda_n)} e^{-\lambda_n |\tilde{x}|} \, .
\end{equation}
The spectral line profile at the boundary is
\begin{equation} \label{eq:sec6_Jx}
  J(\tilde{x}) = \frac{\mathcal{L} \sqrt{6}}{8 \pi f \tau_0 H(\tilde{x})} \sum_{n=1}^\infty Q_n e^{-\lambda_n |\tilde{x}|} \, ,
\end{equation}
which has a normalization of
$\int_{-\infty}^{+\infty} J(x)\,\text{d}x = \frac{3\mathcal{L}}{4 \pi f} \sum_{n=1}^{\infty} Q_n / \lambda_n$.
Using equation~(\ref{eq:sec2_u(r)}), the radiation energy density is
\begin{equation} \label{eq:sec6_u(r)}
  u(\tilde{r}) = \frac{\mathcal{L}}{c} \Gamma\left(\frac{1}{3}\right) \left(\frac{2 a \tau_0}{\sqrt{\pi}}\right)^{1/3} \sum_{n=1}^\infty \frac{Q_n}{\lambda_n^{4/3} \tilde{r}^{\gamma-1}} \frac{J_{\gamma-1}(\lambda_n \tilde{r})}{J_\gamma(\lambda_n)} \, ,
\end{equation}
while $-4\pi \int_0^R r^2 \nabla\vartheta_n(r)\,\text{d}r \approx 4\pi \kappa R^2 \int_0^1 \tilde{r}^{\kappa-1} \vartheta_n(\tilde{r})\,\text{d}\tilde{r}$ gives
\begin{equation}
  M_\text{F} = 8 \pi R^2 \gamma \Gamma\left(\frac{4}{3}\right) \left(\frac{2 a \tau_0}{\sqrt{\pi}}\right)^{1/3} \sum_{n=1}^\infty \frac{Q_n}{\lambda_n^{7/3}} \frac{{}_1F_{2}\left(\frac{\kappa}{2}; \gamma, \frac{\kappa}{2} + 1; -\frac{\lambda_n^2}{4}\right)}{{}_0F_1\left(\gamma + 1, -\frac{\lambda_n^2}{4}\right)} \, ,
\end{equation}
and likewise with $2 \pi R^2 \int \vartheta_n(\tilde{\bmath{r}})\,\text{d}\tilde{V} = 4 \pi R^2 / \lambda_n$ we have
\begin{equation}
  N_{\text{scat}} = \tau_0 \sqrt{24 \pi^3} R^2 \sum_{n=1}^\infty \frac{Q_n}{\lambda_n^2} \, .
\end{equation}
Following the same procedure as previous sections we can derive approximate analytic solutions for each of these quantities for specific vases of $Q_n$. However, the expressions are sufficiently complex and in some cases have very limited accuracy, so we choose to focus on the emergent spectra as a working example. We provide the results of numerical calculations of the exact series expressions in Figures~\ref{fig:sec5_peak_pos_trapping_time} and \ref{fig:sec5_peak_trapping_time}. To allow compact notation in this section we introduce the function
\begin{equation} \label{eq:XiFunction}
  \Xi_s^{\pm}(\tilde{x}) = \pi^{-s} e^{-\pi |\tilde{x}| (\kappa+2)/4}\,\Phi\left(\pm e^{-\pi |\tilde{x}|}, s, \frac{\kappa+2}{4}\right) \, .
\end{equation}

\subsection{Central point source} \label{sec:Power_law_sphere_central_point_source}
For a point source we have $\eta(r) = \delta(r) / 4\pi r^2$, which can be transformed to $\eta(\tilde{r}) = \delta(\tilde{r}) \tilde{r}^{1-3\kappa/2} / (2 \pi \kappa R^3)$, yielding
\begin{equation} \label{eq:sec6_Qn_exact}
  Q_n = \frac{\lim_{\tilde{r} \rightarrow 0} \left[\tilde{r}^{1-\gamma} J_{\gamma-1}(\lambda_n \tilde{r})\right]}{2\pi  R^2 J_\gamma(\lambda_n)} = \frac{(\lambda_n/2)^{\gamma-1}}{2\pi  R^2 \Gamma(\gamma) J_\gamma(\lambda_n)} \, .
\end{equation}
If we assume that $\beta \gtrsim -1/2$ so that $\kappa \lesssim 4$ then we can use the asymptotic expansion to further simplify the normalization factor to $J_\gamma(\lambda_n) \approx \sqrt{2/\pi \lambda_n} \sin(\lambda_n - \pi\kappa/4) \approx (-1)^{n-1} \sqrt{2/\pi \lambda_n}$. This restriction still allows us to explore the main underlying physics of different power-law slopes with concise mathematical expressions. We discuss a strategy for obtaining solutions suitable for all $\kappa$ in Appendix~\ref{app:solution_suitable_for_all_beta}. The spectral line profile at the boundary is then
\begin{align} \label{eq:sec6_analytic_sol_big_beta}
  J(\tilde{x}) &\approx \frac{\mathcal{L} \sqrt{6}}{16 \pi^{3/2} R^2 \Gamma(\gamma) f \tau_0 H(\tilde{x})} \sum_{n=1}^\infty (-1)^{n-1} \left(\frac{\lambda_n}{2}\right)^{\gamma-1/2} e^{-\lambda_n |\tilde{x}|} \notag \\
  &= \frac{\mathcal{L} \sqrt{3}}{2^{\gamma+3} \pi^{3/2} R^2 \Gamma(\gamma) f \tau_0 H(\tilde{x})} \Xi^{-}_{-\kappa/2}(\tilde{x}) \, ,
\end{align}
where the normalization is $3 \mathcal{L} \Xi^{-}_{1-\kappa/2}(0) / [2^{\gamma-5/2} \pi^{3/2} f R^2 \Gamma(\gamma)]$. Although we can derive a peak position from equation~(\ref{eq:sec6_analytic_sol_big_beta}) by setting the derivative to zero, for accuracy we should instead use equation~(\ref{eq:sec6_central_point_source_corrected_sol}). The result is shown in Figure~\ref{fig:sec5_peak_trapping_time} along with results for other quantities for point and uniform sources within power-law density profiles. Due to the singularity in the special functions when $\beta \rightarrow -1$, we restrict the range to well-behaved values while still illustrating the deviation from equation~(\ref{eq:sec6_analytic_sol_big_beta}).

\subsection{Uniform source}
If the emissivity traces the absorption coefficient then $\eta(r) = \eta_0 r^{\beta}$, where $\eta_0 = (\beta+3) / 4\pi R^{\beta+3}$, but $\tau_0 / k_0 = R^{\beta+1} / (1+\beta)$ so
\begin{equation}
  Q_n = \frac{2\tau_0 \eta_0}{k_0} \int_0^1 \tilde{r}^\gamma \frac{J_{\gamma-1}\left(\lambda_n \tilde{r}\right)}{J_\gamma(\lambda_n)}\,\text{d}\tilde{r} = \frac{\gamma}{\pi R^2 \lambda_n} \, .
\end{equation}
It is interesting that in this case there are no longer any Bessel functions in the expression for the spectral quantities, which means the following expressions are accurate for arbitrary $\beta$. The spectral profile at the boundary is
\begin{equation} \label{eq:sec6_J_US}
  J(\tilde{x}) = \frac{\mathcal{L}\sqrt{6}(\kappa+1)}{16 \pi^2 R^2 f \tau_0 H(\tilde{x})} \Xi^{+}_1(\tilde{x}) \, ,
\end{equation}
with a normalization of $3 \gamma \mathcal{L} \psi'(\frac{\kappa+2}{4}) / [4 \pi^4 f R^2]$, where $\psi'(z)$ is the derivative of the digamma function, which is the logarithmic derivative of the gamma function, i.e. $\psi(z) = \Gamma'(z)/\Gamma(z)$. We note that due to the simple form of $Q_n$ in this case it is possible to also write simple expressions for other quantities. For example, the number of scatters becomes $N_\text{scat} = -\tau_0 \sqrt{6} \pi^{-5/2} \gamma \psi''(\frac{\kappa + 2}{4})$.

\subsection{Power-law profile}
We now consider a power-law profile for the emissivity as well. If we choose $\eta(r) = \eta_0 r^\alpha$, where $\eta_0 = (\alpha+3)/4 \pi R^{\alpha+3}$ then
\begin{align}
  Q_n &= \frac{\gamma \delta}{\pi R^2} \int_0^1 \tilde{r}^{\gamma (2 \delta - 1)} \frac{J_{\gamma-1}\left(\lambda_n \tilde{r}  \right)}{J_\gamma\left(\lambda_n\right)}\,\text{d}\tilde{r} \notag \\
  & = \frac{\gamma}{\pi R^2} \left(\frac{\lambda_n}{2}\right)^\gamma \frac{{}_1 F_2 \left(\gamma\delta; \gamma, \gamma\delta + 1; -\lambda_n^2/4\right)}{\Gamma(\gamma+1) \lambda_n J_\gamma(\lambda_n)} \, ,
\end{align}
where $\delta \equiv (\alpha+3) / (\beta+3)$, consistent with the convention for a central point source ($\delta = 0$) and a uniform source ($\delta = 1$). In this case, instead of having a single parameter, both the emissivity $\alpha$ and opacity $\beta$ power-law slopes contribute to $Q_n$ through the relative sourcing $\delta$ and geometric factor $\gamma$. Following the previous sections we retain the first two dominant terms in the series expansion about infinity for an approximate expression:
\begin{equation}
  Q_n \approx \frac{\gamma \delta}{\pi R^2} \left(\frac{1}{\lambda_n} + (-1)^{n-1} C(\gamma,\delta) \lambda_n^{\gamma(1-2\delta) - 1/2} \right) \, ,
\end{equation}
where $C(\gamma,\delta) = \sqrt{\pi} 2^{\gamma(2\delta-1) - 1/2} \Gamma(\gamma\delta) / \Gamma(\gamma-\gamma\delta)$.
The expression for the spectral line profile at the boundary in this case is
\begin{equation} \label{eq:sec6_J(x)_PL}
  J(\tilde{x}) = \frac{\mathcal{L}\sqrt{6} \gamma\delta C(\gamma,\delta)}{8 \pi^2 R^2 f \tau_0 H(\tilde{x}) } \Xi_{2\gamma\delta-\kappa/2}^{-}\left(\tilde{x}\right) + \delta J_{\text{uni}}(\tilde{x}) \, ,
\end{equation}
where $J_\text{uni}(\tilde{x})$ is the result from equation~(\ref{eq:sec6_J_US}) with normalization $3 \mathcal{L} \gamma \delta [\Xi_{2}^{+}(0) + C(\gamma,\delta) \Xi_{2+\gamma\delta-\kappa/2}^{-}(0)] / [4 \pi^2 f R^2]$. Due to the approximations made these results are only valid when $\kappa$ is not too large, otherwise the equations should be viewed as effective solutions. We again note that $\alpha = -3$ exactly corresponds to a point source and $\alpha = \beta$ to the uniform case. Qualitatively, the peak of the spectral line at the boundary shifts to the centre as $\beta$ steepens and $\alpha$ flattens, so the slope of the absorption coefficient has the opposite effect of the emissivity slope. In realistic galaxy environments we expect hierarchical density concentrations ($\beta < 0$) and extended emission ($\alpha > -3$). Collectively this means the Ly$\alpha$ line profile becomes increasingly skewed and narrow as the peak is closer to the centre for a given column density.

\section{Gridless Monte Carlo Method}
\label{sec:Gridless Monte-Carlo Method}
In the previous sections, we derived analytic solutions for a number of idealized models. We now describe the numerical method we employed to validate our new solutions. To ensure the robustness of the results we employ the MCRT method to solve equation~(\ref{eq:general}) directly without the spatial and frequency diffusion approximations leading to equation~(\ref{eq:final_transfer_eq}). Although we focus our discussion on power-law density profiles in slab and spherical geometries, the gridless MCRT method can easily be generalized to other applications. The main idea is to perform exact integration for the optical depth rather than a discretized version based on an arbitrary grid representation. This is particularly useful for idealized models with analytic representations, where it is unnecessary to discretize altogether. In our specific case, the multi-scale nature of power-law profiles along with the special handling of singularities and sharp gradients provides the main motivation for the more accurate GMCRT scheme.

\begin{figure}
  \centering
  \includegraphics[width = .785\columnwidth]{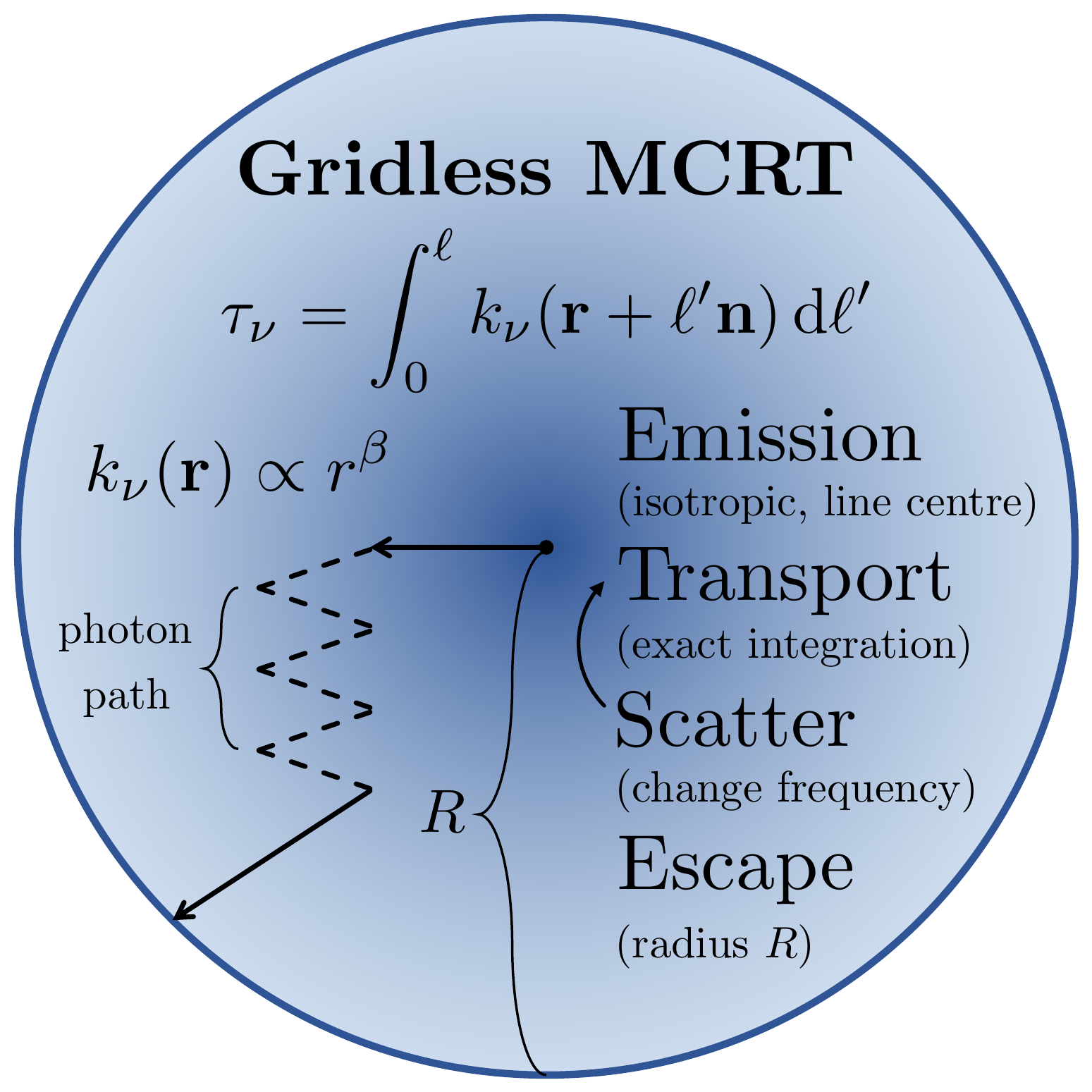}
  \caption{Schematic diagram of the gridless MCRT method, which differs from the standard scheme in that the ray tracing is performed with exact integration as described in Section~\ref{sec:Gridless Monte-Carlo Method}. An individual photon trajectory is otherwise given by the processes of emission followed by iterating between transport and scattering until escaping the computational domain.} \label{fig:sec3_schematic_diagram}
\end{figure}

The MCRT method solves the radiative transfer equation via discrete sampling of individual photon histories to build statistically converged radiation fields and observable properties. The main procedures are illustrated in Figure~\ref{fig:sec3_schematic_diagram}. First, photon packets are generated according to the emission source distribution. Then the trajectory is determined by alternating between ray-tracing and scattering until the photon escapes the computational domain. The Monte Carlo philosophy employs random numbers to decide how far photons move between subsequent scattering events and the change in frequency and direction during each scattering event. After simulating a large number for photon packets, our primary interest is the emergent spectra or distributions of escaped frequencies, as this directly corresponds to our theoretical predictions. The code utilized in this paper is a modified version of the Cosmic Ly$\alpha$ Transfer code (\textsc{colt}), and we refer the reader to \citet{Smith2015} for further details about the numerical prescriptions employed therein.

\subsection{General gridless transport}
The propagation distance for any photon is determined by the traversed optical depth, which in static gas can be defined as
\begin{equation} \label{eq:GMCRT_eq_of_l}
  \tau = H(x) \int_0^\ell k(\bmath{r} + \ell' \bmath{n})\,\text{d}\ell' \, .
\end{equation}
The actual optical depth follows from an exponential distribution accounted for in MCRT by drawing a random number $\xi$ from a uniform distibution in the domain $(0,1)$, i.e.
\begin{equation}
  \tau_\text{scat} = -\ln\xi \, .
\end{equation}
The final step is to determine the path length traveled by the photon. For example, if we assume a homogeneous medium such that $k(\bmath{r}) = k_0$ then the integral is simple and $\ell = -\ln\xi / k_0 H(x)$. The central feature of GMCRT is to extend optical depth calculations to inhomogeneous, anisotropic, or non-static media. In principle, the exact integral can be evaluated during the ray-tracing procedure and inverted if necessary to find the scattering distance. Although in many circumstances the integral might be complex, we still have numerical techniques to handle such cases.

\subsection{Power-law in slab geometry}
\label{sec3_power_law_slab}
We now specialize to the case of an infinite parallel slab geometry. In this setup we assume the absorption coefficient depends on the $z$-axis coordinate and frequency as $k(\bmath{r},x) = k_0 |z|^\beta H(x)$, where the domain is over $z \in (-Z,Z)$ and $\beta > -1$ to ensure a finite optical depth across the central density singularity. We parametrize the setup by the total optical depth at line centre:
\begin{equation}
  \tau_0 = \int_0^Z k(z)\,\text{d}z
  = \frac{k_0}{\beta+1} Z^{\beta+1} \quad \text{or} \quad
  k_0 = \frac{(\beta+1) \tau_0}{Z^{\beta+1}} \, .
\end{equation}
The traversed optical depth for an arbitrary ray depends on the starting position, $z_0 \equiv \bmath{r}_0 \bmath{\cdot} \hat{\bmath{z}}$, and angular cosine, $\mu \equiv \bmath{n} \bmath{\cdot} \hat{\bmath{z}}$, giving
\begin{equation}  \label{eq:slab_tau_integral}
  -\frac{\ln\xi}{k_0 H(x)}
  = \int_0^\ell |z_0 + \mu \ell'|^\beta\,\text{d}\ell' \, .
\end{equation}
To simplify our discussion, we apply a reflective boundary condition at the centre of the slab, which means that some care must be taken to restrict distances to $\ell \leq -z_0 / \mu$ if $\mu < 0$. However, this ensures that $z_0 \geq 0$ and the integral can be inverted to give a distance of
\begin{equation}
  \ell = \frac{1}{\mu} \left[ \left( z_0^{\beta+1} - \frac{(\beta+1) \mu \ln\xi}{k_0 H(x)} \right)^{\frac{1}{\beta+1}} - z_0 \right] \, .
\end{equation}

\begin{figure}
  \centering
  \includegraphics[width=\columnwidth]{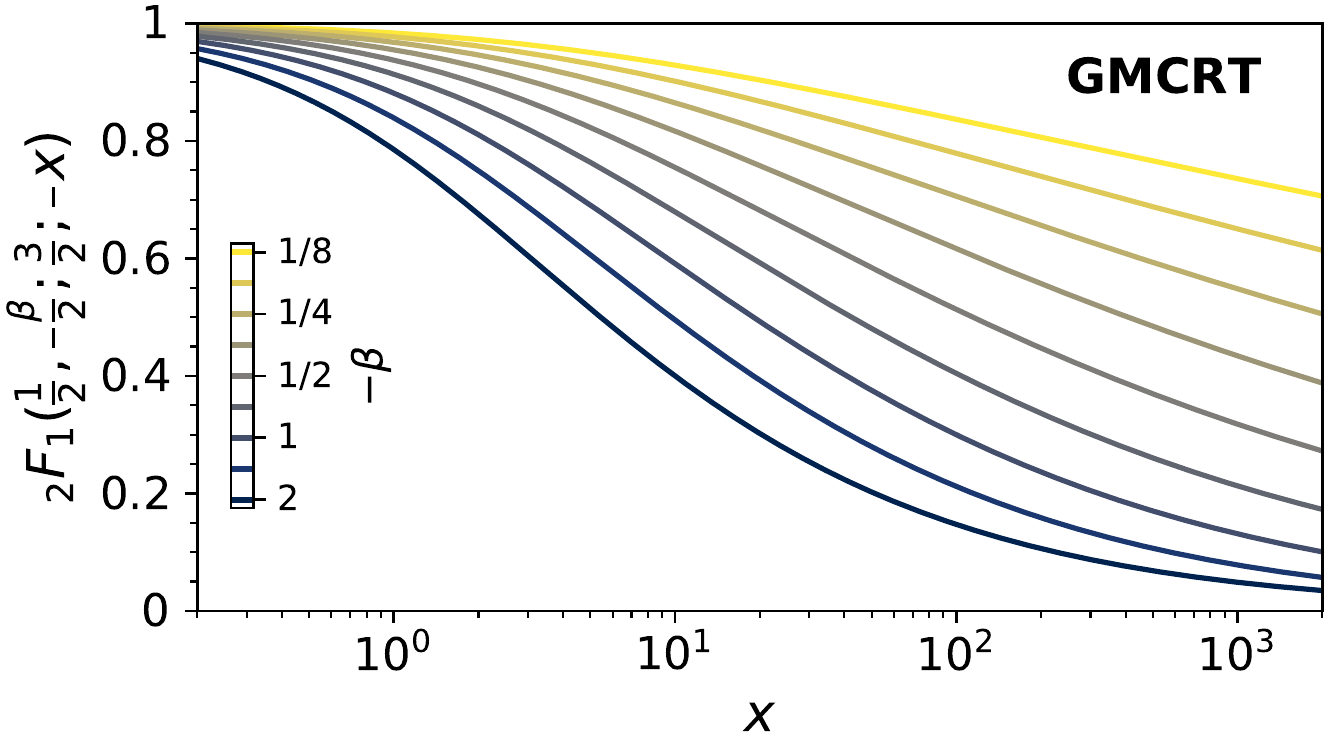}
  \caption{Behaviour of the hypergeometric function derived in equation~(\ref{eq:l-trans}) for gridless transport through power-law density profiles in spherical geometry. The shape is characterized by a departure from unity when $\beta < 0$. The transition occurs earlier for steeper slopes as the contribution to the traversed optical depth is increasingly sensitive to the innermost radii.}
  \label{fig:hypergeometric}
\end{figure}

\subsection{Power-law in spherical geometry} \label{sec3.3_power_law_in_spherical_geometry}
In spherical geometry the absorption coefficient now depends on the radial coordinate as $k(\bmath{r},x) = k_0 r^\beta H(x)$, where the domain is over $r \in (0,R)$ and $\beta > -1$. Similar to the previous case, we parametrize the setup by the total optical depth at line centre:
\begin{equation} \label{eq:spherical_optical_depth}
  \tau_0 = \int_0^R k(r)\,\text{d}r
  = \frac{k_0}{\beta+1} R^{\beta+1} \quad \text{or} \quad k_0 = \frac{(\beta+1) \tau_0}{R^{\beta+1}} \, .
\end{equation}
The traversed optical depth for an arbitrary ray again depends on the starting position, $\bmath{r}_0$, with the radial coordinate given by $r_0 \equiv \| \bmath{r}_0 \|$ and the angular cosine by $\mu \equiv \bmath{n} \bmath{\cdot} \hat{\bmath{r}}_0$. We then define the impact parameter to the origin as $r_\text{min}^2 \equiv (1 - \mu^2) r_0^2$ and change the variable of integration to $\ell'' = \ell' + \mu r_0$ to obtain
\begin{align} \label{eq:l-trans}
  -\frac{\ln\xi}{k_0 H(x)}
  &= \int_0^\ell \left\|\bmath{r}_0 + \ell' \bmath{n}\right\|^{\,\beta}\,\text{d}\ell' \notag \\
  &= \int_{\mu r_0}^{\ell + \mu r_0} \left({\ell''}^2 + r_\text{min}^2\right)^{\beta/2}\,\text{d}\ell'' \notag \\
  &= r_\text{min}^\beta \left[ (\ell + \mu r_0)\;{}_2F_1\left(\frac{1}{2}, -\frac{\beta}{2}; \frac{3}{2}; -\frac{(\ell + \mu r_0)^2}{r_\text{min}^2} \right) \right. \notag \\
  & \qquad\qquad\quad \left. - \mu r_0\;{}_2F_1 \left(\frac{1}{2}, -\frac{\beta}{2}; \frac{3}{2}; -\frac{(\mu r_0)^2}{r_\text{min}^2} \right) \right] \, .
\end{align}
The properties of this hypergeometric function are shown in Figure~\ref{fig:hypergeometric}. Equation~(\ref{eq:l-trans}) is a transcendental equation for $\ell$ that cannot be solved analytically.\footnote{We note that some special cases admit analytically invertible expressions. For example, the right hand side of equation~(\ref{eq:l-trans}) reduces to $\ell$ for the trivial case of $\beta = 0$, $\sinh^{-1}\left(\frac{\ell + \mu r_0}{r_\text{min}}\right) - \sinh^{-1}\left(\frac{\mu r_0}{r_\text{min}}\right)$ for $\beta = -1$, and $\left[\tan^{-1}\left(\frac{\ell + \mu r_0}{r_\text{min}}\right) - \tan^{-1}\left(\frac{\mu r_0}{r_\text{min}}\right)\right]/r_\text{min}$ for $\beta = -2$.} To find the scattering location we implement an iterative root finder based on Halley's method, which converges rapidly as the rate of convergence is cubic. Specifically, given an initial guess for the distance, e.g. $\ell_0 = 0$, the next estimate is
\begin{equation}
  \ell_{n+1} = \ell_n - \frac{f(\ell_n)}{f'(\ell_n)} \left[ 1 - \frac{f(\ell_n)}{f'(\ell_n)} \frac{f''(\ell_n)}{2f'(\ell_n)} \right]^{-1} \, .
\end{equation}
Here $f(\ell)$ is the difference between right and left hand sides of equation~(\ref{eq:l-trans}), such that the first and second derivatives are
\begin{equation}
  f'(\ell) = \left( r_\text{min}^{2} + (\ell + \mu r_0)^2 \right)^{\beta/2}
\end{equation}
and
\begin{equation}
  \frac{f''(\ell)}{f'(\ell)} = \frac{\beta (\ell + \mu r_0)}{r_\text{min}^{2} + (\ell + \mu r_0)^2} \, .
\end{equation}
The optical depth function is monotonic, so the root finder method is quite robust and usually converges after a few iterations. However, as a protection against corner cases we also implemented a brute force bisection method if convergence is not reached within a reasonable number of iterations. Finally, we employ the GSL library when we need to evaluate the hypergeometric functions.

\subsection{Acceleration schemes} \label{sec3.4_Acceleration_schemes}
In optically thick environments Ly$\alpha$ photons spend much of their time undergoing core scatterings with negligible diffusion in physical and frequency space. Such scatterings can be avoided by preferentially selecting atoms with perpendicular velocity components greater than a critical frequency $x_\text{crit}$. Thus, for computational efficiency we employ the dynamical core-skipping scheme with $x_\text{crit} = \frac{1}{5} (a \tau_0)^{1/3}$ following \citet{Smith2015}. The modification required for the GMCRT method is the interpretation of the product $a \tau_0$ as being the minimum value from the photon to escape. For example, with a power-law density profile in spherical geometry this is $a \tau_0 = a k_0 (R^{\beta+1} - r_0^{\beta+1}) / (\beta+1)$. Additionally, we note that the GMCRT ray-tracing with direct hypergeometric function evaluations is more expensive than grid-based MCRT, but this is outweighed by the gain in accuracy and robustness needed for this work. The simulations presented here take advantage of parallel computing resources, which results in highly efficient code because every photon packet is independent.

\subsection{Line-of-sight surface brightness images}
The MCRT method naturally allows the construction of line-of-sight surface brightness images using the next-event estimator method \citep{Yusef-Zadeh1984}. For each scattering, we may calculate the probability that the photon would have been scattered towards the observer. The transmitted flux is attenuated by the traversed optical depth to escape along that sightline, i.e. $e^{-\tau_\text{esc}}$. The contributions from all scattering events over all photon packets can be used to generate observed images and spectra. We therefore briefly discuss how this is done in the context of the GMCRT scheme.

\begin{figure}
  \centering
  \includegraphics[width=\columnwidth]{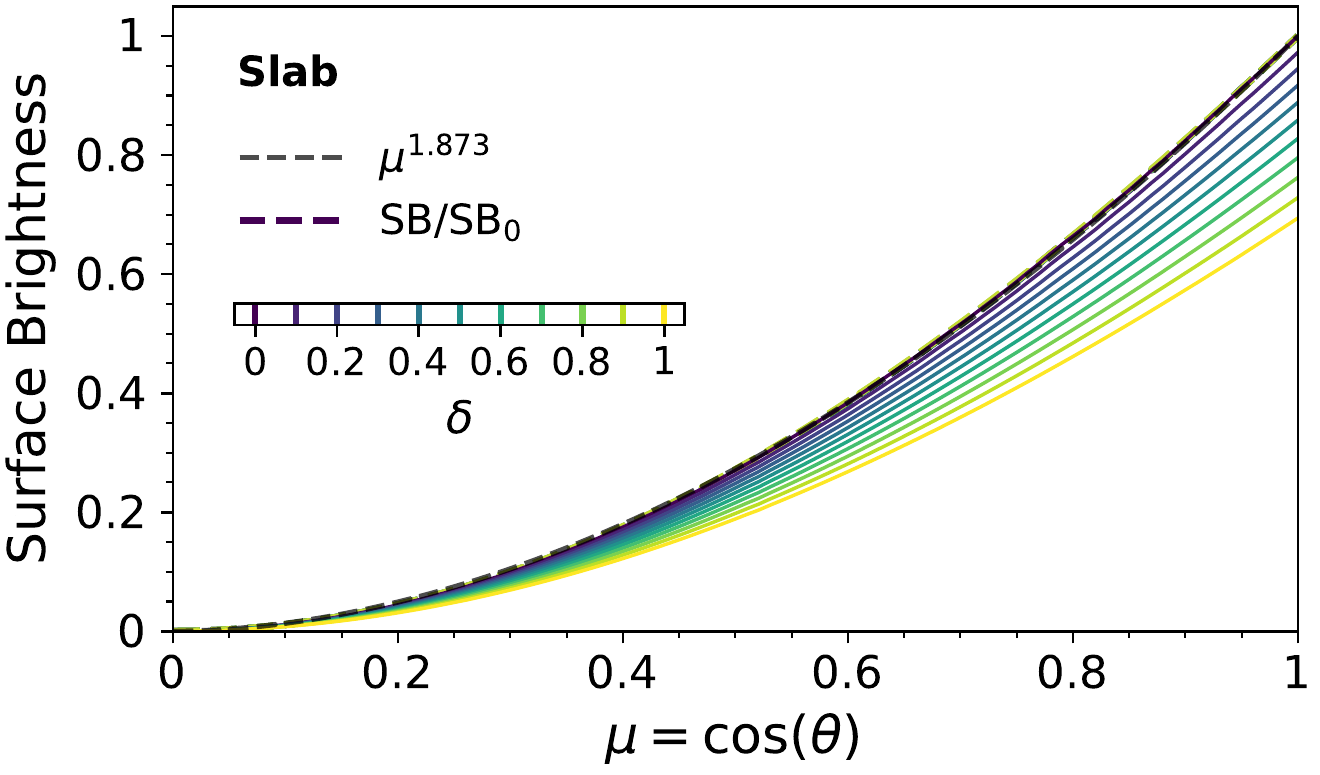}
  \caption{The emergent surface brightness from a plane parallel slab as a function of $\mu = \cos \theta$, where $\theta$ is the observed angle with respect to the normal direction. The curves are calculated from equation~(\ref{eq:sec5_I}) using the derived expressions for the mean intensity from equation~(\ref{eq:sec3_J(z,x)_power_law}). The normalized shapes are essentially identical due to the assumed boundary condition, and are well described by the dipole-like expression $\text{SB}(\mu)/\text{SB}(1) \approx \mu^{1.873}$.}
  \label{fig:sec6_SB_Slab}
\end{figure}

\subsubsection{Slab geometry}
In slab geometry a distant observer is uniquely defined by the angular cosine $\mu$. In this case the optical depth from equation~(\ref{eq:slab_tau_integral}) is
\begin{equation}
  \tau_\text{esc}(\pm \tilde{z}, \mu) = \tau_\text{esc}(\mp \tilde{z}, -\mu) = \frac{\tau_0 H(x)}{\mu} \left(1 \mp \tilde{z}\right) \, .
\end{equation}
Contributions from the opposite half of a mirrored slab setup can be accounted for by averaging positive and negative directed sightlines, i.e. $[e^{\tau_\text{esc}(\tilde{z}, \mu)} + e^{\tau_\text{esc}(\tilde{z}, -\mu)}]/2$.
Finally, the angle dependent flux emanating from the surface is calculated as
\begin{equation}
  I(\mu) = \int_{-\infty}^{+\infty} \int_{-1}^{1} \frac{J(\tilde{z},x)}{\tau_0 H(x)} e^{-\tau_\text{esc}(\tilde{z}, \mu)}\,\text{d}\tilde{z}\,\text{d}x \, ,
\end{equation}
which cannot be evaluated analytically. However, using the expression for intensity given in equation~(\ref{eq:sec3_J(z,x)_power_law}), we can numerically calculate the surface brightness for different values of $\delta$. The result is shown in Figure~(\ref{fig:sec6_SB_Slab}) and leads to the important insight that the normalized profiles are universal. This result likely follows from the assumed boundary conditions as the observed photons last interacted within a few skin depths into the surface. Qualitatively, the result resembles that of a dipole function with $\text{SB}(\mu)/\text{SB}(1) \approx \mu^2$.

\subsubsection{Spherical geometry}
\begin{figure}
  \includegraphics[width=\columnwidth]{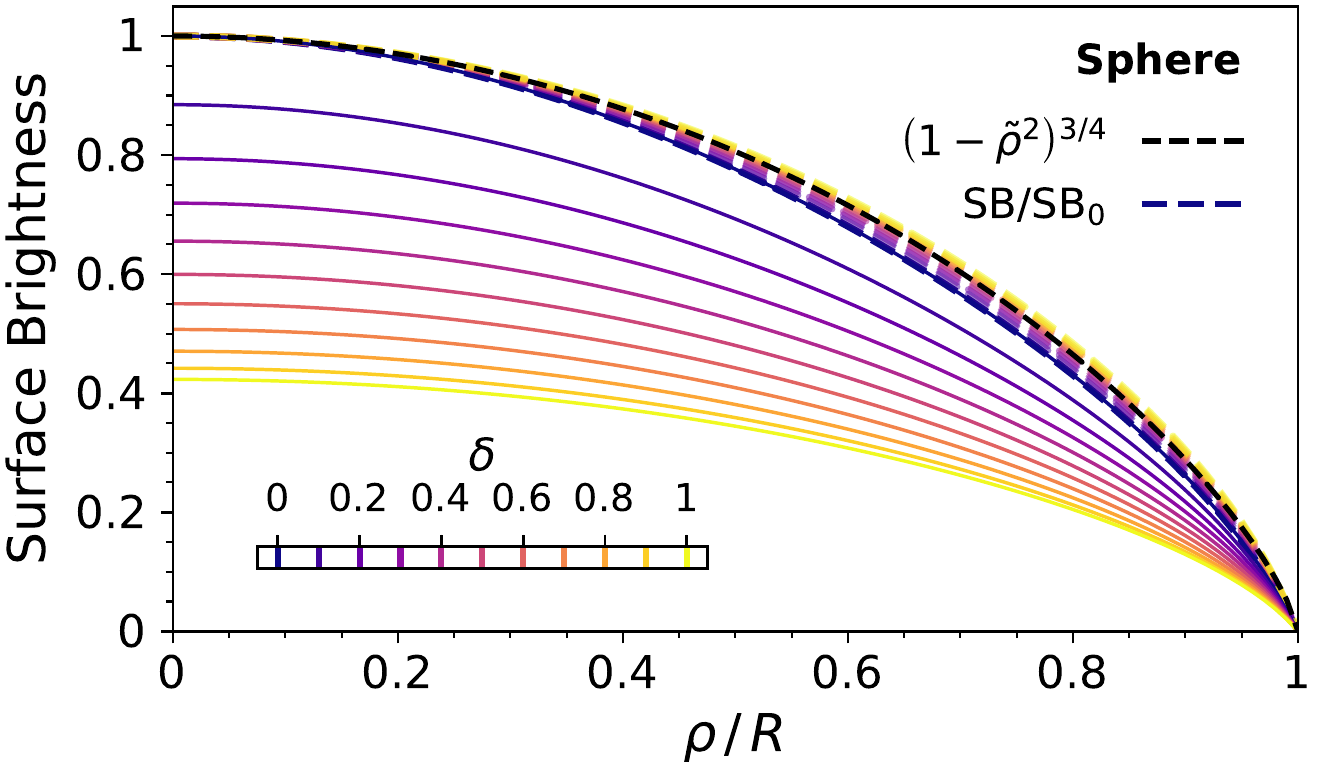}
  \caption{The emergent surface brightness from a homogeneous sphere as a function of polar radius $\rho$. The overall magnitude of the intensity decreases with increasing $\delta$, but the normalized profile follows a universal shape due to the assumed boundary condition. Specifically, we find the radial dependence is well modelled by the simple function $\text{SB}(\tilde{\rho})/\text{SB}(0) \approx (1 - \tilde{\rho}^2)^{0.76}$.}
  \label{fig:sec5_SB}
\end{figure}

In spherical geometry all observers are equivalent and the unique aspect is the polar radius $\rho$. In this case we calculate the optical depth in the $\hat{z}$ direction:
\begin{equation}
  \tau_\text{esc}(\rho) = H(x) \int_{z_0}^{\sqrt{R^2-\rho^2}} k\left(\sqrt{\rho^2 + z^2}\right) \text{d}z \, .
\end{equation}
Using the expression from equation~(\ref{eq:l-trans}), but substituting distances with $\mu r_0 = z_0$, $r_{\text{min}}^2  = \rho^2$, and $\ell = \sqrt{R^2 - \rho^2} - z_0$, results in the following expression:
\begin{align}
  \tau_{\text{esc}} = (\beta+1) \tau_0 H(x) \left(\frac{\rho}{R}\right)^\beta &\left[\sqrt{1 - \left(\frac{\rho}{R}\right)^2} \;{}_2F_1\left(\frac{1}{2}, -\frac{\beta}{2}; \frac{3}{2}; 1 - \frac{R^2}{\rho^2} \right) \right. \notag \\
  -&\left.\frac{z_0}{R} \;{}_2F_1\left(\frac{1}{2}, -\frac{\beta}{2}; \frac{3}{2}; -\frac{z_0^2}{\rho^2} \right)  \right] \, .
\end{align}
In spherical geometry we can also derive an expression for the radial surface brightness. Assuming we know $J(r,x)$ we have
\begin{equation} \label{eq:sec5_I}
  I(\tilde{\rho}) = \int_{-\infty}^{+\infty} \int_{-\tilde{Z}}^{\tilde{Z}} \frac{J(\sqrt{\tilde{\rho}^2 + \tilde{z}^2},x)}{\tau_0 H(x)} e^{-\tau_0 H(x) \left(\tilde{Z} - \tilde{z}\right)}\,\text{d}\tilde{z}\,\text{d}x \, ,
\end{equation}
where $\tilde{\rho}$ denotes the integrated polar radial coordinate and $\tilde{Z} = \sqrt{1-\tilde{\rho}^2}$. Similar to the slab case, we employ the analytical expression for the mean intensity given in equation~(\ref{eq;sec5_general_J}) to numerically calculate the surface brightness for different values of $\delta$. The result is shown in Figure~(\ref{fig:sec5_SB}), which again reveals a universal normalized profile due to the assumed boundary conditions. For utility we provide a simple power-law model fit for the shape as $\text{SB}(\tilde{\rho})/\text{SB}(0) \approx (1 - \tilde{\rho}^2)^{3/4}$.

\subsection{Numerical Verification}
\label{app:numerical_verification}
We now demonstrate the validity of our new analytic solutions by comparing them to the results from numerical calculations. We focus on the emergent spectra as this is easily obtained from simulations and directly corresponds to observational data. We employ the GMCRT method as described above. For numerical efficiency we simulate $10^5$ photons within an optically-thick environment specified by $a \tau_0 = 5000$ at a temperature of $T = 10$\,K. We also require a relative error of less than $10^{-5}$ in the root finding procedure when determining scattering distances.

\subsubsection{Slab geometry}
\begin{figure}
  \centering
  \includegraphics[width=\columnwidth]{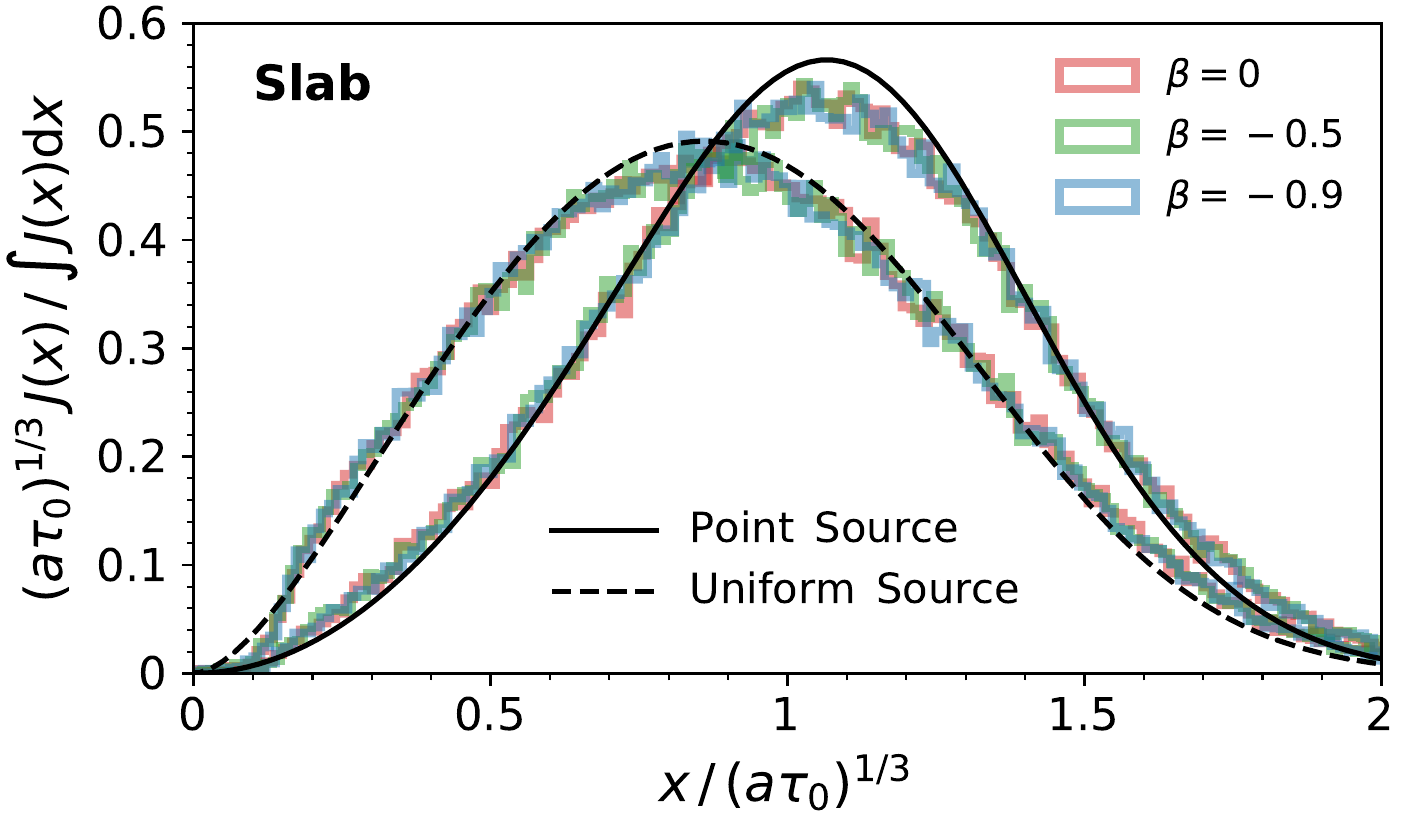}
  \caption{Numerical verification of the central point source and uniform emission analytic solutions in slab geometry. The solid and dashed curves are the analytic solutions from equations~(\ref{eq:sec4_sol_uniform_CPS}) and (\ref{eq:sec4_sol_uniform_source}), respectively, while the histograms are data from simulations. A key feature of this test is that the emergent spectrum is independent of the specific density profile and only depends on the relative distribution of emission to opacity. This is illustrated by adjusting the power-law slope for the spatial part of the absorption coefficient as $\beta = \{0, -0.5, -0.9\}$. The agreement between simulations also showcases the robustness of the GMCRT method. We emphasize that the degeneracy of profiles is an inherent feature of slab geometry.}
  \label{fig:appendixB_Slab_CPS_US}
\end{figure}

In section \ref{sec4.1} we found that in slab geometry only the relative distribution of emission to opacity affects the emergent line profile. We now test this more explicitly by performing simulations with different power-law profiles for the spatial part of absorption coefficient $k(z) \propto z^\beta$, with $\beta = \{0, -0.5, -0.9\}$. We confirm that in all cases the calculated spectrum follows the analytic solution given by equation~(\ref{eq:sec4_sol_uniform_CPS}) when photons are injected as a central point source. We also confirm the analytic expectation that uniform sources follow the prediction given by equation~(\ref{eq:sec4_sol_uniform_source}) independent of density profile in slab geometry. In this case we simply inject photons with random positions according to the density distribution, i.e. with $\eta(z) \propto k(z)$. This is done by inverting the cumulative distribution function such that if $\xi \in [0,1]$ is a uniform random number then the initial emission follows from $z = Z \xi^{1/(\alpha+1)}$ with $\alpha = \beta$ for uniform source. Again, the power-law profile serves as a convenient way to capture the essential physics by representing centralized and extended environments with a single parameter $\delta$. The results for a central point source and uniform emission are shown in Figure~\ref{fig:appendixB_Slab_CPS_US}.

\subsubsection{Spherical geometry}
\begin{figure}
  \centering
  \includegraphics[width=\columnwidth]{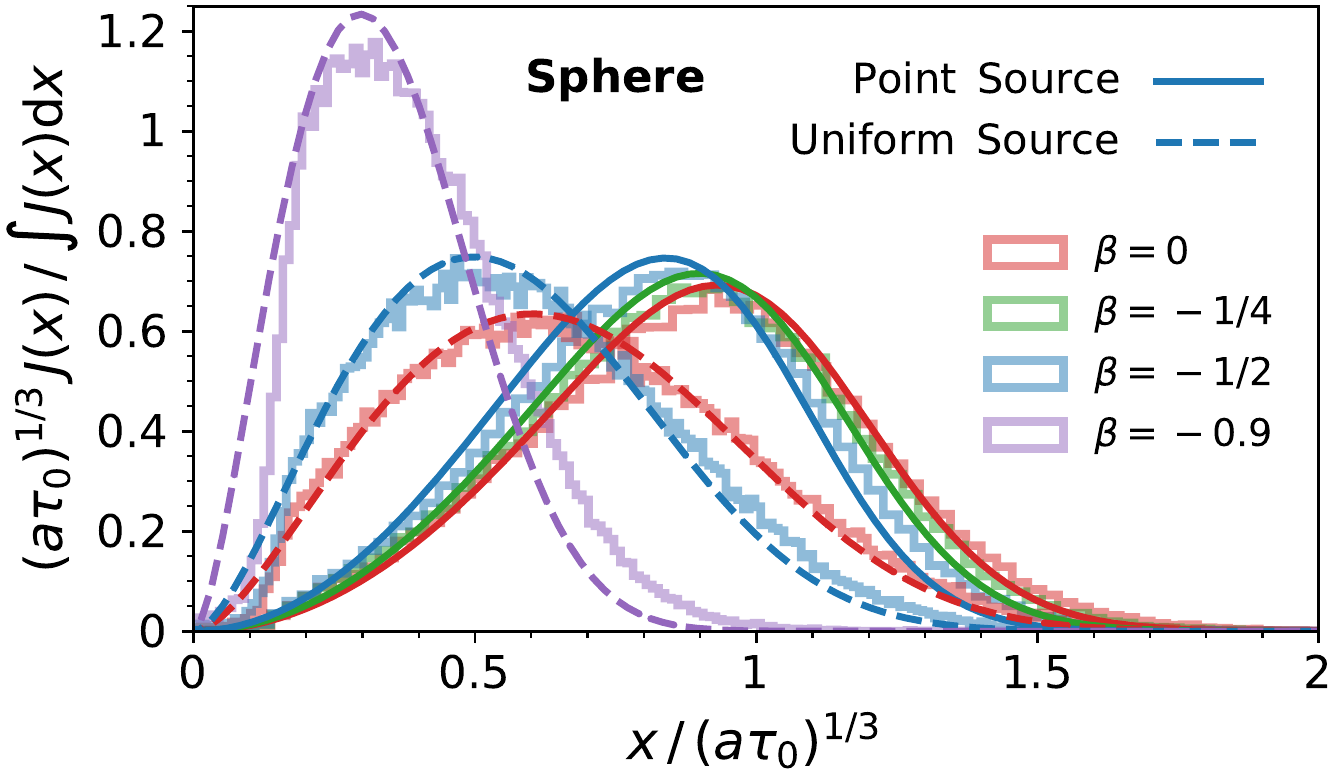}
  \caption{Numerical verification of the analytic solutions in spherical geometry within a power-law density profile $k \propto r^\beta$. The solid and dashed curves are from equations~(\ref{eq:sec6_analytic_sol_big_beta}) and (\ref{eq:sec6_J_US}), respectively, while the histograms are simulated data using the GMCRT method. In this case, there is a geometrical advantage for photons escaping from steeper density profiles. The central point source solution based on the asymptotic approximation becomes inaccurate when $\beta \lesssim -1/2$, but this can be corrected as in Appendix~\ref{app:solution_suitable_for_all_beta}. On the other hand, the solution for the special case of uniform emission remains accurate even as $\beta \rightarrow -1$. Also if $a\tau_0$ is increased, the data better match the predictions due to the approximation that $\lambda_n \ll f \tau_0 H(x)$. In general, as the slope steepens the peak becomes more narrow and shifts towards line centre, which is especially evident when comparing the factor of $\sim 3$ change from concentrated/flat to extended/steep environments.}
  \label{fig:sec6_numerical_test_for_big_beta}
\end{figure}

In spherical geometry the emergent spectra depends on both the emissivity and opacity profiles. Therefore, we provide the same tests as the slab case but notice several differences and complications that deserve special attention. First, the central point source solution from equation~(\ref{eq:sec6_analytic_sol_big_beta}) is only valid for $\beta \gtrsim -1/2$ but it is interesting that in this range the profile does not change very much. This is because each photon experiences the same line centre optical depth from emission to escape independent of the density profile. However, geometric curvature shifts the relative likelihood of either continuing to be trapped or descending rapidly along the opacity gradient. We also note that the approximation of the Bessel function when $\beta \rightarrow -1$ leads to an inaccurate solution, which requires a correction procedure discussed in Appendix~\ref{app:solution_suitable_for_all_beta} that again agrees with the robust GMCRT numerical results. On the other hand, the analytic solution for the case of uniform emission from equation~(\ref{eq:sec6_J_US}) remains surprisingly accurate as $\beta \rightarrow -1$. This is because $Q_n$ can be calculated without approximations, which eliminates a primary source of error in the Bessel function expansion. It is also interesting that in this case the effect of a steeper density profile is more apparent as escape is facilitated by the geometric advantage. Analogous to the slab case, the initial positions of photon packets are drawn from $r = R \xi^{1/(\alpha+3)}$ with $\alpha = \beta$ for a uniform profile. The comparison between analytic and numerical results for both the central point source and uniform emission are shown in Figure~(\ref{fig:sec6_numerical_test_for_big_beta}).

\section{Summary and Discussion}
\label{sec:summary}
Observations of the Ly$\alpha$ line provide valuable information about the formation and evolution of high-redshift galaxies. Although numerical simulations are necessary to fully elucidate the intricacies of radiative transfer processes, it is often beneficial to develop physical intuition for interpreting both observational and theoretical results. In particular, analytic solutions for resonant-line radiation in optically thick media have been crucial for the development of Ly$\alpha$ theory \citep{Harrington1973,Neufeld1990,Dijkstra2006}. In this paper, we successfully extended the existing formalism to obtain new analytic solutions for resonant-line transport focusing on power-law density and emissivity profiles, which are sufficiently general to provide additional context for idealized Ly$\alpha$ modelling. Following previous works we employed the approximation that spatial transport and frequency redistribution are local diffusion processes. However, we relaxed the requirement of a uniform absorption coefficient to obtain a more general differential equation allowing further insights about resonance lines. In particular, we qualitatively and quantitatively determined how geometry, density, and emissivity influence the internal and emergent radiation. The main results of our study are summarized as follows.
\begin{itemize}
  \item[(1)]
  In slab geometry, any density profile in real space is equivalent to a homogeneous representation in optical depth space, which simplifies derivations. This has the side effect that solutions only depend on the relative emissivity-to-opacity profile, i.e. as being either concentrated or extended. This degeneracy means it is impossible to distinguish between static slab density models based on spectra alone. However, by introducing power-law profiles for the emissivity $\eta \propto z^\alpha$ and opacity $k \propto z^\beta$ we found that a single parameter $\delta \equiv (\alpha + 1) / (\beta + 1)$ captures the features of a wide range of setups. Our generalized solutions demonstrate a continuous transition between central point-like ($\delta = 0$) and uniform ($\delta = 1$) sources.
  For convenience we derived a useful estimate for the impact on emergent line profiles within non-uniform media in equation~(\ref{eq:x_esc_estimate}), which can be summarized as $x_\text{esc} \propto (1 + \delta/2)^{-1/6}$.

  \item[(2)] In spherical geometry, the eigenfunction expansion is coupled to the density profile, which complicates derivations. However, this breaks the degeneracy found in the slab model so that line profiles retain information about their host environments. In this case, it is again insightful to explore the solution space with a power-law formalism. In addition to the analogous emissivity-to-opacity parameter $\delta \equiv (\alpha + 3) / (\beta + 3)$ there is a geometric factor $\gamma \equiv (\kappa + 1) / 2$ related to the order of the Bessel eigenfunction with generating weight factor $r^2 \propto \tilde{r}^\kappa$ where $\kappa \equiv 2 / (\beta + 1)$. We first derived solutions for the homogeneous case with $\beta = 0$ as a function of $\delta$ similar to the slab case, but also presented generalized solutions for $\beta < 0$.

  \item[(3)] Overall, the behaviour in both slab and spherical geometries are qualitatively similar with properties consistent within a factor of $\lesssim 2$. The exception is when $\beta \rightarrow -1$ the density becomes highly concentrated and properties start to change rapidly in anticipation of the geometric singularity. For a given optical depth $\tau_0$, increasing $\delta$ robustly shifts the peak frequency $x_\text{peak}$ towards line centre, reduces the number of scatters $N_\text{scat}$, the force multiplier $M_\text{F}$, and the photon trapping time $t_\text{trap}$, and extends the characteristic radius $\langle r \rangle$. In all cases, our results reduce to previously known solutions for homogeneous setups ($\beta = 0$) with special emissivities $\alpha$.

  \item[(4)] We derive several novel solutions that to our knowledge have not appeared previously in the literature. In particular, we wish to highlight the following: (\textit{i}) simple expressions for the full radiation field $J(r,x)$, e.g. equations~(\ref{eq:sec4-J(z,x)-PS}) and (\ref{eq:sec5-J(r,x)-PS}) for a point source in slab and spherical geometries,\footnote{We note that some of these solutions are completions of earlier works \citep[e.g.][]{Harrington1973,Neufeld1990,Dijkstra2006} that were recently independently derived by \citet{Seon2020}.} (\textit{ii}) the emergent spectrum for a uniform source in spherical geometry in equation~(\ref{eq:sec4_US}), and (\textit{iii}) a general expression for the emergent spectrum in spherical geometry assuming power-laws for both the emissivity and opacity in equation~(\ref{eq:sec6_J(x)_PL}). These analytic solutions allow for more general model predictions, observational comparisons, and code validation related to the nature of resonant-line radiative transfer within non-unifom media.

  \item[(5)] We also calculated the surface brightness profiles based on the analytic solutions for homogeneous slabs and spheres with different values of $\delta$. The distributions with incident viewing angle $\mu = \cos \theta$ and polar radius $\rho$ all follow the same regular shapes. Specifically, these are approximately given by $\text{SB} \propto \mu^2$ for slabs and $\text{SB} \propto (1 - \rho/R)^{3/4}$ for spheres, independent of the emissivity power-law slope $\alpha$. We interpret this as being imposed by the assumed boundary conditions which control the conditions at the surface.
\end{itemize}

Beyond this, we also developed a gridless MCRT method to test the validity of the new analytic results. The Monte Carlo method solves the radiative transfer equation by integrating along rays and sampling from probability distributions for emission and scattering. The novelty of GMCRT is that the ray tracing is performed exactly using the underlying density field. It is therefore not necessary to discretize the computational domain as is typically done for MCRT algorithms. The gridless scheme is highly robust and accurate, even in the presence of the singularities that arise with the power-law profiles studies herein. GMCRT is competitive for production simulations and represents a powerful tool for targeted studies of idealized environments. In our case, we were able to successfully verify the theoretical predictions from the new analytic solutions, which is less straightforward than the homogeneous setups. The principles of the GMCRT method are easily generalized to other applications and may thus have broader utility within the radiative transfer community.

Throughout this paper we have not properly considered the effects of internal dust absorption, multiphase media, and non-thermal velocities, including rotation, turbulence, and discrete cloud motions. However, these can be quiet important for realistic galaxy models. While the present work focuses on static, dust-free environments, the analytic formalism can continue to be systematically generalized to further our intuition about absorption, anisotropic covering fractions, and macroscopic velocities. Of course, these complex phenomena have already been explored in numerous discussions throughout the literature and are often incorporated in numerical simulations. For example, a significant effort has gone into studying Ly$\alpha$ radiative transfer in the context of clumpy media \citep{HansenOh2006,DijkstraKramer2012,Laursen2013,GronkeBullDijkstra2015}. Previous studies have also examined the impact of rotation \citep{Garavito2014,Remolina2019} and turbulent velocity structure \citep{Kakiichi2019,Kimm2019} on emergent Ly$\alpha$ spectral profiles. In relation to our work, some of these effects amount to additional or altered terms in the radiative transfer equation itself, or are amenable to gridless MCRT extensions comparable to the ones proposed here. For example, moving media introduces a Doppler term connecting the velocity gradient to the frequency derivative of the radiation field. One may similarly model turbulence based on its statistical properties and clumpy media by the number of surface scatters. The expected Ly$\alpha$ escape fraction can also be derived in a similar fashion as \citet{Neufeld1990}. Finally, we note that piecewise density and emissivity profiles, e.g. shell regions, can also be incorporated into the analytic framework. Such configurations result in a system of equations coupled by appropriate continuity conditions at each interface. Thus, there is ample room for additional work with both the analytic and GMCRT approaches.

We emphasize that the power-law models explored in this work are still too idealized for comparison with realistic galaxy environments. However, the intuition about the nature of resonant-line radiative transfer including the impact of emissivity and opacity distributions provides valuable insights for interpreting theoretical and observational data. In fact, our results can already be applied to observations when discussing deviations from simple model estimates (e.g. see equation~\ref{eq:x_esc_estimate}). In this sense, we are optimistic that analytical and idealized studies will continue to develop even in an era when state-of-the-art hydrodynamics simulations with resonant-line modelling are increasingly available \citep[e.g.][]{Behrens2019,Kakiichi2019,Kimm2019,Laursen2019,Smith2019,Li2020,Michel-Dansac2020}. The methods in this paper are quite general and in the future we plan to extend them further, e.g. to include non-static environments. Such efforts are complementary to the ongoing development of Ly$\alpha$ theory and our broader understanding of high-redshift galaxies.

\section*{Acknowledgements}
We thank the referee for constructive comments and suggestions which have improved the quality of this work.
BL thanks Mark Vogelsberger for hosting him as a visitor at the MIT Kavli Institute for Astrophysics and Space Research.
We thank David Barnes, Hui Li, Stephanie O'Neil, Kaili Cao, Yingtian Chen, and Yuan Wang for insightful discussions related to this work.
AS acknowledges support for Program number \textit{HST}-HF2-51421.001-A provided by NASA through a grant from the Space Telescope Science Institute, which is operated by the Association of Universities for Research in Astronomy, Incorporated, under NASA contract NAS5-26555.

\section*{Data Availability}
The data underlying this article will be shared on reasonable request to the corresponding author.


\bibliographystyle{mnras}
\bibliography{biblio}

\begin{thebibliography}{}
\makeatletter
\relax
\def\mn@urlcharsother{\let\do\@makeother \do\$\do\&\do\#\do\^\do\_\do\%\do\~}
\def\mn@doi{\begingroup\mn@urlcharsother \@ifnextchar [ {\mn@doi@}
  {\mn@doi@[]}}
\def\mn@doi@[#1]#2{\def\@tempa{#1}\ifx\@tempa\@empty \href
  {http://dx.doi.org/#2} {doi:#2}\else \href {http://dx.doi.org/#2} {#1}\fi
  \endgroup}
\def\mn@eprint#1#2{\mn@eprint@#1:#2::\@nil}
\def\mn@eprint@arXiv#1{\href {http://arxiv.org/abs/#1} {{\tt arXiv:#1}}}
\def\mn@eprint@dblp#1{\href {http://dblp.uni-trier.de/rec/bibtex/#1.xml}
  {dblp:#1}}
\def\mn@eprint@#1:#2:#3:#4\@nil{\def\@tempa {#1}\def\@tempb {#2}\def\@tempc
  {#3}\ifx \@tempc \@empty \let \@tempc \@tempb \let \@tempb \@tempa \fi \ifx
  \@tempb \@empty \def\@tempb {arXiv}\fi \@ifundefined
  {mn@eprint@\@tempb}{\@tempb:\@tempc}{\expandafter \expandafter \csname
  mn@eprint@\@tempb\endcsname \expandafter{\@tempc}}}

\bibitem[\protect\citeauthoryear{{Adams}}{{Adams}}{1972}]{Adams1972}
{Adams} T.~F.,  1972, \mn@doi [\apj] {10.1086/151503}, \href
  {https://ui.adsabs.harvard.edu/abs/1972ApJ...174..439A} {174, 439}

\bibitem[\protect\citeauthoryear{Adams}{Adams}{1975}]{Adams1975}
Adams T.~F.,  1975, \mn@doi [ApJ] {10.1086/153891}, 201, 350

\bibitem[\protect\citeauthoryear{Ahn, Lee  \& Lee}{Ahn et~al.}{2002}]{Ahn2002}
Ahn S.-H.,  Lee H.-W.,   Lee H.~M.,  2002, \mn@doi [ApJ] {10.1086/338497}, 567,
  922

\bibitem[\protect\citeauthoryear{Auer}{Auer}{1968}]{Auer1968}
Auer L.~H.,  1968, \mn@doi [ApJ] {10.1086/149705}, 153, 783

\bibitem[\protect\citeauthoryear{{Behrens}, {Pallottini}, {Ferrara},
  {Gallerani}  \& {Vallini}}{{Behrens} et~al.}{2019}]{Behrens2019}
{Behrens} C.,  {Pallottini} A.,  {Ferrara} A.,  {Gallerani} S.,   {Vallini} L.,
   2019, \mn@doi [\mnras] {10.1093/mnras/stz980}, \href
  {https://ui.adsabs.harvard.edu/abs/2019MNRAS.486.2197B} {486, 2197}

\bibitem[\protect\citeauthoryear{{Dijkstra}}{{Dijkstra}}{2014}]{Dijkstra2014}
{Dijkstra} M.,  2014, \mn@doi [\pasa] {10.1017/pasa.2014.33}, \href
  {https://ui.adsabs.harvard.edu/abs/2014PASA...31...40D} {31, e040}

\bibitem[\protect\citeauthoryear{{Dijkstra} \& {Kramer}}{{Dijkstra} \&
  {Kramer}}{2012}]{DijkstraKramer2012}
{Dijkstra} M.,  {Kramer} R.,  2012, \mn@doi [\mnras]
  {10.1111/j.1365-2966.2012.21131.x}, \href
  {https://ui.adsabs.harvard.edu/abs/2012MNRAS.424.1672D} {424, 1672}

\bibitem[\protect\citeauthoryear{{Dijkstra}, {Haiman}  \& {Spaans}}{{Dijkstra}
  et~al.}{2006}]{Dijkstra2006}
{Dijkstra} M.,  {Haiman} Z.,   {Spaans} M.,  2006, \mn@doi [\apj]
  {10.1086/506243}, \href
  {https://ui.adsabs.harvard.edu/abs/2006ApJ...649...14D} {649, 14}

\bibitem[\protect\citeauthoryear{{Garavito-Camargo}, {Forero-Romero}  \&
  {Dijkstra}}{{Garavito-Camargo} et~al.}{2014}]{Garavito2014}
{Garavito-Camargo} J.~N.,  {Forero-Romero} J.~E.,   {Dijkstra} M.,  2014,
  \mn@doi [\apj] {10.1088/0004-637X/795/2/120}, \href
  {https://ui.adsabs.harvard.edu/abs/2014ApJ...795..120G} {795, 120}

\bibitem[\protect\citeauthoryear{{Ge} \& {Wise}}{{Ge} \&
  {Wise}}{2017}]{GeWise2017}
{Ge} Q.,  {Wise} J.~H.,  2017, \mn@doi [\mnras] {10.1093/mnras/stx2074}, \href
  {https://ui.adsabs.harvard.edu/abs/2017MNRAS.472.2773G} {472, 2773}

\bibitem[\protect\citeauthoryear{{Gronke}, {Bull}  \& {Dijkstra}}{{Gronke}
  et~al.}{2015}]{GronkeBullDijkstra2015}
{Gronke} M.,  {Bull} P.,   {Dijkstra} M.,  2015, \mn@doi [\apj]
  {10.1088/0004-637X/812/2/123}, \href
  {https://ui.adsabs.harvard.edu/abs/2015ApJ...812..123G} {812, 123}

\bibitem[\protect\citeauthoryear{{Habetler} \& {Matkowsky}}{{Habetler} \&
  {Matkowsky}}{1975}]{Habetler1975}
{Habetler} G.~J.,  {Matkowsky} B.~J.,  1975, \mn@doi [Journal of Mathematical
  Physics] {10.1063/1.522618}, \href
  {https://ui.adsabs.harvard.edu/abs/1975JMP....16..846H} {16, 846}

\bibitem[\protect\citeauthoryear{{Hansen} \& {Oh}}{{Hansen} \&
  {Oh}}{2006}]{HansenOh2006}
{Hansen} M.,  {Oh} S.~P.,  2006, \mn@doi [\mnras]
  {10.1111/j.1365-2966.2005.09870.x}, \href
  {https://ui.adsabs.harvard.edu/abs/2006MNRAS.367..979H} {367, 979}

\bibitem[\protect\citeauthoryear{{Harrington}}{{Harrington}}{1973}]{Harrington1973}
{Harrington} J.~P.,  1973, \mn@doi [\mnras] {10.1093/mnras/162.1.43}, \href
  {https://ui.adsabs.harvard.edu/abs/1973MNRAS.162...43H} {162, 43}

\bibitem[\protect\citeauthoryear{{Higgins} \& {Meiksin}}{{Higgins} \&
  {Meiksin}}{2012}]{Higgins2012}
{Higgins} J.,  {Meiksin} A.,  2012, \mn@doi [\mnras]
  {10.1111/j.1365-2966.2012.21917.x}, \href
  {https://ui.adsabs.harvard.edu/abs/2012MNRAS.426.2380H} {426, 2380}

\bibitem[\protect\citeauthoryear{Hummer}{Hummer}{1962}]{Hummer1962}
Hummer D.~G.,  1962, \mn@doi [MNRAS] {10.1093/mnras/125.1.21}, 125, 21

\bibitem[\protect\citeauthoryear{{Kakiichi} \& {Gronke}}{{Kakiichi} \&
  {Gronke}}{2019}]{Kakiichi2019}
{Kakiichi} K.,  {Gronke} M.,  2019, arXiv e-prints, \href
  {https://ui.adsabs.harvard.edu/abs/2019arXiv190502480K} {p. arXiv:1905.02480}

\bibitem[\protect\citeauthoryear{{Kimm}, {Blaizot}, {Garel}, {Michel-Dansac},
  {Katz}, {Rosdahl}, {Verhamme}  \& {Haehnelt}}{{Kimm} et~al.}{2019}]{Kimm2019}
{Kimm} T.,  {Blaizot} J.,  {Garel} T.,  {Michel-Dansac} L.,  {Katz} H.,
  {Rosdahl} J.,  {Verhamme} A.,   {Haehnelt} M.,  2019, \mn@doi [\mnras]
  {10.1093/mnras/stz989}, \href
  {https://ui.adsabs.harvard.edu/abs/2019MNRAS.486.2215K} {486, 2215}

\bibitem[\protect\citeauthoryear{Laursen, Razoumov  \& Sommer-Larsen}{Laursen
  et~al.}{2009}]{Laursen2009}
Laursen P.,  Razoumov A.~O.,   Sommer-Larsen J.,  2009, \mn@doi [ApJ]
  {10.1088/0004-637X/696/1/853}, 696, 853

\bibitem[\protect\citeauthoryear{{Laursen}, {Duval}  \& {{\"O}stlin}}{{Laursen}
  et~al.}{2013}]{Laursen2013}
{Laursen} P.,  {Duval} F.,   {{\"O}stlin} G.,  2013, \mn@doi [\apj]
  {10.1088/0004-637X/766/2/124}, \href
  {https://ui.adsabs.harvard.edu/abs/2013ApJ...766..124L} {766, 124}

\bibitem[\protect\citeauthoryear{{Laursen}, {Sommer-Larsen}, {Milvang-Jensen},
  {Fynbo}  \& {Razoumov}}{{Laursen} et~al.}{2019}]{Laursen2019}
{Laursen} P.,  {Sommer-Larsen} J.,  {Milvang-Jensen} B.,  {Fynbo} J. P.~U.,
  {Razoumov} A.~O.,  2019, \mn@doi [\aap] {10.1051/0004-6361/201833645}, \href
  {https://ui.adsabs.harvard.edu/abs/2019A&A...627A..84L} {627, A84}

\bibitem[\protect\citeauthoryear{{Li}, {Gu}, {Yajima}, {Zhu}  \& {Maji}}{{Li}
  et~al.}{2020}]{Li2020}
{Li} Y.,  {Gu} M.~F.,  {Yajima} H.,  {Zhu} Q.,   {Maji} M.,  2020, \mn@doi
  [\mnras] {10.1093/mnras/staa733}, \href
  {https://ui.adsabs.harvard.edu/abs/2020MNRAS.tmp..686L} {}

\bibitem[\protect\citeauthoryear{{Loeb} \& {Rybicki}}{{Loeb} \&
  {Rybicki}}{1999}]{LoebRybicki1999}
{Loeb} A.,  {Rybicki} G.~B.,  1999, \mn@doi [\apj] {10.1086/307844}, \href
  {https://ui.adsabs.harvard.edu/abs/1999ApJ...524..527L} {524, 527}

\bibitem[\protect\citeauthoryear{{Michel-Dansac}, {Blaizot}, {Garel},
  {Verhamme}, {Kimm}  \& {Trebitsch}}{{Michel-Dansac}
  et~al.}{2020}]{Michel-Dansac2020}
{Michel-Dansac} L.,  {Blaizot} J.,  {Garel} T.,  {Verhamme} A.,  {Kimm} T.,
  {Trebitsch} M.,  2020, \mn@doi [\aap] {10.1051/0004-6361/201834961}, \href
  {https://ui.adsabs.harvard.edu/abs/2020A&A...635A.154M} {635, A154}

\bibitem[\protect\citeauthoryear{{Neufeld}}{{Neufeld}}{1990}]{Neufeld1990}
{Neufeld} D.~A.,  1990, \mn@doi [\apj] {10.1086/168375}, \href
  {https://ui.adsabs.harvard.edu/abs/1990ApJ...350..216N} {350, 216}

\bibitem[\protect\citeauthoryear{Osterbrock}{Osterbrock}{1962}]{Osterbrock1962}
Osterbrock D.~E.,  1962, \mn@doi [ApJ] {10.1086/147258}, 135, 195

\bibitem[\protect\citeauthoryear{{Partridge} \& {Peebles}}{{Partridge} \&
  {Peebles}}{1967}]{Partridge1967}
{Partridge} R.~B.,  {Peebles} P.~J.~E.,  1967, \mn@doi [\apj] {10.1086/149079},
  \href {https://ui.adsabs.harvard.edu/abs/1967ApJ...147..868P} {147, 868}

\bibitem[\protect\citeauthoryear{{Remolina-Guti{\'e}rrez} \&
  {Forero-Romero}}{{Remolina-Guti{\'e}rrez} \&
  {Forero-Romero}}{2019}]{Remolina2019}
{Remolina-Guti{\'e}rrez} M.~C.,  {Forero-Romero} J.~E.,  2019, \mn@doi [\mnras]
  {10.1093/mnras/sty3009}, \href
  {https://ui.adsabs.harvard.edu/abs/2019MNRAS.482.4553R} {482, 4553}

\bibitem[\protect\citeauthoryear{Rybicki \& Dell'Antonio}{Rybicki \&
  Dell'Antonio}{1994}]{Rybicki1994}
Rybicki G.~B.,  Dell'Antonio I.~P.,  1994, \mn@doi [ApJ] {10.1086/174170}, 427,
  603

\bibitem[\protect\citeauthoryear{{Rybicki} \& {Lightman}}{{Rybicki} \&
  {Lightman}}{1979}]{RybickiBook}
{Rybicki} G.~B.,  {Lightman} A.~P.,  1979, {Radiative processes in
  astrophysics}.
John Wiley \& Sons, Ltd

\bibitem[\protect\citeauthoryear{{Seon} \& {Kim}}{{Seon} \&
  {Kim}}{2020}]{Seon2020}
{Seon} K.-I.,  {Kim} C.-G.,  2020, arXiv e-prints, \href
  {https://ui.adsabs.harvard.edu/abs/2020arXiv200500238S} {p. arXiv:2005.00238}

\bibitem[\protect\citeauthoryear{{Smith}, {Safranek-Shrader}, {Bromm}  \&
  {Milosavljevi{\'c}}}{{Smith} et~al.}{2015}]{Smith2015}
{Smith} A.,  {Safranek-Shrader} C.,  {Bromm} V.,   {Milosavljevi{\'c}} M.,
  2015, \mn@doi [\mnras] {10.1093/mnras/stv565}, \href
  {https://ui.adsabs.harvard.edu/abs/2015MNRAS.449.4336S} {449, 4336}

\bibitem[\protect\citeauthoryear{{Smith}, {Bromm}  \& {Loeb}}{{Smith}
  et~al.}{2017}]{Smith2017}
{Smith} A.,  {Bromm} V.,   {Loeb} A.,  2017, \mn@doi [\mnras]
  {10.1093/mnras/stw2591}, \href
  {https://ui.adsabs.harvard.edu/abs/2017MNRAS.464.2963S} {464, 2963}

\bibitem[\protect\citeauthoryear{{Smith}, {Tsang}, {Bromm}  \&
  {Milosavljevi{\'c}}}{{Smith} et~al.}{2018}]{Smith2018}
{Smith} A.,  {Tsang} B. T.~H.,  {Bromm} V.,   {Milosavljevi{\'c}} M.,  2018,
  \mn@doi [\mnras] {10.1093/mnras/sty1509}, \href
  {https://ui.adsabs.harvard.edu/abs/2018MNRAS.479.2065S} {479, 2065}

\bibitem[\protect\citeauthoryear{{Smith}, {Ma}, {Bromm}, {Finkelstein},
  {Hopkins}, {Faucher-Gigu{\`e}re}  \& {Kere{\v{s}}}}{{Smith}
  et~al.}{2019}]{Smith2019}
{Smith} A.,  {Ma} X.,  {Bromm} V.,  {Finkelstein} S.~L.,  {Hopkins} P.~F.,
  {Faucher-Gigu{\`e}re} C.-A.,   {Kere{\v{s}}} D.,  2019, \mn@doi [\mnras]
  {10.1093/mnras/sty3483}, \href
  {https://ui.adsabs.harvard.edu/abs/2019MNRAS.484...39S} {484, 39}

\bibitem[\protect\citeauthoryear{Tasitsiomi}{Tasitsiomi}{2006a}]{Tasitsiomi2006}
Tasitsiomi A.,  2006a, \mn@doi [ApJ] {10.1086/504460}, 645, 792

\bibitem[\protect\citeauthoryear{{Tasitsiomi}}{{Tasitsiomi}}{2006b}]{Tasitsiomi2006b}
{Tasitsiomi} A.,  2006b, \mn@doi [\apj] {10.1086/505682}, \href
  {https://ui.adsabs.harvard.edu/abs/2006ApJ...648..762T} {648, 762}

\bibitem[\protect\citeauthoryear{Unno}{Unno}{1952}]{Unno1952}
Unno W.,  1952, PASJ, 4, 100

\bibitem[\protect\citeauthoryear{Verhamme, Dubois, Blaizot, Garel, Bacon,
  Devriendt, Guiderdoni  \& Slyz}{Verhamme et~al.}{2012}]{Verhamme2012}
Verhamme A.,  Dubois Y.,  Blaizot J.,  Garel T.,  Bacon R.,  Devriendt J.,
  Guiderdoni B.,   Slyz A.,  2012, \mn@doi [A\&A]
  {10.1051/0004-6361/201218783}, 546, A111

\bibitem[\protect\citeauthoryear{{Yusef-Zadeh}, {Morris}  \&
  {White}}{{Yusef-Zadeh} et~al.}{1984}]{Yusef-Zadeh1984}
{Yusef-Zadeh} F.,  {Morris} M.,   {White} R.~L.,  1984, \mn@doi [\apj]
  {10.1086/161780}, \href
  {https://ui.adsabs.harvard.edu/abs/1984ApJ...278..186Y} {278, 186}

\bibitem[\protect\citeauthoryear{Zheng \& Miralda-Escud\'{e}}{Zheng \&
  Miralda-Escud\'{e}}{2002}]{Zheng2002}
Zheng Z.,  Miralda-Escud\'{e} J.,  2002, \mn@doi [ApJ] {10.1086/342400}, 578,
  33

\makeatother
\end{thebibliography}

\appendix

\section{Boundary Conditions}
\label{app:derivation_of_f}
Here we discuss the unknown factor $f$ that appears in the boundary conditions of equation~(\ref{eq:sec2_boundary condition}). Following \citet{RybickiBook}, we employ the two-stream approximation which describes both free-streaming and diffusion in one-dimensional geometries to reasonable accuracy. It is assumed that the entire radiation field can be represented by rays traveling at two angles, $\mu = \cos \theta = \pm 1/\sqrt{3}$, and we denote the outward an inward intensities as $I_\nu^+ \equiv I_\nu(\mu = +1/\sqrt{3})$ and $I_\nu^- \equiv I_\nu(\mu = -1/\sqrt{3})$. Therefore, in terms of $I^+$ and $I^-$ the relevant angular moments of the specific intensity become
\begin{equation}
  J_\nu \equiv \frac{1}{2} \int_{-1}^1 I_\nu\,\text{d}\mu = \frac{1}{2} \left(I_\nu^+ + I_\nu^-\right)
\end{equation}
and
\begin{equation}
  H_\nu \equiv \frac{1}{2} \int_{-1}^1 \mu^2 I_\nu\,\text{d}\mu = \frac{1}{6} \left(I_\nu^+ + I_\nu^-\right) = \frac{1}{3} J_\nu \, .
\end{equation}
Using Fick's law from equation~(\ref{eq:sec2_Ficks-Law}), under the diffusion approximation in slab and spherical geometries we have
\begin{equation}
  H_\nu = -\frac{1}{3 k_\nu} \frac{\partial J_\nu}{\partial r} = -\frac{1}{3} \frac{\partial J_\nu}{\partial \tau_\nu} \, .
\end{equation}
where $\tau_\nu$ represents the optical depth. After substitution we have
\begin{equation}
  I^+_\nu = J_\nu - \frac{1}{\sqrt{3}} \frac{\partial J_\nu}{\partial \tau_\nu} \qquad \text{and} \qquad
  I^-_\nu = J_\nu + \frac{1}{\sqrt{3}} \frac{\partial J_\nu}{\partial \tau_\nu} \, .
\end{equation}
If the medium extends over $r \in [0, R]$ and there is no incident radiation, then $I^-_\nu|_{r=R} = 0$, which gives the boundary condition
\begin{equation}
  \left[ \frac{\partial J_\nu}{\partial \tau_\nu} + \sqrt{3} J_\nu \right]_{r=R} = 0 \, .
\end{equation}
Thus, in the two-stream approximation we have $f = \sqrt{3}$. Other methods for obtaining boundary conditions result in equations of the same form but with different values of $f$, e.g. the `asymptotic diffusion limit' \citep{Habetler1975}. In fact, they convey the same physics and result in equivalent solutions. Therefore, in this paper we propagate $f$ throughout our derivations. In principle, one might consider calibrating the value as a function of system parameters (e.g. $a\tau_0$) by comparing to numerical solutions.

\section{Errors from approximations}
\label{app:errors}
\begin{figure}
  \centering
  \includegraphics[width=\columnwidth]{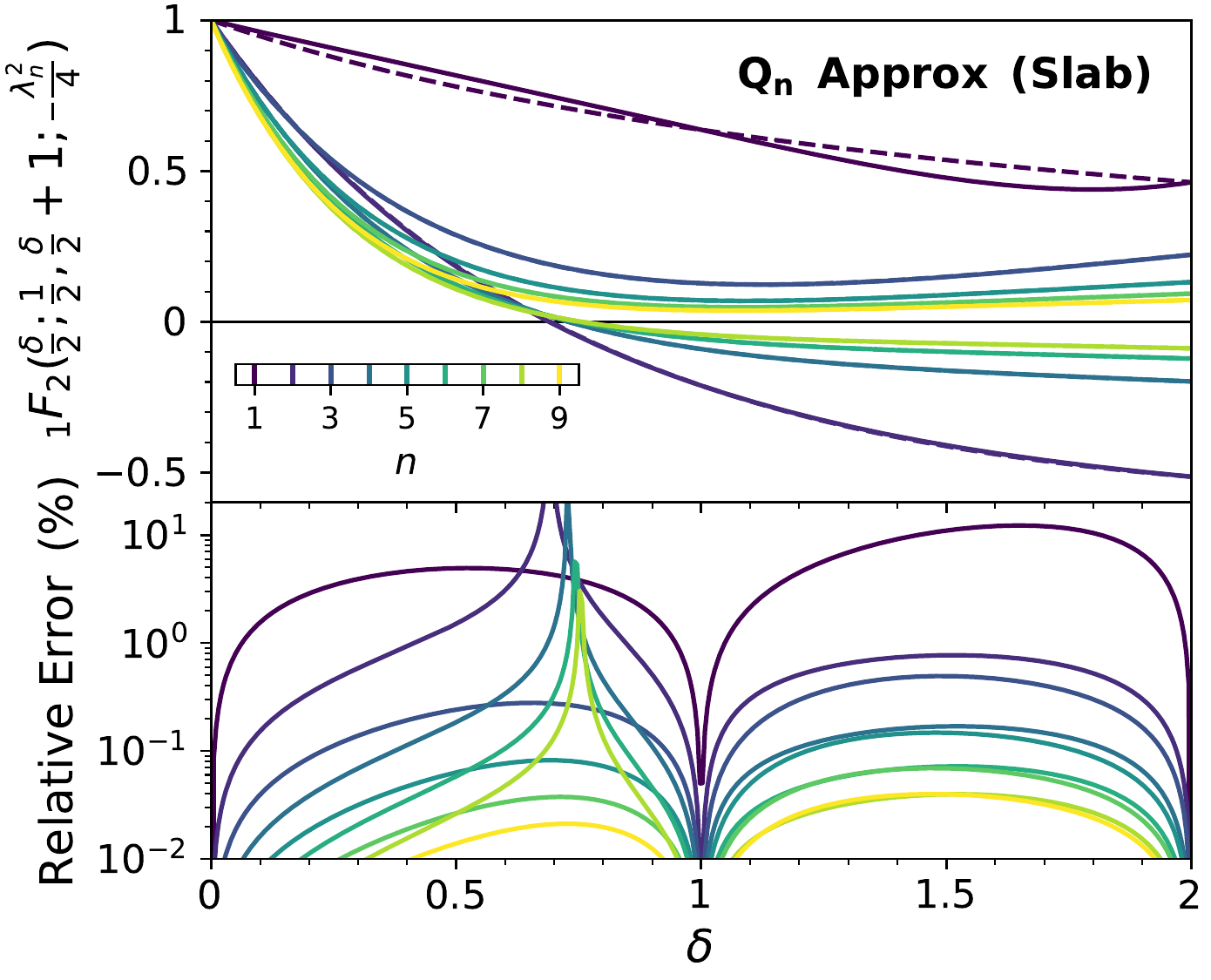}
  \caption{The error from successive terms introduced by the approximation in equation~(\ref{eq:sec4_Qn_generalization_approximation}) for the hypergeometric function in equation~(\ref{eq:sec4_Qn_generalization}). In the upper panel, the dashed line is the accurate value and the solid line is the approximation. As shown in the plot, the first term contributes most of the error. The approximation is highly accurate when $\delta$ is 0 (point source), 1 (uniform source), or 2 (linear source), but becomes quite poor for $\delta > 2$.}
  \label{fig:appendix_slab_error}
\end{figure}

\begin{figure}
  \centering
  \includegraphics[width=\columnwidth]{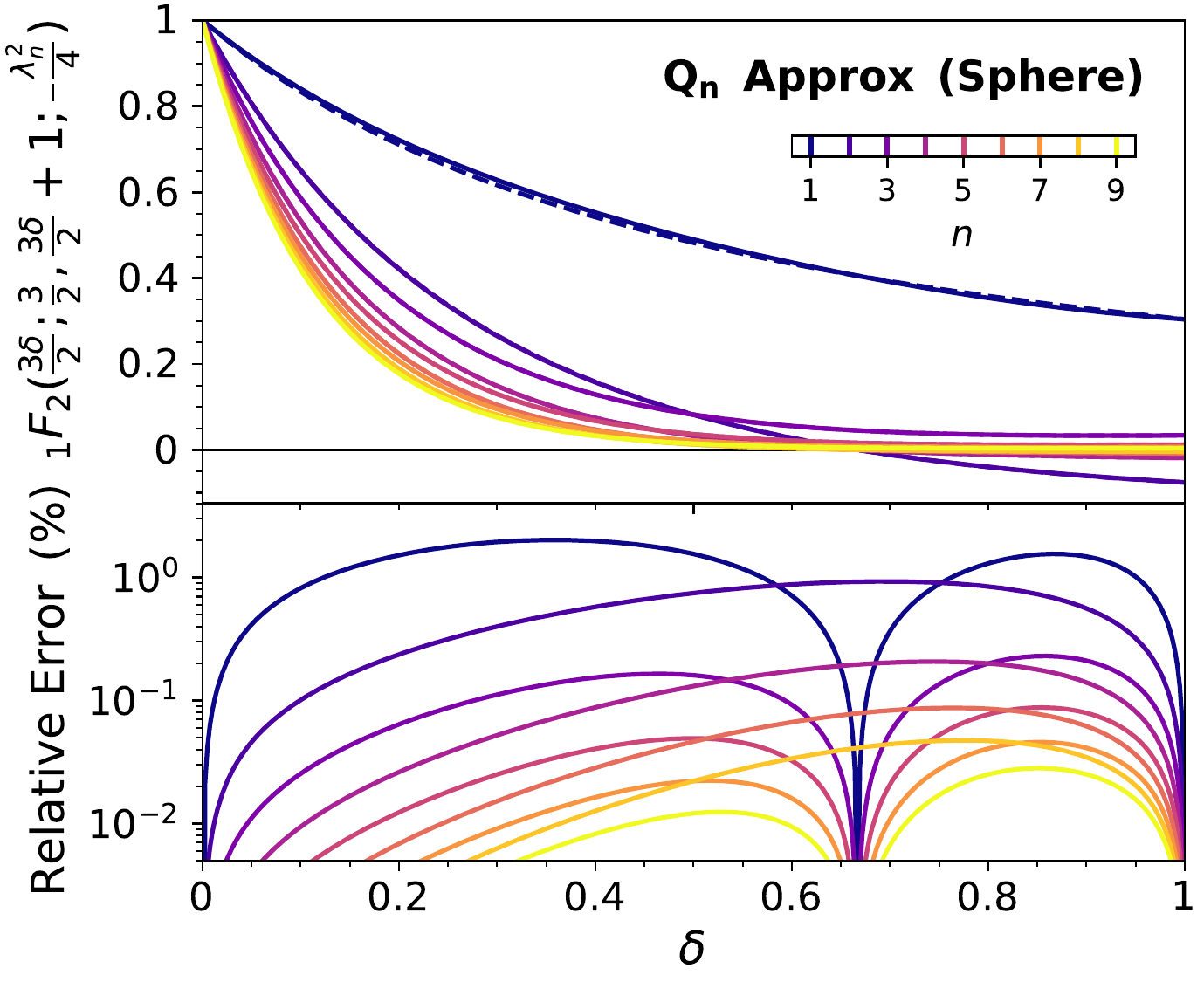}
  \caption{The error from successive terms introduced by the approximation in equation~(\ref{eq:sec5_approx_1F2}) for the hypergeometric function in equation~(\ref{eq:sec5_1F2}). In the upper panel, the dashed line is the accurate value and the solid line is the approximation. As shown in the plot, the first term contributes most of the error. The approximation is highly accurate when $\delta \in [0,1]$, and is exact when $\delta$ is $0$ (point source), $2/3$ (when $\eta \propto 1/r$), or 1 (uniform source).}
  \label{fig:appendix_sphere_error}
\end{figure}

Throughout this work we introduced a number of approximations in the analytic derivations. These were discussed in detail in the text where they were introduced. However, for clarity we now illustrate the errors introduced by the approximations for a power-law emissivity profile in slab and spherical geometries in Figures~\ref{fig:appendix_slab_error} and \ref{fig:appendix_sphere_error}, respectively.

\section{Solutions suitable for all Power}
\label{app:solution_suitable_for_all_beta}
\begin{figure}
    \centering
    \includegraphics[width=\columnwidth]{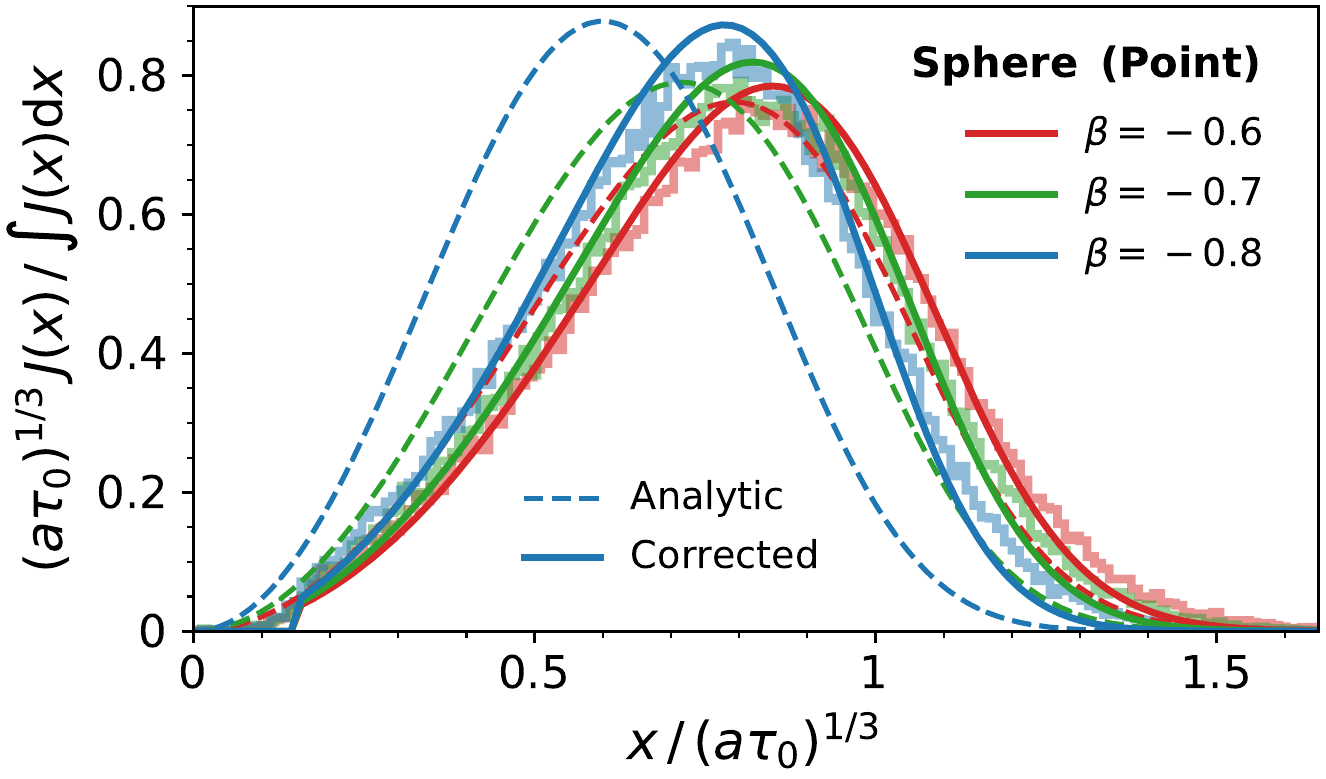}
    \caption{Numerical verification of the correction procedure for density profiles with steep power-law slopes $\beta \lesssim -1/2$. The simulation setup is the same as before but we choose $\beta = \{-0.6, -0.7, -0.8\}$ to test the analytic theory. The dashed lines are from equation~(\ref{eq:sec6_analytic_sol_big_beta}), which is inaccurate in these extreme cases. The solid curves are from the corrected solution from equation~(\ref{eq:sec6_central_point_source_corrected_sol}) with $m = 10^4$ to ensure accuracy. We note that we also have removed an artificial singularity at $x = 0$ to reproduce the proper normalization. As the plot demonstrates, the corrected solution fits the simulated data calculated with the GMCRT method (shown as histograms).}
    \label{fig:sec6_power_law_low_beta}
\end{figure}

In Section~\ref{sec:Power_law_sphere_central_point_source} we provided an approximate form of the point source solution in spherical geometry within a power-law density profile. However, when $\beta \rightarrow -1$ the centre-to-edge optical depth starts to diverge and the asymptotic form of the Bessel function in equation~(\ref{eq:sec6_Qn_exact}) is no longer accurate. The error arises from the condition $\lambda_n \gg \kappa(\kappa-2)/4$, which is is no longer satisfied when $\kappa$ is large. The full expression of the spectral line outside the sphere is
\begin{equation}
  J(\tilde{x}) = \frac{\mathcal{L}\sqrt{6}}{2^{\gamma+3} \pi^2 R^2 \Gamma(\gamma) f \tau_0 H(\tilde{x}) } \sum_{n=1}^{\infty}\frac{e^{-\lambda_n |\tilde{x}| }\lambda_n^{\gamma-1}}{J_{\gamma}(\lambda_n)} \, .
\end{equation}
One way to improve the accuracy of the solution is to retain the first few problematic terms and derive a closed form solution for the remaining terms. Let us set the first $m-1$ terms to the exact values, and then use the approximations in equation~(\ref{eq:sec6_eq_of_approx_Bessel})~and~(\ref{eq:sec6_power_law_eigenvalue}) for which $\lambda_n$ is sufficiently large to provide reasonable accuracy. Thus, the expression reduces to
\begin{align} \label{eq:sec6_central_point_source_corrected_sol}
  J(\tilde{x}) = &\frac{\mathcal{L}\sqrt{6}}{2^{\gamma+3} \pi^2 R^2 \Gamma(\gamma) f \tau_0 H(\tilde{x}) }\left[\sum_{n=1}^{m-1}\frac{e^{-\lambda_n |\tilde{x}| }\lambda_n^{\gamma-1}}{J_{\gamma}(\lambda_n)} +\right. \notag \\
  &\left. \sqrt{\frac{\pi}{2}} (-1)^{m-1}\pi^{\kappa/2}\frac{\Phi\left(-e^{-\pi |\tilde{x}|},-\frac{\kappa}{2},m-\frac{1}{2}+\frac{\kappa}{4}\right)}{e^{(m-1/2+\kappa/4)\pi |\tilde{x}|}}  \right] \, .
\end{align}
In this case we also need to find the first $m-1$ eigenvalues using the exact condition that $J_{\gamma-1}(\lambda_n) = 0$ instead of the approximate eigenvalues in equation~(\ref{eq:sec6_power_law_eigenvalue}). A similar approach may be used for the trapping time or other properties.
In Figure~\ref{fig:sec6_power_law_low_beta} we provide numerical verification that this correction procedure extends the range of validity even with extreme power-law slopes approaching $\beta = -1$. However, there is still a detailed problem with equation~(\ref{eq:sec6_central_point_source_corrected_sol}) as the solution does not converge at $\tilde{x} = 0$. The severe oscillations introduce a numerical singularity that produces errors in the normalization. For this reason, we choose $m$ to be high enough that the remainder is small and then ignore the result in a small neighborhood around zero. We then renormalize the profile for an accurate analytic solution that matches the results from simulations, as shown in Figure~(\ref{fig:sec6_power_law_low_beta}).

\bsp 
\label{lastpage}
\end{document}